\documentclass[prd, aps, reprint,preprintnumbers, nofootinbib, floatfix]{revtex4-1}

\usepackage{mathrsfs}

\usepackage{amssymb}
\usepackage{amsmath}
\usepackage[dvipsnames]{xcolor}
\usepackage[colorlinks=true,linkcolor=blue,citecolor=blue, urlcolor=blue]{hyperref}   
\usepackage{graphicx}
\usepackage[utf8]{inputenc}
\usepackage{slashed}
\usepackage{bm}
\usepackage{soul}
\usepackage{upgreek}
\newcommand{\kompost}{K{\o}MP{\o}ST}
\usepackage[shortlabels]{enumitem}
\definecolor{uibred}{RGB}{167, 38, 47}
\def\Eq#1{Eq.~(\ref{#1})}
\def\Eqs#1{Eqs.~(\ref{#1})}
\def\eq#1{(\ref{#1})}

\def\Fig#1{Fig.~\ref{#1}}

\def\Sec#1{Sec.~\ref{#1}}
\def\Ref#1{Ref.~\cite{#1}}

\def\p{\mathbf{p}}
\newcommand{\e}{\mathcal{E}}
\def\x{\mathbf{x}} 
\newcommand{\xt}{\mathbf{x}}

\def\k{\mathbf{k}}

\def\bm{\mathbf}

\def\p{\bm{p}}
\def\x{\bm{x}}
\def\wtilde{\tilde{w}}

\newcommand{\TBg}{\overline{T}}

\newcommand{\tauekt}{\tau_\textsc{ekt}}
\newcommand{\tauhydro}{\tau_\text{hydro}}

\begin{document}
\title{QCD thermalization: \emph{Ab initio} approaches and interdisciplinary connections}
\author{J\"{u}rgen Berges}
\email[]{berges@thphys.uni-heidelberg.de}
\affiliation{Institute for Theoretical Physics, Heidelberg University, Philosophenweg 16, D-69120 Heidelberg, Germany}
\preprint{CERN-TH-2020-080}

\author{Michal P. Heller}
\email{michal.p.heller@aei.mpg.de}
\altaffiliation{\emph{On leave from:} National Centre for Nuclear Research,
Pasteura 7, Warsaw, PL-02093, Poland}
\affiliation{Max Planck Institute for Gravitational Physics\\ (Albert Einstein Institute),\\ Am M{\"u}hlenberg 1, D-14476 Potsdam, Germany}

\author{Aleksas Mazeliauskas}
\email[]{aleksas.mazeliauskas@cern.ch}
\affiliation{Theoretical Physics Department, CERN, CH-1211 Geneva 23, Switzerland}

\author{Raju Venugopalan}
\email[]{rajuv@bnl.gov}
\affiliation{Physics Department, Brookhaven National Laboratory, Upton, New York 11973-5000, USA}
\begin{abstract}
Heavy-ion collisions at BNL's Relativistic Heavy Ion Collider and CERN's Large Hadron Collider provide strong evidence for the formation of a quark-gluon plasma, with temperatures extracted from 
relativistic viscous hydrodynamic simulations shown to be well above the transition temperature from hadron matter.  
Outstanding problems in QCD include how the strongly correlated quark-gluon matter forms in a heavy-ion collision, its properties off-equilibrium,  and the thermalization process in the plasma.
 We review here the theoretical progress in this field in weak coupling QCD effective field theories and in strong coupling holographic approaches based on gauge-gravity duality. We outline the interdisciplinary connections of different stages of the thermalization process to non-equilibrium dynamics in other systems across energy scales ranging from inflationary cosmology, to strong field QED, to ultracold atomic gases, with emphasis on the universal dynamics of non-thermal and  hydrodynamic attractors.  We survey measurements in heavy-ion collisions that are sensitive to the early non-equilibrium stages of the collision and discuss the potential for future measurements. We summarize the current state-of-the art in thermalization studies and identify promising avenues for further progress. 
\date{\today}

\end{abstract}

\maketitle
\onecolumngrid
\vspace{-1.0cm}
\begin{center}
\includegraphics[width=0.95\linewidth]{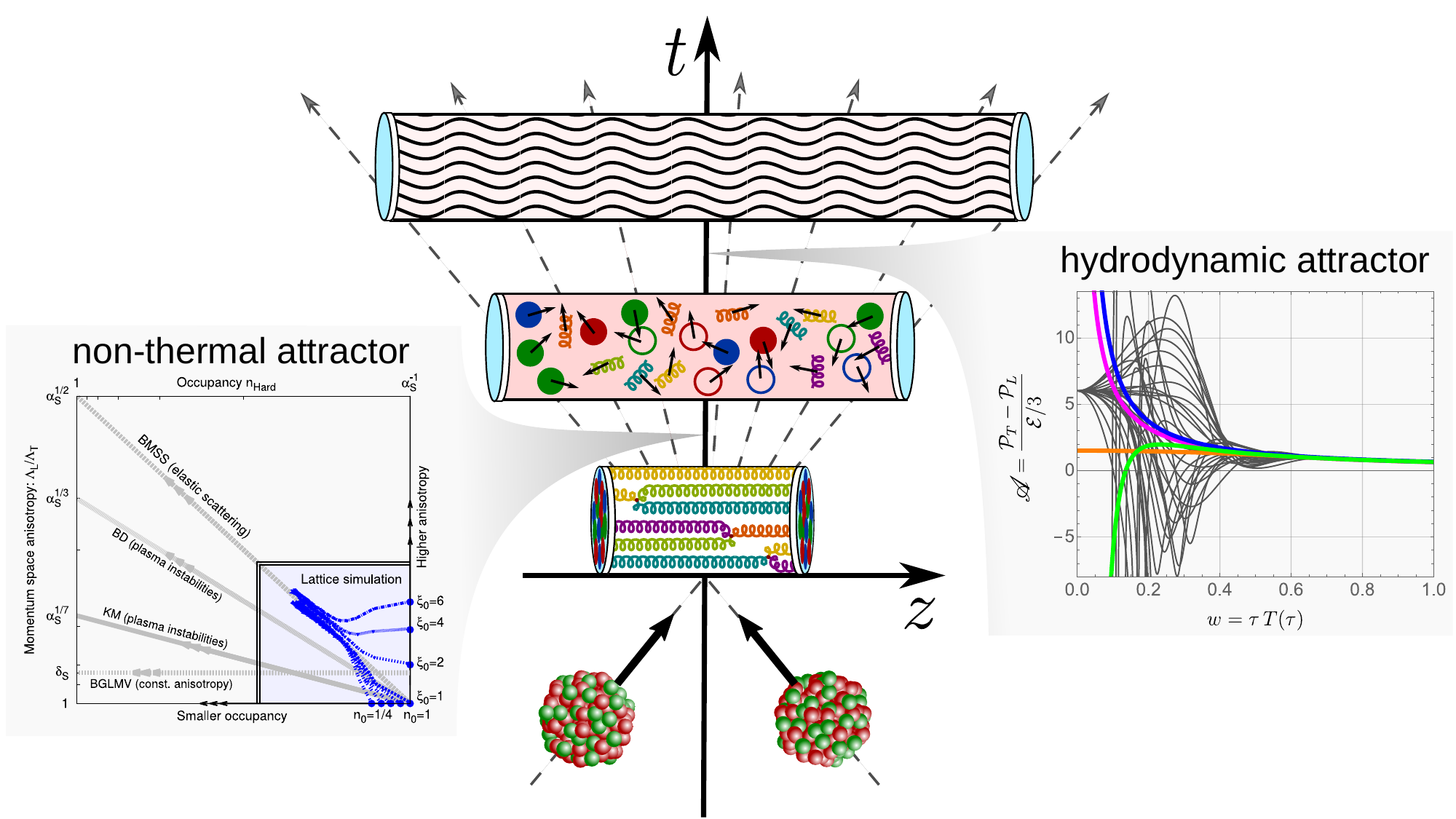}
\end{center}
\vspace{-1.0cm}
\twocolumngrid
\clearpage
\tableofcontents

\section{Big picture questions and outline of the review\label{sec:introduction}}

Ultrarelativistic collisions of heavy nuclei at the BNL Relativistic Heavy Ion Collider (RHIC) and the CERN Large Hadron Collider (LHC) produce several thousand particles in 
each event, generating the hottest and densest matter on Earth~\cite{Arsene:2004fa,Adcox:2004mh,Alver:2006wh,Adams:2005dq, Muller:2012zq, Roland:2014jsa,Foka:2016zdb,Foka:2016vta}. At the highest LHC energies, temperatures of the order of five trillion Kelvin are attained~\cite{Adam:2015lda}. Temperatures on this scale previously existed only at the earliest instants of our Universe, a 10th of a microsecond after the Big Bang. Lattice gauge theory studies~\cite{Bazavov:2018mes} show strongly interacting matter at these temperatures to be well over a crossover temperature from hadron matter to a regime where the degrees of freedom describing bulk thermodynamic quantities are the fundamental quark and gluon fields of Quantum Chromodynamics (QCD). The results of experimental and theoretical studies indicate that shortly after the heavy-ion collision, the produced quark-gluon fields form a strongly correlated state of matter, widely known as the quark-gluon plasma (QGP)~\cite{Shuryak:1980tp}. 

The heavy-ion experiments at RHIC and LHC therefore provide us with a unique opportunity to study terrestrially the spacetime evolution of this non-Abelian QGP. A striking finding from the RHIC  and LHC experiments is that the experimental data are consistent with a description of the QGP as a nearly perfect fluid with a very low value of 
shear viscosity to entropy density ratio of $\upeta/s\leq 0.2$ (in natural units)~\cite{Romatschke:2017ejr}. These values are very close to  $\upeta/s= 1/(4\pi)$, a universal property of a class of gauge theories with a large number of degrees of freedom at infinite coupling~\cite{Policastro:2001yc,Buchel:2003tz,Kovtun:2004de,Iqbal:2008by} that is described in terms of a dual gravity picture~\cite{Maldacena:1997re}.

While our understanding of the thermal properties of QGP matter and the flow of the nearly perfect fluid
has developed significantly, progress in theoretical descriptions of the early stages of heavy-ion collisions has been made relatively recently.  In particular, there is a growing realization that the far-from-equilibrium dynamics that characterizes early time physics is extremely important for understanding collective phenomena in the heavy-ion experiments~\cite{Mrowczynski:2016etf,Busza:2018rrf,Schlichting:2019abc}. This review summarizes our perspective on the theoretical and phenomenological progress in this active research area and places these developments in a wider interdisciplinary context.

The QCD thermalization process represents an initial value problem in quantum field theory (QFT). It requires understanding the many-body correlations in the colliding hadrons,  how such correlations influence multi-particle production as the collision occurs, and  the subsequent effective loss of information of these many-body correlations during the thermalization process of the matter produced. 
While von Neumann entropy is conserved in the unitary quantum evolution of a nuclear collision in isolation, observables of interest may nevertheless approach (local) thermal equilibrium.
The characteristic time scales for the corresponding effective loss of information and the extent to which the dynamics finally leads to an approach to (local) thermal equilibrium for key observables are the central topics of this review.

In particular, we will focus on the following key questions prompted by the dynamics of each stage of the spacetime evolution of quark-gluon matter in heavy-ion collisions\footnote{For a complementary perspective on open questions in heavy-ion collisions, we refer the reader to Ref.~\cite{Busza:2018rrf}.}:
\begin{itemize}
\item{\it What are the many-body correlations of strongly interacting matter in the colliding nuclei?}  

The colliding nuclei produce the initial state for the subsequent thermalization process. In principle, there can be different thermalization scenarios for different initial conditions. Although many details of the quantum evolution are lost quickly, it is crucial to classify the range of initial conditions (such as underoccupied versus\ overoccupied) leading to a certain class of dynamical processes.

In QCD, a proton (or any other nucleus) must be viewed as a collection of short or long lived configurations of 
partons (quarks, antiquarks, and gluons), where each configuration carries the quantum numbers of the proton. 
When the proton or nucleus is boosted to high energies,  short lived configurations typically containing large numbers of partons live much longer due to time dilation. It is therefore more likely that a probe of the hadron at high energies will scatter off such many-body configurations of partons and that their decay will dominate the physics of multi-particle production in ultrarelativistic nuclear collisions. Learning how precisely multi-particle production occurs requires a deep knowledge of the spatial and momentum distributions of partons in the boosted nuclei, the nature of their correlations, and how these correlations change with system size and with collision energy.

\item{\it What is the physics of the first yoctosecond  $(10^{-24}\,\text{seconds})$ of the collision?}

Parton configurations in a boosted nucleus have their momenta distributed between a few fast modes and more plentiful soft modes. In a heavy-ion collision, these fast modes in each of the  nuclei interact relatively weakly with the other nucleus and populate the ``fragmentation regions" corresponding to 
polar angles very close to the beam axes~\cite{VanHove:1974wa}.  The slower degrees of freedom interact more strongly with each other and produce strongly interacting gluon matter outside the fragmentation regions. 

This spacetime picture of nuclear collisions was developed in a groundbreaking paper by Bjorken to describe the subsequent hydrodynamic flow of the quark-gluon plasma~\cite{Bjorken:1982qr}, albeit he did not address 
how thermalization occurs in this scenario. An interesting question in this regard is whether the strong interactions of the soft modes with each other are due to strong coupling or whether they may be due to the large occupancy of these soft modes. The answer to this question may also influence the degree of transparency of the fast modes, in particular a ``limiting fragmentation" scaling phenomenon seen in data. 

A spacetime scenario in which both soft and hard modes in the nuclei interact very strongly and generate hydrodynamic flow was suggested by Landau. It is conceivable that there is a transition between these two spacetime pictures with energy~\cite{Gelis:2006tb,Casalderrey-Solana:2013aba}; if so, can they be distinguished by phenomena such as limiting fragmentation \cite{Goncalves:2019uod}?

\item{\it Is there a unifying theoretical description of quark-gluon matter off-equilibrium?}

The quark-gluon matter formed in the first few yoctoseconds of the heavy-ion collision is very far from equilibrium. A key question in its description is whether 
weak and strong coupling extrapolations to realistic values can lead to similar phenomenology.

A potentially rich line of inquiry is to isolate which features of the non-equilibrium evolution of strongly correlated or coupled quark-gluon matter are universal. One example is universal dynamics in the approach to local thermal equilibrium governed by viscous hydrodynamics.
Another example is universality in time dependence across a class of non-equilibrium states for certain observables. In a weak coupling scenario, at high occupancies, these include far-from-equilibrium attractors associated with non-thermal fixed points~\cite{Berges:2008wm,Berges:2013eia,Berges:2013fga}. 

Far-from-equilibrium hydrodynamic attractors are observed to emerge in both strong and weak coupling~\cite{Heller:2015dha,Romatschke:2017vte}. A related important set of questions concerns the use of effective theories like hydrodynamics for systems that are far away from equilibrium. Yet another line of inquiry is to determine how features of the dynamics evolve between the weakly coupled and strongly coupled regimes. An intriguing possibility to consider is whether the topological properties of strongly correlated systems may help provide unifying descriptions at both weak and strong coupling.

\item{\it Can we cleanly isolate signatures of quark-gluon matter off-equilibrium?}

If matter in bulk locally equilibrates in heavy-ion collisions, the only information of the non-equilibrium evolution that survives is what is imprinted as initial conditions 
for its subsequent hydrodynamic evolution. The exceptions are electroweak and so-called ``hard probes"; both of these are sensitive to the full history of the spacetime evolution of QCD matter. 

A significant development in recent years is the vastly improved ability of the RHIC and LHC experiments to perform ``event engineering" whereby final states can be studied by varying the ``control parameters" corresponding to nuclear size, centrality of collision impact, and final state multiplicities (triggered thereby on typical versus rare event configurations) across wide ranges in energy and system size~\cite{Schukraft:2012ah}. A challenging question is whether we can constrain the current state-of-the-art computational techniques to accurately reflect the systematics of this event engineering and, further, to use these to isolate empirically the out-of-equilibrium dynamics.

\item {\it Interdisciplinary connections}

The study of the out-of-equilibrium dynamics of strongly correlated systems is an important topic of significant contemporary interest in a number of sub-fields of physics. As we later discuss, the ideas and methods outlined in this review have significant overlap with these fields. Can one exploit these interdisciplinary connections to make progress? 

\end{itemize}

We will address the previously listed outstanding questions in two {\it ab initio} theoretical approaches to the problem of thermalization in QCD. One approach, the Color Glass Condensate Effective Field Theory (CGC EFT), is applicable at very high energies corresponding to a  regime of very weak coupling $\alpha_S \ll 1$ and  very high gluon occupancies $f_g$ satisfying $\alpha_S f_g\sim 1$. This regime of weak coupling and high occupancies in QCD is characterized by a large emergent ``saturation" scale that is much larger than the intrinsic non-perturbative scales corresponding to color confinement and chiral symmetry breaking. 

The CGC EFT employs weak coupling many-body methods to separate (or factorize) these soft 
non-perturbative modes from the harder modes of the order of the saturation scale. Specifically, the requirement that physics be independent of the scale separation between soft and hard modes leads to renormalization group equations that describe how such non-perturbative information provided as an input at a given energy scale changes as it evolves. As one approaches asymptotic energies, the factorization of the hard and soft scales becomes increasingly robust and many of the properties of quark-gluon matter can be computed systematically. The quark-gluon matter in this limit is called the Glasma~\cite{Lappi:2006fp,Gelis:2006dv}. 

The other {\it ab initio} approach to thermalization is in the limit of strong ’t Hooft coupling of $\alpha_S N_c\rightarrow \infty$, as the number of colors $N_c\rightarrow \infty$. In this limit, holographic approaches based on gauge-gravity duality~\cite{Maldacena:1997re,Gubser:1998bc,Witten:1998qj} are robust and can be used to obtain exact results in non-Abelian gauge theories, with the 
 best understood example being  ${\cal N} = 4$ superconformal Yang-Mills theory.

 Neither of these theoretical approaches to the problem of thermalization are directly applicable to real world heavy-ion collisions at RHIC and LHC energies, where the relevant couplings are likely neither particularly weak nor infinitely strong. Thus data-theory comparisons  rely on phenomenological descriptions characterized by extrapolations of {\it ab initio} approaches well beyond their strict regimes of validity. By anchoring such phenomenological models in fundamental theory in well controlled limits, their success or failure in comparisons to data can then be traced to a particular set of assumptions in the extrapolations.  We will clarify throughout the review whenever such phenomenological extrapolations are made. 

We will begin in Section~\ref{sec:hadrons} by discussing the structure of matter within the colliding hadrons and heavy nuclei at high energies. After a brief introduction to QCD and the associated parton picture of hadrons at high energies, we will focus our attention on what happens when the phase space density of partons in the wavefunctions of the colliding hadrons becomes large. Driving this physics is  an emergent energy-dependent close packing ``saturation" scale $Q_S$~\cite{Gribov:1984tu}, which grows with energy and nuclear size, allowing for a systematic weak coupling description of the properties of saturated partons in high energy QCD. 
Specifically, we will discuss 
the CGC EFT, wherein the high energy hadron is described as a coherent state of static color sources and dynamical gluon fields. The saturation scale is manifest in the CGC EFT, allowing one to describe strongly correlated many-body parton correlations in the hadron wavefunctions~\cite{Gelis:2010nm,Kovchegov:2012mbw}. 

Non-perturbative soft modes of the high energy nuclei, their color charge distributions, and many-body correlations thereof, are represented by a density matrix at a given energy scale that is much smaller than those of the hard weakly coupled modes. While this non-perturbative density matrix has to be parametrized at the initial scale by physically plausible assumptions, 
a renormalization group (RG) framework~\cite{JalilianMarian:1997dw,Iancu:2000hn} allows one to study systematically the energy evolution of parton many-body correlations  as the hadron is boosted to higher energies. 

In Section~\ref{sec:Glasma}, we will outline the problem of multi-particle production in quantum field theory in the presence of strong fields and discuss how this leads to a first principles description of the very early time evolution of 
the Glasma.
Inclusive quantities such as multiplicities or energy densities, and their spacetime correlations, can be computed systematically in the Glasma in powers of the coupling $\alpha_S\ll 1$ at sufficiently high energy. At leading order in this power counting, the Glasma fields are highly occupied classical fields, with magnitude $1/\alpha_S$. 

At next-to-leading order (NLO), we discuss how quantum fluctuations, co-moving with the colliding nuclei, can be absorbed into the density matrices describing their non-perturbative many-body color charge distributions. In contrast, non-comoving quantum fluctuations produced after the collision 
 in the Glasma are unstable and display quasi-exponential dynamical growth~\cite{Romatschke:2005pm}. We later describe how the physics of these unstable modes at very early proper times $\tau\lesssim \frac{1}{Q_S}\log^2(1/\alpha_S)$ is captured in a classical-statistical approximation of the quantum evolution with given quantum initial conditions.

Section~\ref{sec:classicalstatistical} describes the non-linear time evolution of far-from-equilibrium quark gluon matter for weak couplings relevant at very high energies. The range of validity of classical-statistical field theory descriptions for the evolution is discussed in terms of the two-particle-irreducible (2PI) quantum effective action, which motivates fully 3+1-dimensional numerical simulations of the expanding Glasma fields. 

The lattice field theory simulations demonstrate the emergence of a non-thermal attractor described by a self-similar gluon distribution, whose dependence on momentum, and an overall cooling rate, are characterized by universal numbers independent of the initial conditions. Because the numerical simulations correctly describe dynamics in the infrared, the attractor solution helps one to identify the right effective kinetic theory among several competing options. 

Kinetic theory increasingly captures the relevant dynamics of the thermalization process as the system expands and cools. In Section~\ref{sec:kinetictheory}, we discuss the leading order kinetic theory framework, going progressively from elastic $2\leftrightarrow 2$ scatterings to effective collinear $1\leftrightarrow 2$ processes and taking special note of interference and plasma instability effects. 
Phenomenological extrapolations to realistic couplings can also be explored in the language of hydrodynamic attractors, where 
the dependence on the coupling is replaced by the kinematic viscosity $\upeta/s$. For values of the kinematic viscosity extracted from hydrodynamic simulations of heavy-ion collision, reasonable predictions are obtained for entropy production~\cite{Giacalone:2019ldn}, as well as for hydrodynamic and chemical equilibration times~\cite{Kurkela:2018xxd,Kurkela:2018wud}.

In~Section~\ref{sec:strongcoupling}, we provide an overview of holography based  strong coupling approaches to thermalization in gauge theories. Our focus is on the conceptual features, universal mechanisms, and predictions from these studies.
In particular, {\it ab initio} holographic computations predict the applicability of hydrodynamics over a time scale set by the local energy density, when the expanding matter in heavy-ion collisions settings is characterized by a large spatial anisotropy in its energy-momentum tensor~\cite{Chesler:2009cy,Chesler:2010bi,Heller:2011ju}. This is at variance with the common presumption of local thermal equilibrium in applying  hydrodynamics; in a paradigm shift, the transition to hydrodynamic flow is now referred to as hydrodynamization rather than thermalization~\cite{CasalderreySolana:2011us}.

We will discuss, in particular, phenomenological attempts to apply these ideas to 
model heavy-ion collisions in the context of (1+1)-dimensional  boost invariant flow 
where hydrodynamization and hydrodynamic attractors were first discovered. We will also cover work on more realistic holographic
descriptions of heavy-ion collisions that model confinement, the breaking of conformal invariance, the running of the coupling, and large $N_c$ suppressed non-local correlations. 

Section~\ref{sec:signatures} is devoted to a discussion of signatures of non-equilibrium dynamics in heavy-ion data. While electromagnetic and high transverse momentum strongly interacting final states are sensitive to early time dynamics,  significant contributions to their rates accrue from all stages of the spacetime evolution of the system. Measurements of long range correlations among high momentum final states offer promise in isolating the early time non-equilibrium dynamics of the Glasma from the late stage hydrodynamic flow. This can be achieved by  ``event engineering" the response of these final states to variations in energy and system size. We will also discuss how bulk observables, in combination with these final states, can constrain thermalization scenarios. 

A striking example of the role of topology in heavy-ion collisions is the Chiral Magnetic Effect (CME)~\cite{Kharzeev:2007jp} corresponding to a vector current along the direction of an external magnetic field that is induced by topological transitions. The CME is primarily an early time effect; in this case as well, event engineering of multi-particle correlations offers the possibility of uncovering its role.

In Section~\ref{sec:interdisciplinary}, we will address the question of the interdisciplinary connections of the thermalization process in heavy-ion collisions to that of other strongly correlated systems across energy scales. A striking similarity of strongly correlated flow in heavy-ion collisions to that of unitary Fermi gases was already noted shortly after the discovery of the QGP perfect fluid. 
The Glasma likewise shares common features with other overoccupied systems across energy scales, from inflationary dynamics in the early Universe to a quantum portrait of black holes as highly occupied graviton states to those of overoccupied ultracold Bose gases. 

A concrete example of the influence of interdisciplinary ideas is that of the turbulent thermalization process underlying the non-thermal attractor in the Glasma, which has been widely discussed in the context of reheating in the early Universe following inflation~\cite{Micha:2004bv,Berges:2008wm}. The latter in turn is, in the perturbative high-momentum regime, a relativistic generalization of weak wave turbulence in fluids~\cite{zakharov2012kolmogorov}. In the non-perturbative infrared regime, the Glasma attractor is nearly identical to that of overoccupied cold atomic gases, sharing the same scaling functions and exponents in a wide spectral range~\cite{Berges:2014bba}. This is suggestive of a classification of far-from-equilibrium systems into universality classes analogous to those for critical phenomena close to equilibrium~\cite{Hohenberg:1977ym}. An exciting development with cross-disciplinary potential is the use of state-of-the-art cold atom experiments to provide deep insight into such universal dynamics~\cite{Prufer:2018hto,Erne:2018gmz,glidden2020bidirectional}. 

The search for effective theories far from equilibrium is also a major research direction in the theory of complex systems ranging from 
understanding entanglement to information loss and thermalization of closed quantum many-body systems, with insights to be gained from ``tabletop" atomic and condensed matter systems~\cite{Eisert:2014jea}. On the other end of the energy scale are the connections to black holes and string theory with respect to general questions regarding the scrambling of information~\cite{Lashkari:2011yi,Maldacena:2015waa} and the unitary dynamics underlying black hole formation and evaporation~\cite{Hawking:1974rv,Hawking:1976ra,Page:1993wv,Penington:2019npb,Almheiri:2019psf,Almheiri:2019hni}.

Finally, the role of topology in heavy-ion collisions has interdisciplinary connections in the chiral magnetic effect which is now observed in condensed matter systems~\cite{Li:2016vlc}. Continual advances in laser technology also offer great promise in the precision study of anomalous currents off equilibrium.

We end the review in Section~\ref{sec:conclusions} with a brief summary and outlook toward future developments in our understanding of thermalization in QCD. As the outline suggests, thermalization in QCD is a rich field with many research directions and we have had to make choices in our presentation due to space limitations. 
An important topic that we do not address is the off-equilibrium dynamics of QCD matter in the vicinity of a critical point~\cite{Akamatsu:2018vjr,Bzdak:2019pkr,Bluhm:2020mpc}. Another is the related topic of hydrodynamic fluctuations~\cite{Akamatsu:2016llw, An:2019osr}.  Other noteworthy omissions in our presentation include the discussion of holographic deep inelastic scattering~\cite{Polchinski:2002jw,Hatta:2007cs,Shuryak:2017phz}, holographic hard probes~\cite{Herzog:2006gh,Gubser:2006bz,Chesler:2008wd,Chesler:2008uy,Chesler:2013cqa} and features of  linear response theory~\cite{Son:2002sd,Herzog:2002pc,Kovtun:2005ev}. Some aspects of holographic approaches that we omit or treat only partially were discussed 
in \cite{CasalderreySolana:2011us,DeWolfe:2013cua,Chesler:2015lsa,Heller:2016gbp,Florkowski:2017olj}.

\section{Hadron structure at high energies\label{sec:hadrons}}
The initial value problem of the thermalization process in hadron-hadron collisions requires a deep understanding of the structure of QCD matter in the wavefunctions of the colliding hadrons. The spacetime picture since the early days of QCD is that the highly Lorentz contracted large $x$ valence partons in the ultrarelativistic hadron wavefunctions go through unscathed in the collision, while their accompanying small $x$  ``fur coat of wee-parton vacuum fluctuations"~\cite{Bjorken:1976mk}  interacts strongly to form hot and dense matter~\cite{Bjorken:1982qr}. The wee parton phase space distributions evolve with energy and nuclear size; their properties  determine key features of the bulk properties of the matter produced after the collision. 

In this section, after a brief introduction to QCD and the parton model of hadrons at high energies we will discuss significant developments in the description of hadron wavefunction in the CGC EFT.
In particular, we will address how the semi-hard saturation scale $Q_S$ arises in the nuclear wavefunctions, which justifies their description as highly occupied gluon shockwaves. As the largest scale in the problem, it not only sets the scale for many-body correlations in these shockwaves, and in the Glasma matter produced after the collision, but subsequently also determines the thermalization time and the initial temperature of the quark-gluon plasma. 

\subsection{Quantum Chromodynamics\label{sec:firstprinciples}}
Quantum Chromodynamics (QCD), the modern theory of the strong force in nature, is a nearly perfect theory, with the only free parameters being the quark masses~\cite{Wilczek:1999id}. The Lagrangian of the theory can be written compactly as 
\begin{equation}
{\cal L}_{\rm QCD}= -\frac{1}{4} F_{\mu\nu}^a F^{\mu\nu,a}+ \sum_f {\bar \Psi}_i^f\left(i\gamma^\mu D_{\mu,ij} -m_f\delta_{ij}\right)\Psi_j^f \,. \label{eq:QCD-Lagrangian}
\end{equation}
Here $F_{\mu\nu}^a=\partial_\mu A_\nu^a -\partial_\nu A_\mu^a-g f^{abc} A_\mu^b A_\nu^c$ is the QCD field strength tensor for the color gauge fields $A_\mu^a$ that live in the adjoint representation of $SU(3)$,  with $a=1,\cdots, 8$ and $f^{abc}$ the structure constants of the gauge group. The quark fields live in the fundamental representation of $SU(3)$ and are labeled with their color and flavor indices $\Psi_i^f$ where the color index $i=1,\cdots,3$ and $f$ denotes the flavors of quarks with masses $m_f$. Finally,  the Dirac matrix $\gamma^\mu$ is contracted with the covariant derivative $D_{\mu,ij}=\partial_\mu \delta_{ij} + ig t_{ij}^a A_\mu^a$, with $t_{ij}^a$ the generators of $SU(3)$ in the fundamental representation.

The theory is rich in symmetry. The structure of the Lagrangian is dictated by the invariance of the quark and gluon fields under local $SU(3)$ color gauge transformations. In addition, for massless quarks the theory has a global chiral $SU(3)_L\times SU(3)_R$ symmetry, global baryon number $U(1)_V$ and axial charge $U(1)_A$ symmetries, and the quark and gluon fields are invariant under scale transformations. The Lagrangian is invariant under discrete parity, charge, and time reversal symmetries. 

All of these symmetries, except that of  local $SU(3)$ color, are broken by vacuum or quantum effects that give rise to all the emergent phenomena in the theory, including confinement, asymptotic freedom, quantum anomalies and the spontaneous breaking of chiral symmetry.

Because QCD is a confining theory, it is not analytically tractable in general and numerical methods are essential to uncover its properties. Euclidean lattice Monte Carlo methods can be applied to compute, with good accuracy, ``static" properties of the theory such as the mass spectrum of hadrons, magnetic moments, and thermodynamic properties of QCD at finite temperature \cite{Lin:2017snn,Detmold:2019ghl}. 

These methods are, however, very limited in determining dynamical ``real time" features of theory because of the contributions of a large number of paths to the QCD path integral in Minkowski spacetime. There are promising approaches to surmount this difficulty such as steepest descent  Lefshetz thimble methods but they are currently applicable only to problems in 1+1 dimensions~\cite{Alexandru:2017lqr}. Likewise, quantum computing offers an alternative paradigm for computing real time dynamics, but its applicability to QCD likely remains far in the future~\cite{Preskill:2018fag}. 

One should note that the production of high transverse momentum and massive particles (jets and  heavy quarkonia being two notable examples) can be computed with high precision in perturbative QCD (pQCD)~\cite{Collins:1989gx}. This is because these processes correspond to very short transverse distances and asymptotic freedom tells us that the QCD coupling $\alpha_S$ is weak at these scales.

\subsection{QCD at small $x$ and high parton densities}
\label{sec:saturation}
A great success of pQCD is the QCD parton model~\cite{Bjorken:1969ja}, wherein the complex dynamics of quark and gluon fields in hadrons can, at high energies and large momentum resolutions, be viewed as that of a weakly interacting gas of partons (single-particle quark, antiquark, and gluon states). 
The cleanest way to access this sub-nucleon structure is in the deeply inelastic scattering 
(DIS) of electrons or other leptons off nucleons and nuclei, wherein a virtual photon emitted by the electron strikes a quark or antiquark inside the hadron. 

For the thermalization process of interest in this review, the asymptotic high energy  (or ``Regge") limit of DIS is most relevant.  This limit corresponds to the Bjorken DIS variable $x_{\rm Bj}\sim Q^2/s \rightarrow 0$ where $Q^2$ is the squared four-momentum transfer and $s$ is the squared center-of-mass energy. In the parton model, $x_{\rm Bj}\approx x$, where $x$ is the light cone fraction of the momentum of the hadron carried by the struck parton\footnote{In hadron-hadron collisions, it is more appropriate to speak in terms of momentum fractions, so we  henceforth use $x$ instead of $x_{\rm Bj}$.}. At small $x$, or equivalently at high energies, the number of partons in the hadron proliferate rapidly, as first observed in DIS experiments at the HERA collider in Germany~\cite{Abt:1993cb,Ahmed:1995fd,Derrick:1993fta,Derrick:1994sz,Martin:1994kn,Lai:1994bb}. This growth is consistent with the predictions of the pQCD Dokshitzer-Gribov-Lipatov-Altarelli-Parisi (DGLAP)~\cite{Gribov:1972ri,Lipatov:1974qm,Altarelli:1977zs,Dokshitzer:1977sg} evolution equations.
 
The mathematical basis of the parton model in QCD follows from the observation that if one picks a lightcone\footnote{Lightcone coordinates are $k^\pm = (k^0\pm k^3)/\sqrt{2}$ and lightcone fields are defined as $A^\pm = (A^0\pm A^z)/\sqrt{2}$; we  work here in the metric $g_{\pm,\mp} = 1; g_{i,j} = -1$, where $i$ and $j$ represent the two transverse coordinates.} gauge $A^+=0$ and quantizes the quark and gluon fields of QCD along a  light front surface (say, $x^+=0$), the Hamiltonians of the free quark and gluon fields share the same vacuum\footnote{
In lightcone quantization, this argument requires a careful treatment of  $k^+=0$ vacuum modes~\cite{Nakanishi:1976vf}. For a perturbative treatment of lightcone wavefunctions, it may be sufficient to project out such modes~\cite{Collins:2018aqt,Fitzpatrick:2018ttk}.} as the fully interacting theory~\cite{Brodsky:1997de}. This allows one to construct the hadron wavefunction as a linear combination of a complete set of multi-parton eigenstates, each of which is an eigenstate of the free QCD Hamiltonian. 

In this lightcone framework, the parton distribution functions measured in DIS experiments can be interpreted as one-body states of quarks and gluons that carry a lightcone momentum fraction $x= k^+/P^+$, where $k^+$ is the parton's lightcone momentum and $P^+$ is the lightcone momentum of the hadron. 
As first argued in \cite{Gribov:1984tu,Mueller:1985wy},  two-body ``higher twist" gluon distributions, in a lightcone operator product expansion (OPE),\footnote{In OPE language, these higher twist contributions are suppressed by powers of $1/Q^2$.} grow as the square $[xG_A(x,Q^2)]^2$ of the leading twist gluon distribution. For a fixed $Q^2$, these two-body distributions  become as large as the leading twist one-body distribution as $x\rightarrow 0$. 

\begin{figure}
\centering
\includegraphics[width=0.6\linewidth]{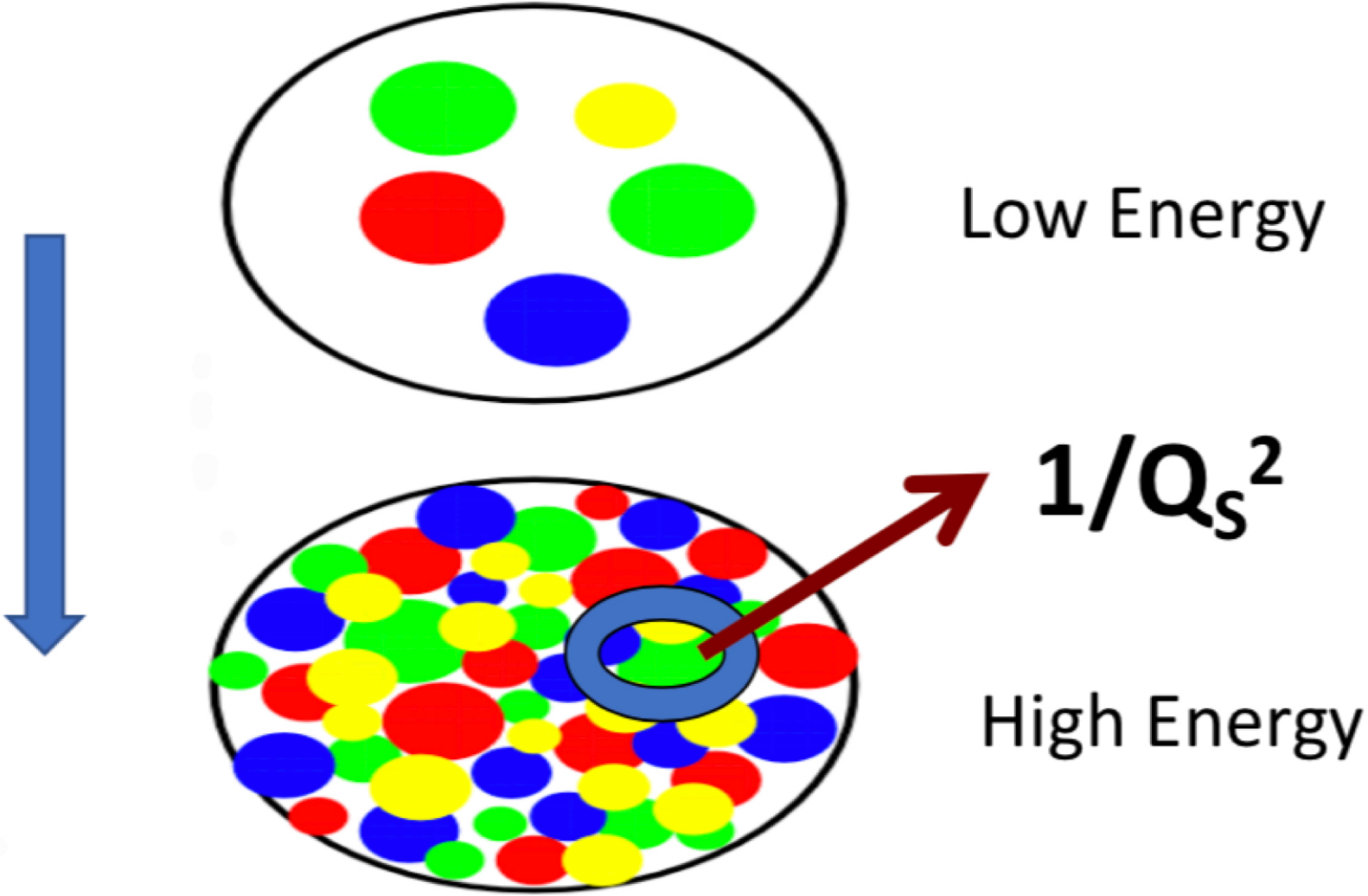}
\caption{Transverse hadron profile resolved in scattering with fixed squared momentum transfer $Q^2$ and increasing center-of-mass energy $\sqrt{s}$. 
The requirement for proliferating soft gluons to have maximal occupancy $1/\alpha_S$ generates the close packing saturation scale $Q_S$. Adapted from \cite{Iancu:2003xm}.}
\label{fig:glue-sat}
\end{figure}

Importantly, the net effect of such many-body contributions\footnote{These include the screening of bremsstrahlung gluons by real and virtual gluons, as well as the recombination of softer gluons into harder gluons.} is opposite that of the leading term, softening the growth in the gluon distribution. When the gluon phase space density is maximal, of the order of $1/\alpha_S$, all $n$-body lightcone distributions contribute equally. This saturation of gluon distributions in a nucleus of radius $R_A$, corresponds to the generation of the saturation scale $Q_S$, 
where parametrically, for $Q^2=Q_S^2$ the maximal occupancy is equated to the gluon phase space density as 
\begin{equation}
\frac{1}{\alpha_S(Q_S)} = \frac{ xG_A(x,Q_S^2)}{2(N_c^2-1) \pi R_A^2 Q_S^2} \, .
\label{eq:sat-condition} 
\end{equation}
\Fig{fig:glue-sat} illustrates the gluon saturation phenomenon and the interpretation of $Q_S$ as the emergent ``close packing" scale. 

\subsection{Effective Field Theory for high parton densities: the Color Glass Condensate}
\label{sec:CGC-EFT}
Since the usual formalism of pQCD relies on two-body and higher twist distributions being small, an alternative framework is necessary to understand the physics of gluon saturation and the emergence of the saturation scale in the nuclear wavefunction at high energies. Fortuitously, the problem of high parton densities can be formulated as a classical effective field theory on the 
light front, which as noted greatly simplifies the problem of heavy-ion collisions at high energies. 

To understand this better, we  will outline here an explicit construction performed for nuclei with large atomic number $A \gg1$~\cite{McLerran:1993ni,McLerran:1993ka,McLerran:1994vd}. 
An important ingredient in this construction in the infinite momentum frame (IMF) $P^+\rightarrow \infty$ of the nucleus is a Born-Oppenheimer separation in time scales between the Lorentz contracted large $x$ ($k^+\sim P^+$) ``valence" modes and the noted ``wee fur" of small $x$ ($k^+ \ll P^+$) gluons and ``sea" quark-antiquark pairs. For partons of transverse momentum $k_\perp$, lightcone lifetimes are given by 
\begin{eqnarray}
\tau_{\rm wee} &=& {1\over k^-} = {2k^+\over k_\perp^2} \equiv {2xP^+\over k_\perp^2} \,\,\,\nonumber \\
\tau_{\rm valence} &\approx& {2P^+ \over k_\perp^2}\, \longrightarrow\, \tau_{\rm wee} \ll \tau_{\rm valence}\,,
\label{eq:valence-wee}
\end{eqnarray}
suggesting that the valence parton modes are static over the times scales over which wee modes are probed. However, one cannot integrate the valence sources completely 
out of the theory because they are sources of color charge for wee partons and must couple to these in a gauge invariant manner. 

Note further that, since wee partons have large lightcone wavelengths ($\lambda_{\rm wee} \sim 1/k^+ = 1/xP^+$), they can resolve a lot of color charge provided that their transverse wavelength is not too large. The inequality
\begin{equation}
\lambda_{\rm wee} \sim {1\over k^+} \equiv {1\over xP^+} \gg \lambda_{\rm valence}\equiv {R_A\,m_N\over P^+}\,,
\label{eq:coherence}
\end{equation}
where on the r.h.s the Lorentz contraction factor is $P^+/m_N$ (with $m_N$  the nucleon mass),  suggests that wee partons with $x\ll A^{-1/3}$ resolve partons\footnote{Wee partons with wavelength $k_\perp \leq \Lambda_{\rm QCD}\sim 1$ fm$^{-1}$ see no color charge at all since color is confined (in nucleons!) on 
this scale. It is only wee partons with $k_\perp \gg \Lambda_{\rm QCD}$ that see color charges from different nucleons along the longitudinal direction.} all along the longitudinal extent  $2 \,R_A\sim A^{1/3}$ in units of the inverse nucleon mass. 

These charges will be random since they are confined to different nucleons and do not know about each other. A wee parton with
momentum $k_\perp$ resolves an area in the transverse plane $(\Delta x_\perp)^2 \sim 1/k_\perp^2$. The number of valence partons that it
interacts simultaneously with is 
\begin{equation}
k \equiv k_{(\Delta x_\perp)^2} = {N_{\rm valence}\over \pi R_A^2}\, (\Delta
x_\perp)^2
\, ,
\label{eq:color-charge}
\end{equation}
which  is proportional to $A^{1/3}$ since $N_{\rm valence}=3A$ in QCD.  For a large nucleus with $k \gg1$, one can show for $N_c\geq 2$ that the most likely color charge representation that the wee gluons couple to is a higher dimensional  classical representation of the order of $\sqrt{k}$~\cite{Jeon:2004rk}. 

Thus, wee partons couple to $\rho$, the classical color charge per unit transverse area of large $x$ sources. On average, since the charge distributions are random, the wee partons will couple to zero charge; however, fluctuations locally can be large. These conditions can be represented as 
\begin{equation}
\langle\rho^a(x_\perp)\rangle = 0\,,\,
\langle{\rho^a(x_\perp)\rho^b(y_\perp)}\rangle
= \mu_A^2\, \delta^{ab}\,\delta^{(2)}(x_\perp -y_\perp) \, ,
\label{eq:Gaussian} 
\end{equation}
where $a=1,\cdots, N_c^2-1$ and $\mu_A^2 = {g^2 A\over 2\pi R_A^2}$ is the color charge squared per unit area. For a large nucleus ($A \gg 1$), $\mu_A^2 \propto A^{1/3} \gg
\Lambda_{\rm QCD}^2$ is a large scale. Since it is the largest scale in the problem, $\alpha_S(\mu_A^2) \ll 1$. This result is remarkable because it provides a concrete example suggesting that QCD 
at small $x$ is a weakly coupled EFT wherein systematic computations of its many-body properties are feasible. 

We can now combine the previous kinematic and dynamical arguments and write 
the generating functional for the small $x$ effective action as 
\begin{equation}
{\cal Z}[j] = \int [d\rho]\,W_{\Lambda^+}[\rho]\,\left\{{\int^{\Lambda^+}[dA]\delta(A^+)e^{iS_{\Lambda^+}[A,\rho]-\int j\cdot A}}
\over {\int^{\Lambda^+}[dA]\delta(A^+)e^{iS_{\Lambda^+}[A,\rho]}}\right\} \, . 
\label{eq:path-integral}
\end{equation}
Here $\Lambda^+$ denotes the longitudinal momentum scale that separates the static color sources from the dynamical gauge fields and the gauge invariant weight functional $W_{\Lambda^+}[\rho]$ describes the distribution of these sources at the scale $\Lambda^+$,  with its path integral over $\rho$ normalized to unity. 

The CGC  effective action can be written in terms of the sources $\rho$ and the fields $A$ as 
\begin{eqnarray}
& &S_{\Lambda^+}[A,\rho] = {1\over 4}\int d^4 x\, F_{\mu\nu}^a \,F^{\mu\nu,a}\nonumber\\
&+& {i\over N_c}\int d^2 x_\perp dx^- \delta(x^-){\rm Tr} \left(\rho\, U_{-\infty,
\infty}[A^-]\right) \, .
\label{eq:effective-action}
\end{eqnarray}
The first term in \Eq{eq:effective-action} is the Yang-Mills action in the QCD Lagrangian given in \Eq{eq:QCD-Lagrangian}. The dynamics of wee gluons in the CGC is specified by this term. The second term\footnote{This  term can alternatively~\cite{JalilianMarian:2000ad} be written as ${\rm Tr}\left[\rho \log(U_{-\infty,\infty})\right]$.
}  denotes the  coupling of the wee gluon fields to the large $x$ color charge densities $\rho$, which we have argued are static lightcone sources. Because the sources are eikonal sources along the lightcone, their  gauge invariant coupling to the wee fields is described by the path ordered exponential along the lightcone time direction 
$U_{-\infty,\infty}={\cal P}\exp\left(ig \int dx^+ A^{-,a}T^a\right)$. Physically, $U$ corresponds to the color rotation of the color sources in the background of the wee gluon fields. 

The weight functional in the effective action (or the Gaussian random color charges in 
\Eq{eq:Gaussian}, in what is now called the McLerran-Venugopalan (MV) model~\cite{McLerran:1993ni,McLerran:1993ka,Kovchegov:1999yj}, can equivalently be written as\footnote{Sub-leading terms 
were discussed in \cite{Jeon:2005cf,Dumitru:2011ax}.} 
\begin{equation}
W_{\Lambda^+}[\rho] = \exp\left(-\int d^2 x_\perp {\rho^a(x_\perp)\rho^a (x_\perp)\over 2 \mu_A^2}\right) \,.
\label{eq:Gaussian-weight-functional}
\end{equation}

For each configuration of $\rho$'s in \Eq{eq:path-integral}, the saddle point of the effective action is given by the Yang-Mills (YM) equations
\begin{equation}
D_\mu F^{\mu\nu,a} = \delta^{\nu +}\,\delta(x^-)\,\rho^a(x_\perp) \, ,
\label{eq:CYM-nucleus}
\end{equation}
whose solution is the non-Abelian analog of the Weiz\"{a}cker-Williams (WW) 
fields in classical electrodynamics.
The chromo-electromagnetic gluon field strengths are singular on the nuclear sheet of width $\Delta x^- 
\sim 2 R m_N/P^+$ and zero (pure gauge) outside. 

The gauge field solutions in lightcone gauge are given by $A^-=0$ and 
\begin{equation}
A_{\rm cl}^k = {1\over ig} V(x^-,x_\perp) \nabla^k V^\dagger (x^-,x_\perp)\, ,
\label{eq:WW}
\end{equation}
where $k=1,2$ are the transverse coordinates and $V= {\cal P}\exp\left[\int_{-\infty}^{x^-} dz^- {1\over \nabla_\perp^2}{\tilde \rho}(z^-,x_\perp)\right]$.  This solution of the equations of motion requires path ordering of the sources in $x^-$~\cite{JalilianMarian:1996xn,Kovchegov:1999yj}. 
Further, the ${\tilde \rho}$ that appears in the solution is the color charge density in Lorenz gauge $\partial_\mu {A^\prime}^\mu=0$, where one has the solution ${A_{\rm cl}^\prime}^+ = {1\over \nabla_\perp^2}{\tilde \rho}(x^-,x_\perp)$, ${A_{\rm cl}^\prime}^- = {A_{\rm cl}^\prime}_\perp =0$. 
In fact, since the Jacobian of the transformation $[d\rho] \rightarrow [d{\tilde \rho}]$ is simple~\cite{JalilianMarian:1996xn}, many-body distributions in the lightcone gauge can be computed  in terms of color charges in Lorentz gauge, a natural choice from the analogy to WW  fields~\cite{Jackson:1998nia}. 

As a simple example, the correlator of gauge fields in a large nucleus can be computed analytically in the MV model by averaging the solution in \Eq{eq:WW} with the weight functional $W$:
\begin{equation}
\langle AA\rangle =\int [d{\tilde \rho}] A_{\rm cl.}[\tilde \rho]A_{\rm cl.}[{\tilde \rho}] W_{\Lambda^+}[{\tilde \rho}] \, .
\label{eq:WW-correlator}
\end{equation}
Equation \eq{eq:WW-correlator} can be further Fourier decomposed to extract the number distribution of wee gluons $\frac{dN}{d^2 k_\perp}$ and expressed\footnote{\label{ft:MV}In the MV model, this defines $Q_S^2 = c_A \mu_A^2$, where the coefficient  $c_A$ is determined numerically~\cite{Lappi:2007ku}.} in terms of $Q_S$. Specifically, for the  occupation number $\phi = \frac{(2\pi)^3}{2(N_c^2-1)}\frac{dN}{\pi R^2 d^2 k_\perp}$ one obtains $\phi\propto\frac{Q_S^2}{k_\perp^2}$ for $k_\perp \gg Q_S$,   However, for $k_\perp \ll Q_S$ the distribution is modified substantially from the WW distribution: $\phi\sim \frac{1}{\alpha_S}\log(Q_S/k_\perp)$. 
This softened infrared distribution in the CGC EFT provides a simple explanation of gluon saturation. 

We are now in a position to understand the term Color Glass Condensate (CGC)~\cite{Iancu:2003xm,Gelis:2010nm}, which is used to describe the ground state properties of a hadron or nucleus at very high 
energies. Color is obvious since the state is composed primarily of a large number of gluons and sea quark-antiquark pairs.  It is a 
glass because these small $x$ gluons and sea quarks are generated by random sources with lifetimes much longer than the characteristic time scales of the scattering. This explains the structure of \Eq{eq:path-integral}, where the path integral over the curly brackets is performed first for fixed color charge distributions and then averaged over an ensemble of such distributions. Finally, the state is a condensate because gluons have occupation numbers 
$\phi\sim 1/\alpha_S$, with momenta peaked at $k_\perp\sim Q_S$.

\begin{figure}
\centering
\includegraphics[width=\linewidth]{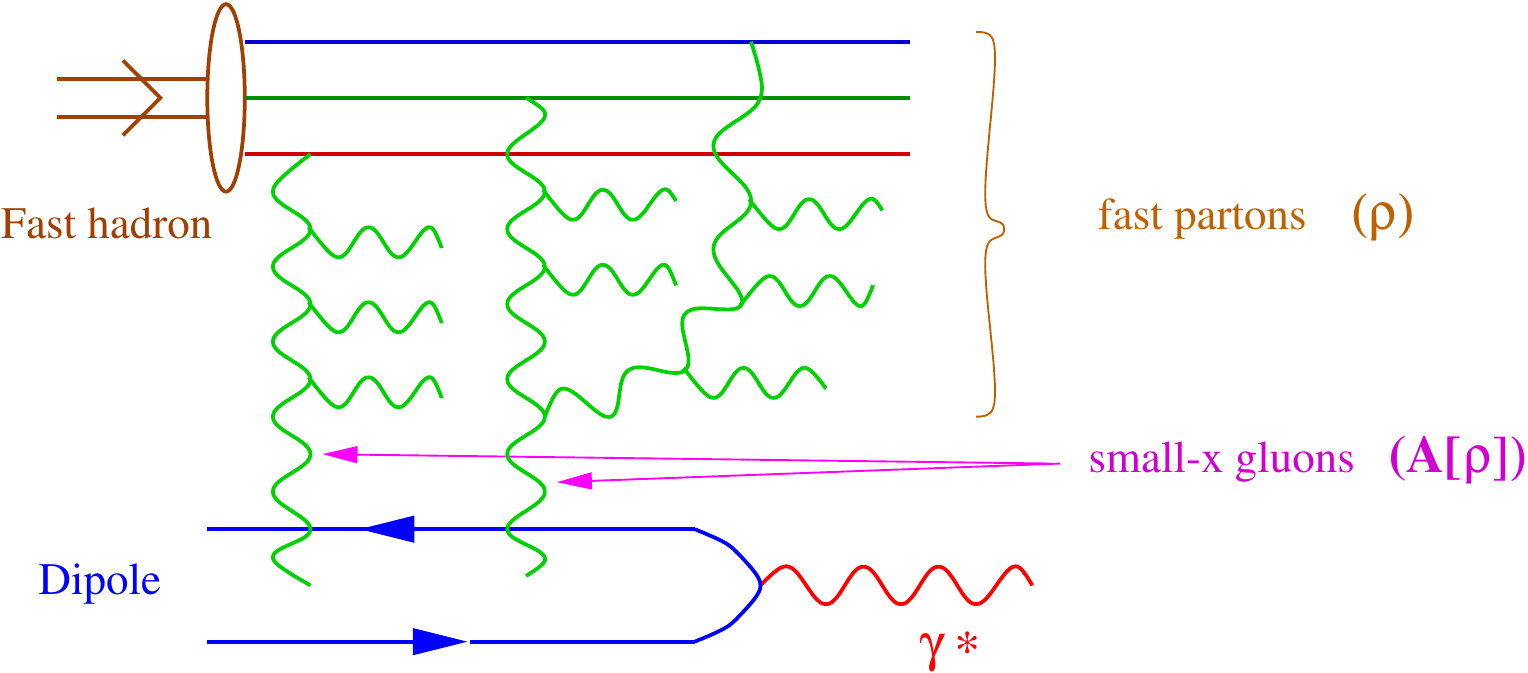}
\caption{DIS in the dipole picture. The virtual photon emitted by the electron splits into a $q{\bar q}$ dipole that scatters off dynamical small $x$ gauge fields coupled 
to the static large $x$ lightcone sources. From \cite{Iancu:2003xm}.}
\label{fig:dipole-amp}
\end{figure}

To take a specific example, consider the inclusive cross-section in the DIS scattering of a virtual photon on the nucleus, illustrated in \Fig{fig:dipole-amp}. In the CGC EFT, it is 
expressed as the cross-section for a fixed distribution of sources convoluted with an ensemble of such sources: 
\begin{equation}
\langle \mathrm{d} \sigma \rangle = \int [\mathcal{D} {\tilde \rho}_{A} ] \, W_{\Lambda^{+}} [{\tilde \rho}_{A}]\, \mathrm{d} \hat{\sigma}[{\tilde \rho}_{A}]\,.
\label{eq:dsigma-inc}
\end{equation}
Thus, on the time scale $t\sim 1/Q$ of the probe, it resolves a condensate of gluons with a well-defined number density of longitudinal modes down to $x\sim x_{\rm Bj} \ll 1$. Because of time dilation [see \Eq{eq:valence-wee}] the averaging over $\rho_A$ with $W$ takes place on much larger time scales. This two-stage averaging process clarifies how one reconciles gauge invariance with the presence of a colored condensate. 

The CGC classical equations possess a ``color memory" effect~\cite{Pate:2017vwa} corresponding to the large gauge transformation $V$ of a quark after interacting with the gluon shockwave, generating a transverse momentum kick $p_\perp\sim Q_S$ to the quark that can be measured in DIS experiments~\cite{Ball:2018prg}. Remarkably, this is exactly analogous to the inertial displacement of detectors after the passage of a gravitational shockwave~\cite{Strominger:2014pwa}. This gravitational memory is deeply related to asymptotic spacetime symmetries and soft theorems in gravity and may also hold useful lessons for QCD.\footnote{An ``infrared triangle" between asymptotic symmetries, memory and soft theorems in gravity~\cite{Strominger:2017zoo} also allows for an elegant interpretation of the infrared structure of QED~\cite{Bieri:2013hqa,Kapec:2017tkm}. While color confinement implies that such universal features do not apply to QCD in general, an emergent $Q_S\gg \Lambda_{\rm QCD}$ suggests that they may be applicable in the Regge limit.}

\subsection{Renormalization group evolution in the CGC EFT}
\label{sec:RG-EFT}
We have discussed a classical EFT for large nuclei and Gaussian sources where the separation between fields (wee partons) and sources (valence sources) was picked randomly to be at the momentum scale $\Lambda^+$. Physical 
observables such as the inclusive cross-section in \Eq{eq:dsigma-inc} should not depend on $\Lambda^+$. This invariance is the essence of the renormalization group and we will later sketch how it is realized in the EFT; a detailed discussion can be found in \cite{Iancu:2003xm}.

The important point to note is that real and virtual quantum fluctuations in the classical background field of the target, while apparently suppressed by ${\mathcal O}(\alpha_S)$ are actually $\alpha_S \log(\Lambda^+/{\Lambda^\prime}^+)\sim {\mathcal O}(1)$ from the phase space integration of these modes  when ${\Lambda^\prime}^+ = \Lambda^+ e^{-1/\alpha_S}$ (or, equivalently, when $x_{\rm wee}= x_{\rm val.}\,e^{-1/\alpha_S}$). These large NLO contributions can be absorbed into the form of the LO cross-section in \Eq{eq:dsigma-inc} at the scale ${\Lambda^\prime}^+$  by redefining the weight functional $W_{\Lambda^+}[\tilde \rho]\rightarrow W_{{\Lambda^\prime}^+}[\tilde \rho^\prime]$. Here ${\tilde \rho}^\prime = {\tilde \rho} + \delta {\tilde \rho}$ is the new color source density at ${\Lambda^\prime}^+$ that incorporates the color charge density $\delta {\tilde \rho}$ induced by quantum fluctuations between $\Lambda^+$ and ${\Lambda^\prime}^+$. 

One can thus write  
\begin{equation}
\langle \mathrm{d} \sigma_{\text{LO+NLO}} \rangle = \int [\mathcal{D}{\tilde \rho}_{A}]\, W_{{\Lambda^\prime}^+}[{\tilde \rho}_{A}]\,\mathrm{d} \hat{\sigma}_{\text{LO}} [{\tilde \rho}_{A}] \, , 
\label{eq:inclusive-LLx}
\end{equation}
where 
\begin{equation}
W_{{\Lambda^\prime}^+}[{\tilde \rho}_{A}] =  \Big[  1+ \log({\Lambda^+}/ {\Lambda^\prime}^+) \mathcal{H}_{\text{LLx}} \Big] W_{\Lambda+}
[{\tilde \rho}_{A}]  \, ,
 \label{eq:W-LLx}
\end{equation}
with the quantum fluctuations absorbed
as we discuss shortly. 

Since the l.h.s of \Eq{eq:inclusive-LLx} should not depend on the arbitrary ``factorization scale" $\Lambda^+$, the derivative of both l.h.s and r.h.s with respect to it should be zero. From \Eq{eq:W-LLx}, one can therefore deduce the Jalilian-Marian–Iancu–McLerran–Weigert–Leonidov–Kovner (JIMWLK) RG equation~\cite{JalilianMarian:1997gr,JalilianMarian:1997dw,Iancu:2000hn} 
\begin{equation}
\frac{\partial}{\partial Y} W_Y [{\tilde \rho}_{A}] = \mathcal{H}_{\text{LLx}}\, W_Y [{\tilde \rho}_{A}] \,,
\label{eq:JIMWLK-RG}
\end{equation}
where the JIMWLK Hamiltonian~\cite{Weigert:2000gi} 
\begin{equation}
 {\mathcal H}_{\rm LLx}= {1\over 2}\,\int_{x_\perp,y_\perp} 
 {\delta \over \delta {\tilde \rho}^a(x_\perp)}
 \chi^{ab}(x_\perp,y_\perp)[{\tilde \rho}]
 {\delta \over {\delta {\tilde \rho}^b(y_\perp)}} \,,
\label{eq:JIMWLK}
\end{equation}
describes the evolution of the gauge invariant weight functional $W$ with rapidity $Y= \log(\Lambda_0^+/\Lambda^+)\equiv \log(x_0/x)$ once the non-perturbative initial conditions for $W$ are specified at an initial $x_0$. 

The Hamiltonian is computed in the CGC EFT, with  $\chi^{ab}(x_\perp,y_\perp)[{\tilde \rho}]=\langle\delta {\tilde \rho}^a(x_\perp)\delta{\tilde \rho}^b(y_\perp)\rangle_{{\tilde \rho}}$  the two-point function of induced charge densities\footnote{Note that here and henceforth in this section, 
$\int_{x_\perp} = \int d^2 x_\perp$ and $\int_{x_\perp,y_\perp} = \int d^2 x_\perp d^2 y_\perp$.} in the classical background field of the hadron.  Note that with this computation of ${\mathcal H}_{\rm LLx}$, the solution of \Eq{eq:JIMWLK-RG} resums leading logarithms 
$\alpha_S \log(x_0/x)$  (LLx) to all orders in perturbative theory. Thus this powerful RG procedure extends the accuracy of computations of the cross-section from $\langle \mathrm{d} \sigma_{\text{LO+NLO}}\rangle \rightarrow \langle \mathrm{d} \sigma_{\text{LO+LLx}}\rangle $.

The JIMWLK RG equation can equivalently be expressed as a hierarchy of equations (the Balitsky-JIMWLK hierarchy independently derived in~\cite{Balitsky:1995ub}) for the expectation value of an operator $O$:
\begin{equation}
{\partial \langle O\rangle_Y\over dY} = \Bigg<{1\over 2} \int_{x_\perp,y_\perp}\!{\delta \over \delta \alpha^a(x_\perp)}
\chi^{ab}(x_\perp,y_\perp){\delta\over \delta \alpha^b(y_\perp)}O[\alpha]\Bigg>_Y ,
\label{eq:operator-smallx}
\end{equation}
where $\alpha^a = \frac{1}{\nabla_\perp^2}{\tilde \rho}^a $. Remarkably,  \Eq{eq:operator-smallx} has the form of a generalized Fokker-Planck equation in functional space, where $Y$ is ``time" and $\chi$ is the diffusion coefficient~\cite{Weigert:2000gi}. 

There is no known analytical solution to the JIMWLK equation; as we later discuss, it can be solved numerically. However, good approximations exist in different limits. In a ``weak field"  (and leading twist) limit $g\alpha \ll 1$, one recovers for the number distribution (and the corresponding occupation number $\phi$) extracted from \Eq{eq:WW-correlator}, the celebrated LLx Balitsky-Fadin-Kuraev-Lipatov (BFKL) equation~\cite{Kuraev:1977fs,Balitsky:1978ic} of pQCD. Another mean field ``random phase" approximation~\cite{Iancu:2001md,Weigert:2000gi} allows one to evaluate the occupation number $\phi$ in the ``strong field" limit of $g\alpha\sim 1$.

The longitudinal extent of the wee gluon cloud generated by the RG evolution has a width $x^- = \frac{1}{k^+}\sim \frac{1}{Q_S}$. This is much more diffuse relative to the width $e^{-1/\alpha_S}\frac{1}{Q_S}$ of the valence modes. The RG evolution also predicts that the width of the wee gluon cloud will shrink with increasing boost (or rapidity) relative to an ``observer"  quark-antiquark pair, albeit at a slower rate than their larger $x$ counterparts. Thus, in the CGC EFT the scale for the overlap of the wave functions in the thermalization process is set by $\frac{1}{Q_S}$ rather than the Lorentz contracted width of the valence quarks given by $\frac{1}{P^+}$.

\subsection{DIS and the dipole model}
\label{sec:DIS-dipole}

Here and in Section~\ref{sec:RG-GS} we will concretely relate the CGC EFT to the structure functions that are measured in DIS. These comparisons are essential for precision tests of the CGC EFT picture of high energy nuclear wavefunctions. They also play an important role in constraining the 
saturation scale and the shadowing of nuclear distributions that are key to determining the initial conditions for early time dynamics in heavy-ion collisions. These connections will become more evident in Section~\ref{sec:Glasma-Evolution}.

The inclusive cross-section can be expressed in full generality as $\langle \mathrm{d} \sigma \rangle = L_{\mu\nu} W^{\mu\nu}$, where $L_{\mu\nu}$ is the well-known 
lepton tensor~\cite{Peskin:1995ev} representing the squared amplitude for the emission of a virtual photon with four-momentum $q^\mu$ and $W^{\mu\nu}$ is the spin-averaged DIS hadron tensor that, 
for a nucleus in the IMF, can be reexpressed as~\cite{McLerran:1998nk,Venugopalan:1999wu} 
\begin{align}
&W^{\mu\nu} = \frac{1}{2\pi} \frac{P^+}{m_N} {\rm Im} \int d^2 X_\perp dX^-
\int d^4 x \,e^{iq\cdot x} \times\nonumber \\
& \langle {\rm Tr} \big[\gamma^\mu S_A(X^-+\frac{x}{2},X^- - \frac{x}{2})\gamma^\nu S_A(X^-- \frac{x}{2},X^- + \frac{x}{2})\big]\rangle\,,
\label{eq:Wmunu}
\end{align}
where $S_A(x,y)= -i\langle \psi(x) {\bar \psi}(y)\rangle_A$ is the quark propagator in the gauge fields $A^\mu$ of the nucleus.\footnote{The second average in \Eq{eq:Wmunu} corresponds to averaging over  ${\tilde \rho}$. We employ the relativistic normalization $\langle P|P\rangle = \frac{P^+}{m_N}(2\pi)^3\delta^{3}(0) \equiv \frac{P^+}{m_N} \int d^2 X_\perp d X^- $, where $X^-$ and $X_\perp$ are center-of-mass coordinates.} 

In the CGC, the leading contribution is obtained by replacing the full QCD background field with the saturated classical background field:
$A^\mu \rightarrow A_{\rm cl}^\mu$, where $A_{\rm cl}^\mu$ are the non-Abelian WW fields in \Eq{eq:WW}. In $A^-=0$ gauge\footnote{The solution of the YM equations
 is identical in this case to the solution in Lorenz gauge.}, the momentum space quark propagator in the classical background field is remarkably simple, given by~\cite{McLerran:1998nk}
$S_{\rm A_{\rm cl}}(p,q) = S_0(p) {\cal T}_q(p,q) S_0(q)$,
where the free Dirac propagator is $S_0=\frac{i \slashed{p}}{p^{2}+i\varepsilon}$ and ${\mathcal{T}}_{q}(q,p)= \pm (2 \pi)\delta(p^{-}-q^{-}) \gamma^{-} \int_{z_\perp}e^{-i(\bm{q}_{\perp} - \bm{p}_{\perp})\cdot \bm{z}_{\perp}} V^{\pm 1}(\bm{z}_{\perp})$ is the effective vertex corresponding to the multiple scattering of the quark (or antiquark) off the shockwave background field represented by the eikonal path ordered phase $V$ 
introduced after \Eq{eq:WW}. 

The DIS structure function is simply related to the inclusive cross-section. Plugging the dressed CGC propagator into \Eq{eq:Wmunu}, one can show, to this order of accuracy, that it can be expressed as~\cite{McLerran:1998nk} 

\begin{equation}
F_2(x,Q^2) = \frac{Q^2}{4\pi^2 \alpha_{\rm em}} \int_0^1\! \! \!dz \!\int_{r_\perp} \!\! \!|\Psi_{\gamma^*\rightarrow q{\bar q}}|^2 \sigma_{q{\bar q}A} (x,Q^2) \,.
\label{eq:dipole}
\end{equation} 
Equation \eq{eq:dipole} can be simply interpreted to be the convolution of the probability of the virtual photon to split into a quark-antiquark pair (which can be computed in QED~\cite{Bjorken:1970ah}) with the ``dipole" scattering cross-section of the quark-antiquark pair to scatter off the nucleus. 
For the impact parameter $b_\perp = (x_\perp+y_\perp)/2$, the cross-section is given by 
\begin{equation}
\sigma_{q{\bar q}A} = 2 \int d^2 b_\perp {\cal N}_Y (b_\perp, r_\perp) \,,
\label{eq:dipole-cross-section}
\end{equation}
where the forward scattering amplitude ${\cal N}_Y(b_\perp,r_\perp) = 1 - {\cal S}_Y(b_\perp,r_\perp)$, with the S-matrix 
\begin{equation}
{\cal S}_Y(r_\perp) = \frac{1}{N_c} \langle {\rm Tr}\left[V(x_\perp) V^\dagger(y_\perp)\right]\rangle_Y \,.
\label{eq:dip-corr}
\end{equation}

One can compute the S-matrix  
explicitly in the MV model, which gives\cite{McLerran:1998nk,Kovchegov:1999yj,Venugopalan:1999wu},
\begin{equation}
{\cal S}_Y (r_\perp) = \exp\left[-\alpha_S \frac{\pi^ 2}{2N_c} \frac{r_\perp^2 A x G_N(x, 1/r_\perp^2)}{\pi R_A^2}\right] \,,
\label{eq:K-T}
\end{equation}
where $G_N$ denotes the gluon distribution
in the proton at the scale $\frac{1}{r_\perp^2}$. 

One can expand the exponential for very small values of $r_\perp$, and one observes that the dipole cross-section is nearly transparent to the color of the small dipoles. As $r_\perp$ grows, the S-matrix decreases; the saturation scale is defined as the value of $r_\perp$ at which the S-matrix has a value that is significantly smaller than what one would anticipate in pQCD. While there is some freedom in setting this scale, its growth with decreasing $x$ is determined by the growth in the gluon distribution.

The MV result in \Eq{eq:K-T} is the QCD Glauber model~\cite{Mueller:1989st} which gives the survival probability of a dipole after multiple independent scatterings off the nucleus. It can be refined by introducing an impact parameter distribution inside the proton~\cite{Bartels:2002cj}, the so-called impact-parameter-dependent
saturation (IP-Sat) model, which can be further extended to model the S-matrix for the nuclei~\cite{Kowalski:2003hm,Kowalski:2007rw}. 

The IP-Sat model provides very good agreement with a wide range of small $x$ DIS data on e+p scattering at HERA~\cite{Rezaeian:2012ji}. The latter constrains the parameters of this model, which in turn is an essential ingredient of the 
IP-Glasma model of the initial conditions for heavy-ion collisions. We will discuss the IP-Glasma model in \Sec{sec:ipglasma}.

An advantage of the MV model formulation is that one can compute with relative ease~\cite{Fujii:2002vh,Blaizot:2004wv,Dominguez:2011wm,Dusling:2017aot,Fukushima:2017mko} not only the dipole Wilson line correlator but also the quadrupole and higher point correlators that appear in semi-inclusive final states in e+A and p+A collisions.

\subsection{RG evolution and geometric scaling}
\label{sec:RG-GS}

The MV model is valid for a large nucleus at rapidities when the bremsstrahlung of soft gluons is not significant, namely, for $\alpha_S Y \leq 1$. The classical expressions we derived have no $x$ dependence. For moderate $x$, one can introduce $x$ dependence in framework along the lines of the IP-Sat model that we discussed. However when $\alpha_S Y \gg 1$, the model is no longer applicable. In this regime, the RG evolution of the S-matrix in \Eq{eq:dip-corr} is described by 
the Balitsky-JIMWLK hierarchy in \Eq{eq:operator-smallx} which,  in addition to the coherent multiple scattering effects in the MV model, captures the previously discussed real and virtual quantum corrections. 

Substituting the expectation value of the correlator of the Wilson lines in \Eq{eq:dip-corr} into the Balitsky-JIMWLK hierarchy in \Eq{eq:operator-smallx}, leads, for $N_c, A\gg 1$, to the closed form
Balitsky-Kovchegov (BK)~\cite{Balitsky:1995ub,Kovchegov:1999yj} equation for the RG evolution in the rapidity of the dipole scattering amplitude: 
\begin{eqnarray}
&&{\partial {\cal N}_Y ({x_\perp,y_\perp})\over \partial Y}={\bar\alpha_S}\int_{z_\perp}{(x_\perp-y_\perp)^2\over (x_\perp-z_\perp)^2(z_\perp-y_\perp)^2}\nonumber\\
&\times&\Big[{\cal N}_Y(x_\perp,z_\perp)+{\cal N}_Y(y_\perp,z_\perp)-{\cal N}_Y(x_\perp,y_\perp)\nonumber \\
&&-{\cal N}_Y(x_\perp,z_\perp) {\cal N}_Y(z_\perp,y_\perp)\Big] \, .
\label{eq:BK}
\end{eqnarray}

The BK equation is the simplest RG equation to capture the physics of gluon saturation. For  ${\cal N}_Y\ll 1$, the non-linear term in the last line of \Eq{eq:BK} can be ignored and the equation reduces to the linear BFKL equation as anticipated previously.
In this limit, the amplitude has the solution
\begin{eqnarray}
{\cal N}_Y(r_\perp) 
\approx \exp\left(\omega {\bar \alpha}_s Y-{\rho\over 2}  -{\rho^2\over 2\beta {\bar \alpha}_S Y} \right) \, ,
\label{eq:Amp-BFKL}
\end{eqnarray}
where $\omega = 4 \log 2\approx 2.77$, $\beta=28\,\zeta(3) \approx 33.67$ and $\rho = \log(1/r_\perp^2 \Lambda_{\rm QCD}^2)$. This solution gives the rapid ``Markovian" growth of the dipole cross-section in rapidity due to the copious production of softer and softer gluons. 

However, when ${\cal N}_Y\sim 1$ the non-linear term arising from the fusion and screening of soft gluons completely saturates the growth of the dipole cross-section. If we impose a saturation condition ${\cal N}_Y = 1/2$ for $r_\perp = 2/Q_S$ on \Eq{eq:Amp-BFKL},  the argument of the exponential vanishes for 
$\rho_s=\log (Q_S^2/\Lambda_{\rm QCD}^2)$, with 
\begin{equation}
Q_S^2 = \Lambda_{\rm QCD}^2 \,\, e^{c {\bar \alpha}_S Y} \,\,\,\,{\rm where}\,\,\,\, c=4.88 \, .
\label{eq:QS-LO}
\end{equation}
Further, if we write $\rho = \rho_S+\delta \rho$, 
where $\delta \rho = \log(1/r_\perp^2 Q_S^2)$, we find that~\cite{Iancu:2002tr} 
\begin{equation}
{\cal N}_Y \approx \left(r_\perp^2 Q_S^2\right)^{\gamma_s}\,,
\label{eq:amp-GS}
\end{equation}
for $Q^2 < Q_S^4/\Lambda_{\rm QCD}^2$, where $\gamma_s=0.63$ is the BK anomalous dimension.

This ``geometrical scaling" of the forward scattering amplitude means that 
\Eq{eq:dipole-cross-section} scales  with $Q^2/Q_S^2(x)$ alone instead of with $x$ and $Q^2$ separately. Remarkably, this phenomenon was observed at HERA, providing a strong hint for the saturation picture~\cite{Stasto:2000er}. Moreover, the wider scaling window $Q^2 < Q_S^4/\Lambda_{\rm QCD}^2$ stretching beyond $Q_S$ provides a first principles explanation for a so-called ``leading twist shadowing" of nuclear parton distributions relative to those in the proton~\cite{Frankfurt:2011cs}. Such shadowed parton distributions are used to compute the rates of hard processes in heavy-ion collisions; understanding their microscopic origins is therefore important for quantifying hard probes of thermalization. 

\begin{figure}
\centering
\includegraphics[width=0.9\linewidth]{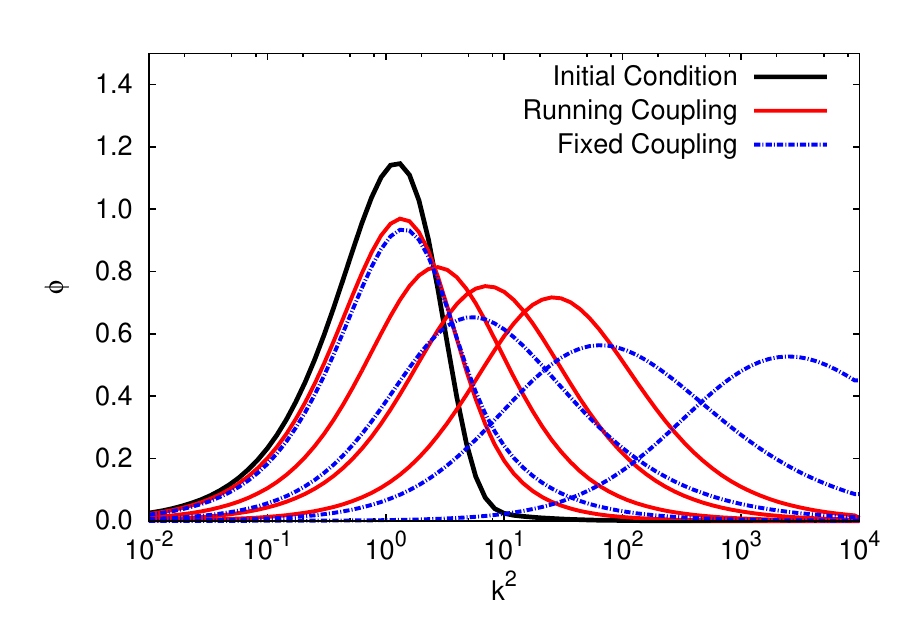}
\caption{Unintegrated gluon distribution (in units of the transverse area) vs the squared transverse momentum (normalized to its value at the peak of the initial condition curve) from the  solution of the Balitsky-Kovchegov equation. The different curves represent increasing rapidities (left to right) for fixed and 
running coupling. From \cite{Dusling:2009ni}.}
\label{fig:rc-BK}
\end{figure}

The BK equation in a reaction-diffusion approximation can be formally mapped into a well-known equation in statistical physics, the Fischer-Kolmogorov-Petrovsky-Piscounov (FKPP) equation~\cite{Munier:2003vc}. In this context, geometrical scaling appears as a late-time solution of a non-linear equation describing a traveling wavefront of constant velocity. In \Fig{fig:rc-BK}, we show numerical results for the 
unintegrated gluon distribution $\phi(k_\perp^2) = \frac{\pi N_c k_\perp^2}{2\alpha_S}\int_0^{+\infty} d^2 r_\perp e^{ik_\perp\cdot r_\perp}\left[1-{\cal N}_Y(r_\perp) \right]^2$, which  displays this traveling wave front structure, with the evolution of the peaks of the wavefronts representing the evolution of  $Q_S^2$ with rapidity. The correspondence of high energy QCD to reaction-diffusion processes is very rich; 
specific applications to DIS have been discussed recently~\cite{Mueller:2018ned,Mueller:2018zwx}. 

 $Q_S^2$ in \Eq{eq:QS-LO} [and the amplitude in \Eq{eq:amp-GS}] grows very rapidly with rapidity, much faster than in the HERA data. However this is significantly modified by running coupling corrections, which are part of the next-to-leading logarithms in $x$ (NLLx) contributions to QCD evolution. The significant effect of these running coupling corrections is clearly seen in \Fig{fig:rc-BK}.

These give~\cite{Mueller:2002zm}, 
\begin{equation}
Q_{s,{\rm running}\,\alpha_S}^2 = \Lambda_{\rm QCD}^2 \exp\left[\sqrt{2b_0 c (Y+Y_0)}\right] \, ,
\label{eq:QS-RC}
\end{equation}
where $b_0$ is the coefficient of the logarithm in the one loop QCD $\beta$-function\footnote{Sub-leading corrections in $Y$ to $Q_S$ have been computed to high order~\cite{Beuf:2010aw}.}. The running coupling results are well approximated by a  power law increase of the amplitude consistent with the HERA data.  Further, qualitative features of geometric scaling persist, 
albeit the window for geometrical scaling is significantly smaller~\cite{Triantafyllopoulos:2008yn}. 

For a large nucleus at the saturation boundary  $Y_0\propto \log^2(A^{1/3})$, one recovers the $A^{1/3}$ scaling of the 
saturation scale in the MV model  from \Eq{eq:QS-RC} for $Y_0 \gg Y$.  A striking result for  $Y\gg Y_0$ is that the saturation scale for a fixed impact parameter becomes independent of $A$. {\it In the asymptotic Regge limit, strongly correlated gluons in the nuclear wavefunctions lose memory of the initial conditions whereby they were generated.}

\subsection{The state of the art in the CGC EFT}
\label{sec:CGC-art}
In this section, we have outlined a description of the wavefunction of a high energy nucleus in the CGC EFT, emphasizing a qualitative understanding of gluon saturation and key related analytical results. There have been significant developments since in the CGC EFT. 

On the formal side, the Balitsky-JIMWLK framework for the LLx evolution of  $n$-point Wilson line correlators has been extended to NLLx~\cite{Balitsky:2013fea,Kovner:2013ona,Kovner:2014lca,Balitsky:2014mca,Caron-Huot:2015bja}. For the two-point dipole correlator, which satisfies the LLx BK equation, the formalism has been extended to NLLx~\cite{Balitsky:2008zza} and for ${\cal N}=4$ supersymmetric Yang-Mills even to NNLLx in a recent {\it tour de force} computation~\cite{Caron-Huot:2016tzz}. The BFKL or BK kernel, however, receives large collinear contributions that need to be resummed in so-called small $x$ resummation schemes for quantitative predictions~\cite{Salam:1998tj,Ciafaloni:1999yw,Iancu:2015vea,Ducloue:2019ezk}. 

While as we have discussed there are good analytical approximations, a full analytical solution of the BK equation does not exist. Numerical simulations have, however, been known for some time for 
the LLx BK equation~\cite{Albacete:2003iq}, the LLx+running coupling BK equation~\cite{Albacete:2004gw,Albacete:2007yr}, and even more recently the full NLLx equation implementing collinear resummation~\cite{Lappi:2016fmu,Ducloue:2019jmy}. In particular, it was shown in \cite{Ducloue:2019jmy} that this NLLx framework provides very good agreement with the HERA data. 

\begin{figure}
\centering
\includegraphics[width=0.9\linewidth]{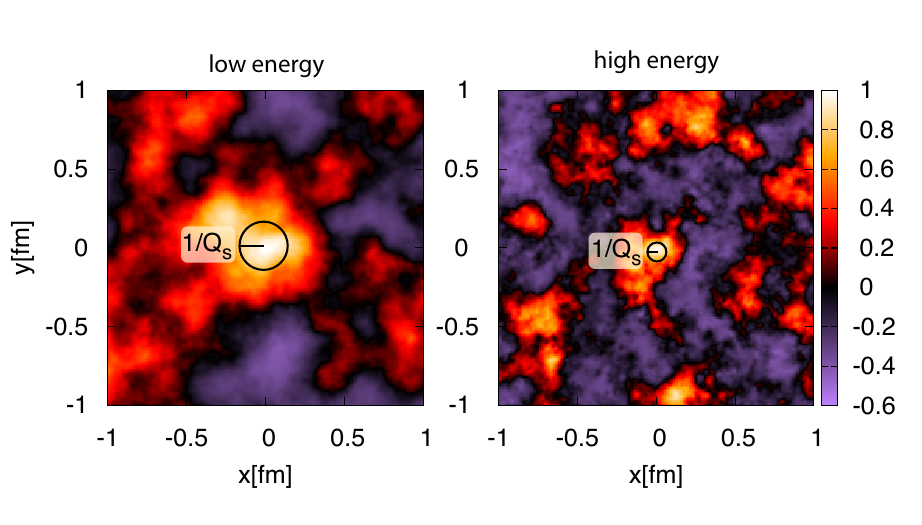}
\caption{Solution of the JIMWLK equation for the correlator of Wilson lines $V(x_\perp) V^\dagger(y_\perp)$ probed by the DIS dipole~\cite{Dumitru:2011vk}. As the nucleus is boosted from low energy (or rapidity) to high
energy, the regions with large values of these correlator shrink spatially, corresponding to larger values of $Q_S$. }
\label{fig:JIMWLK-soln}
\end{figure}

Numerical simulations  of higher point correlators in the Balitsky-JIMWLK hierarchy have also been performed. As noted, \Eq{eq:operator-smallx} has the form of a functional 
Fokker-Planck equation. This can therefore be reexpressed as a Langevin equation in the space of Wilson lines~\cite{Weigert:2000gi,Blaizot:2002np}, allowing one to 
simulate the rapidity evolution of two-point Wilson line correlators~\cite{Rummukainen:2003ns} as well as four-point quadrupole and sextupole\footnote{These are probed in semi-inclusive DIS~\cite{Dominguez:2011wm} and in proton-nucleus collisions~\cite{Kovner:2010xk,Dusling:2017aot,Dusling:2017dqg}.} correlators~\cite{Dumitru:2011vk,Lappi:2019kif,Lappi:2012vw}. Figure \ref{fig:JIMWLK-soln} shows a result for the dipole correlator from these simulations. Unfortunately, a similar Langevin representation is not known at present for the NLLx JIMWLK Hamiltonian. 

Precision computations require not just higher order computations of the JIMWLK kernel but higher order computations of  process dependent ``impact factors" analogous to pQCD computations of coefficient functions that are convoluted, order-by-order, with the DGLAP  splitting functions~\cite{Vermaseren:2005qc}. For inclusive DIS, analytical expressions exist for the virtual photon impact factor $|\Psi_{\gamma^*\rightarrow q{\bar q}}|^2$ in \Eq{eq:dipole}~\cite{Balitsky:2012bs}. Recently, NLO impact factors were computed for DIS exclusive diffractive light vector meson production~\cite{Boussarie:2016bkq} and DIS inclusive photon+dijet production~\cite{Roy:2019cux,Roy:2019hwr}. Numerical implementation of these results remains a formidable task and an essential component of precision studies of gluon saturation at the future Electron-Ion Collider (EIC)~\cite{Accardi:2012qut,Aschenauer:2017jsk}.

An outstanding problem at small $x$ is the impact parameter dependence of distributions. The BFKL kernel at large impact parameters contributes a Coulomb tail 
$\sim 1/b_\perp^2$; the conformal symmetry of the kernel and geometric scaling suggest a particular dependence of the saturation scale on the impact parameter~\cite{Gubser:2011qva}. 
The Coulomb tail is however not regulated by saturation and violates the Froissart bound on the asymptotic behavior of total cross-sections~\cite{Kovner:2002yt}. This is  cured non-perturbatively only by the generation of a mass gap in QCD. 
The Coulomb tail may be less of a problem in large nuclei  with $\Lambda_{\rm QCD} R_A \gg 1$  because the contribution of the Coulomb tail may be suppressed relative 
to protons, for which $\Lambda_{\rm QCD} R_A \sim 1$.

\section{Non-equilibrium QCD matter at high occupancy\label{sec:Glasma}}

The CGC EFT provides us with powerful tools to address multi-particle production in heavy-ion collisions from first principles; the key organizing principle is the kinematic separation in the hadron wavefunction between static color sources at large $x$ and small $x$ gauge fields. In the following, we will sketch the elements of the formalism to follow the thermalization process through the overlap of two CGCs.

To apply this EFT framework to thermalization, one first needs to understand how to compute from first principles  multi-particle production in the presence of strong fields\footnote{A  well-known example of such a formalism is $e^+ e^-$ pair production in strong electromagnetic fields~\cite{Gelis:2015kya}; another is that of Hawking radiation from the Black Hole horizon~\cite{Parikh:1999mf}.}. The quark-gluon matter formed in this process is the Glasma~\cite{Lappi:2006fp,Gelis:2006dv}, a nonequilibrium state with high occupancy [$f\sim {\mathcal O}(1/\alpha_S)$]; this state decays and eventually thermalizes. The description of the temporal evolution of the Glasma can be classified systematically in weak coupling into LO, NLO, etc.

Following our discussion of multi-particle production, we will describe the temporal evolution of the Glasma at LO. This corresponds to the solution of  classical Yang-Mills equations with CGC initial conditions for the fields  using both analytical approaches (valid for transverse momenta greater than the saturation scale) and a nonperturbative real time approach employing Hamilton's equation on the lattice. The LO solutions are independent of rapidity, with the dynamics of the corresponding ``flux tube" structures occuring entirely in the transverse plane of the collision.  We will next discuss the IP-Glasma model of heavy-ion collisions, which combines the LO classical solutions with constraints on $Q_S$ from DIS experiments on the proton and on nuclei. 

However the LO description of the Glasma is limited because the classical fields are unstable to NLO  quantum fluctuations that break boost invariance, growing exponentially in the square root of the proper time. As we later discuss, a careful treatment of such NLO modes shows that the dominant contributions can be resummed and absorbed into a classical-statistical description of the evolution. A key difference from the prior LO description is that the resummed classical-statistical evolution is now in 3+1 dimensions, involving both transverse and longitudinal degrees of freedom. This distinction is of fundamental importance in the subsequent description of the thermalization process in weak coupling. 

In Section~\ref{sec:classicalstatistical}, we will discuss how this classical-statistical description fits into the general weak coupling classification of the evolution of quantum fields and shall outline the power counting that delineates the applicability of this approximation and its subsequent matching to kinetic theory. We will also describe in Section~\ref{sec:classicalstatistical} universal features of the Glasma that makes its study interesting in its own right.

\subsection{Multi-particle production in strong fields}
\label{sec:multi-particle}
To compute multi-particle production systematically in the collision of the CGC gluon ``shockwaves", we will begin with the first principles Lehmann-Symanzik-Zimmermann (LSZ) formalism in QFT. For simplicity, we consider here a self-interacting $\phi^3$ scalar theory; our discussion extends straightforwardly to the Yang-Mills case. 

\begin{figure}
\centering
\includegraphics[width=0.6\linewidth]{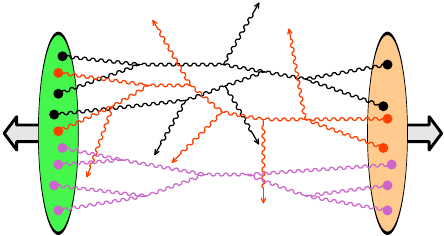}\\
\vspace{1em}
\includegraphics[width=0.6\linewidth]{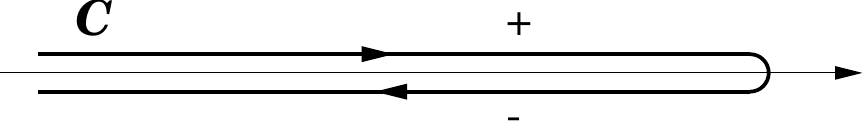}
\caption{Top panel: multi-particle production from cut ``vacuum-vacuum" graphs connecting time dependent sources of the two nuclei after the collision. 
From \cite{Gelis:2010nm}. Bottom panel: the Schwinger-Keldysh closed time 
contour on which the sources and fields are defined.}
\label{fig:SK}
\end{figure}

In the LSZ formalism, the amplitude for $n$-particles in the ``out" state generated from the ``in-vacuum" can be expressed as 
\begin{eqnarray}
&\langle&\! p_{1, {\rm out}}\cdots p_{n,{\rm out}}|0_{\rm in} \rangle = \frac{1}{Z^{n/2}}\int
\Bigg[\prod_{i=1}^n d^4 x_i 
e^{ip_i\cdot x_i}\nonumber \\
&\times&\left(\partial_{x_i}^2+m^2\right){\delta\over \delta J(x_i)}\Bigg] \exp\left(i {\cal V}\right) \, .
\label{eq:LSZ-amp}
\end{eqnarray}
Here $p_1\cdots p_n$ denote the momenta of the produced particles and the ``in-out" vacuum-amplitude $\langle 0_{\rm out}|0_{\rm in}\rangle = \exp(i{\cal V})$, where ${\cal V}$ is the sum of all connected vacuum-vacuum diagrams coupled to external sources. An illustration of multi-particle production for the problem at hand is shown in \Fig{fig:SK}.

In QFT computations, one usually sets $J=0$ after the functional differentiation and $\langle 0_{\rm out}|0_{\rm in}\rangle$ is a pure phase. When $J$ is physical, 
$|\langle 0_{\rm out}|0_{\rm in}\rangle|^2 = \exp(- 2\, {\rm Im\,} {\cal V})\neq 1$. In computing multi-particle production in this context, it is useful to employ\footnote{For other discussions of the SK formalism in the context of the CGC and the Glasma, see \cite{Jeon:2013zga,Wu:2017rry,Leonidov:2018zdi}. For a recent discussion in the context of thermal field theory, see \cite{Ghiglieri:2020dpq}.} the Schwinger-Keldysh (SK) QFT formalism~\cite{Schwinger:1960qe,Keldysh:1964ud}. One introduces $+$ and $-$ vertices with opposite signs of 
the coupling in Feynman diagrams, and likewise for the sources $J_{\pm}$. The  corresponding ``$+$" and ``$-$" fields live on the upper and lower segments of a closed time contour 
ranging forward in time from $t=-\infty$ on the upper contour and back to $-\infty$ on the lower contour, as shown in \Fig{fig:SK}. Time ordered ``$++$"  (anti-time ordered ``$--$") Green's functions ``live" on the upper (lower) contour, and the mixed $+-$ ``Wightman" functions connect the upper and lower contours. 

Following the LSZ formalism, the probability to produce $n$-identical particles is 
\begin{equation}
P_n = \frac{1}{n!} \prod_{i=1}^n {d^3 p_i\over (2\pi)^3 2 E_{p_i}} |\langle p_{1, {\rm out}}\cdots p_{n,{\rm out}}|0_{\rm in} \rangle|^2 \, ,
\end{equation}
where $E_{p_i}^2 = p_i^2 +m^2$. Plugging the expression for the amplitude in \Eq{eq:LSZ-amp} into the r.h.s, one can express the result as~\cite{Gelis:2006yv}
\begin{eqnarray}
P_n = \frac{1}{n!} {\cal D}^n \exp\left( i {\cal V}[J_+] - i{\cal V}[J_-] \right)|_{J_+=J_- = J} \, ,
\label{eq:prob}
\end{eqnarray}
with 
\begin{equation}
{\cal D} = \int_{x,y} \!\!\!\!\!\!Z\,G_{+-}^0(x,y)\frac{\left(\partial_{x_i}^2+m^2\right)}{Z}\frac{\left(\partial_{y_i}^2+m^2\right)}{Z}{\delta\over \delta J_+(x)}{\delta\over \delta J_-(y)}\,. 
\end{equation} 
Here $\int_x = d^4 x$, $G_{+-}^0(x,y) = \int {d^3 p_i\over (2\pi)^3 2 E_{p_i}} e^{ip\cdot (x-y)} \equiv \theta(p^0) \delta^{(3)}(x-y)$ and $Z$ is the residue of the pole of the renormalized propagator. 

The action of the operator ${\cal D}$ can be understood as follows. The ``$+$" piece with $\frac{\left(\partial_{x_i}^2+m^2\right)}{Z}{\delta\over \delta J_+(x)}$ acts on a particular diagram in the connected sum of vacuum-vacuum connected diagrams ${\cal V}[J_+]$ by removing a source $J_+$ and then amputating the renormalized propagator to which it is attached. The same procedure is followed for the ``$-$" piece; the two amputated propagators are then sewn together by the renormalized ``cut" propagator $Z G_{+-}^0$. 

Computing $P_n$ in a theory with physical sources is difficult because one also has to compute the disconnected vacuum-vacuum graphs for each $n$. However if we define a generating functional $F(z) = \sum_n z^n P_n$, \Eq{eq:prob} gives 
\begin{equation}
F(z) = \exp\left(z {\cal D}\right) \exp\left( i {\cal V}[J_+] - i{\cal V}[J_-] \right)|_{J_+=J_- = J} \,,
\label{eq:gen-functional}
\end{equation}
and successive differentiation of \Eq{eq:gen-functional} with respect to $z$ (and setting $z=1$) generates the $n$-particle correlators $\langle n(n-1)(n-2)\cdots\rangle$. These moments do not require one to compute 
the disconnected vacuum-vacuum graphs, since they also appear in the normalization of $P_n$ and therefore cancel out\footnote{Such cancellations are seen in the Abramovsky-Gribov-Kancheli (AGK) rules~\cite{Abramovsky:1973fm} that implement the combinatorics of cut/uncut vacuum-to-vacuum graphs in Reggeon field theory~\cite{Gelis:2006ye}.} in the moments. 

This is illustrated by expressing the r.h.s of \Eq{eq:gen-functional} for $z=1$ as 
\begin{equation}
\exp\left(i{\cal V_{\rm SK}}[J_+,J_-]\right) = \exp\left({\cal D}\right) \exp\left( i {\cal V}[J_+] - i{\cal V}[J_-] \right)\,,
\end{equation}
where now $i{\cal V_{\rm SK}}[J_+,J_-]$ represents the sum over all vacuum--to--vacuum connected graphs that live on the SK closed time contour. One can then 
express the inclusive multiplicity as~\cite{Gelis:2006yv}
\begin{equation}
\langle N\rangle = \int_{x,y} ZG_{+-}^0(x,y) \left[\Gamma_+(x) \Gamma_-(y) + \Gamma_{+-}(x,y)\right]_{J_\pm=J} \,,
\label{eq:inclusive-mult}
\end{equation}
with the amputated one-point and two-point Green's functions in the Schwinger-Keldysh formalism defined, respectively, as 
\begin{equation}
\Gamma_{\pm}(x)= \Delta_x^{\rm R}\frac{\delta i{\cal V}_{\rm SK}}{\delta J_\pm (x)}\,;\,
 \Gamma_{+-}(x,y) = \Delta_x^{\rm R}\Delta_y^{\rm R}
\frac{\delta^2 i{\cal V}_{\rm SK}}{\delta^2 J_+ (x) J_-(y)}\,,
\end{equation}
with $\Delta_x^{\rm R} = \frac{\partial_x^2+m^2}{Z}$. 

In summing over all the nodes of all the trees connecting $\Gamma_+(x)$ to the sources,  the time (anti-time) ordered Feynman propagators in each tree  on the upper (lower) SK contour 
are recursively converted to retarded propagators: $G_R=G_{++}-G_{+-}\equiv G_{-+}-G_{--}$. This is equivalent to solving the classical equations of motion with retarded boundary conditions when $J_\pm = J$ ! A further important result is that the renormalized cut propagator $\Gamma_{+-}$ is obtained by solving the small fluctuation equations of motion in the classical background, also as an initial value problem with retarded boundary conditions. 

As  previously discussed, the classical fields, and sources thereof, of the colliding CGCs are static shockwaves; as such, they do not spontaneously decay and are thus part of the nuclear wavefunction.  After the collision, the colored sources become time dependent. Thus, 
$\Gamma_\pm$  in \Eq{eq:inclusive-mult} corresponds to $\partial_x^2 {\cal A}_{\pm,{\rm cl.}}^\mu$ where ${\cal A}_{\pm,{\rm cl.}}^\mu$ is the time dependent ${\mathcal O}(1/g)$ Glasma field in the forward lightcone. The two-point function $\Gamma_{+-}(x,y)$ in \Eq{eq:inclusive-mult} is ${\mathcal O}(1)$ and therefore NLO in the power counting for the inclusive multiplicity in the Glasma. The formalism can be extended to higher orders in $\alpha_S$. Its generalization to higher multiplicity moments was developed in \cite{Gelis:2006cr}.

\subsection{The LO Glasma: classical gluon fields from shockwave collisions} 
\label{sec:LO-Glasma}

Since at LO in our power counting only the product $\Gamma_+(x) \Gamma_-(y)\equiv \partial_x^2 {\cal A}_+^\mu \partial_x^2 {\cal A}_-^\nu$ in \Eq{eq:inclusive-mult} contributes, one obtains for a fixed distribution of lightcone sources $\rho_{\pm,1,2} = \rho_{1,2}$ (where $1,2$ denote the two nuclei)~\cite{Gelis:2007kn}
\begin{equation}
{d\langle N\rangle_{\rm LO} \over dY d^2 p_\perp}[\rho_1,\rho_2]=\frac{1}{16\pi^3}\int_{x,y}\! \!\!\Delta_x^{\rm R} \Delta_y^{\rm R}\varepsilon_\lambda^\mu \varepsilon_\lambda^\nu 
{\cal A}_\mu (x) {\cal A}_\nu (y) \,,
\label{eq:incl-mult-LO}
\end{equation}
where repeated indices are summed over. Note also that ${\cal A} (x)\equiv {\cal A}_\mu [\rho_1,\rho_2](x)$ and $m=0$ in $\Delta_{x,y}^{\rm R}$. An integration by parts 
\begin{equation}
\int d^4 x \,e^{ip\cdot x} \partial_x^2 {\cal A}_\mu (x) =  \int_{x^0\rightarrow +\infty}\!\!\!\! \!\!\!\!\!\!\!\!\!\!\!\!d^3 x \,e^{ip\cdot x}\left(\partial_0 - i E_p\right) {\cal A}_\mu (x) \,,
\label{eq:classical-evolution}
\end{equation}
shows that \Eq{eq:incl-mult-LO} can be computed by solving the classical YM equations in \Eq{eq:CYM-nucleus} [with $J^\mu = \delta^{\mu+}\delta(x^-)\rho_1(x_\perp) + \delta^{\mu-}\delta(x^+)\rho^2(x_\perp)$ and ${\cal A}_{\mu}(x)|_{x^0=-\infty}=0$] to determine ${\cal A}_{\mu}(x)$. 

\begin{figure}
\centering
\includegraphics[width=0.7\linewidth]{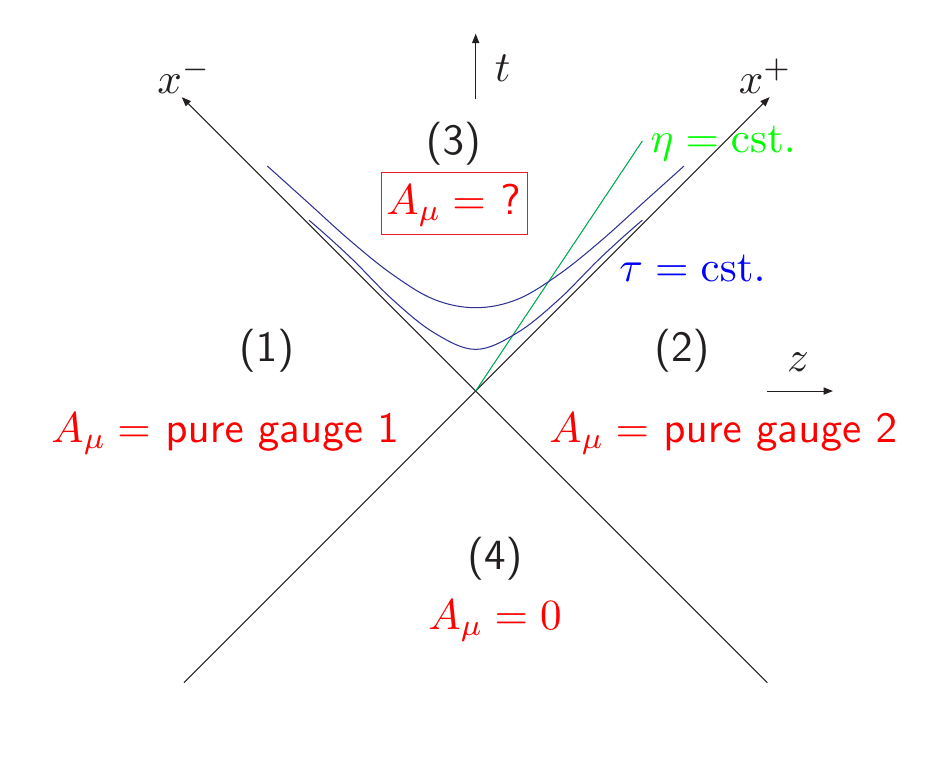}
\caption{Spacetime diagram of gauge field configurations. Before the collision, the gauge fields are pure gauge solutions 
with zero field strength. (In the text, the pure gauge solution of the right moving nucleus is denoted by $A_{1,{\rm cl.}}^i$ and that of the left moving nucleus is denoted by $A_{2,{\rm cl.}}^i$.)  After the collision, 
the gauge field solution (${\cal A}^{i,\eta}$ in the text) 
correspond to finite field strengths in the Glasma.
From \cite{Lappi:2006fp}.}
\label{fig:shockwaves}
\end{figure}

For the following discussion, it will be convenient to introduce the $(\tau,\eta,x_\perp)$ coordinate system, where the proper time $\tau=\sqrt{(x^0)^2 - (x^3)^2}$ and the spacetime rapidity $\eta = \frac{1}{2}\log(\frac{x^0+x^3}{x^0-x^3})$, and $g_{\mu\nu} = {\rm diag}(1,-\tau^2,-1,-1)$. 
A convenient gauge to solve the YM equations in the forward lightcone is the Fock-Schwinger gauge ${\cal A}^\tau \equiv x^+ {\cal A}^- +x ^- {\cal A}^+=0$. 
In this gauge\footnote{A perturbative solution was also found in Lorenz gauge $\partial_\mu {\cal A}^\mu=0$
~\cite{Kovchegov:1997ke}.}, the solution to the YM equations are manifestly boost invariant: ${\cal A}^{\mu}(\tau,\eta,x_\perp)\equiv {\cal A}^{\mu}(\tau,x_\perp)$ and one obtains~\cite{Kovner:1995ja,Kovner:1995ts,Gyulassy:1997vt},
\begin{equation}
{\cal A}^{i} = A_{1,{\rm cl.}}^i + A_{2,{\rm cl.}}^i \,;\, {\cal A}^{\eta} = \frac{ig}{2} [A_{1,{\rm cl.}}^i,A_{2,{\rm cl.}}^i]\,,
\label{eq:KMW}
\end{equation}
with $\partial_\tau {\cal A}^i=0$ and $\partial_\tau {\cal A}^{\eta}=0$ at $\tau=0^+$. This solution is obtained by matching the delta-functions on the lightcone wedges in \Fig{fig:shockwaves}. 

Since the gauge fields are functionals of $\rho_{1,2}$, the full average inclusive multiplicity in the Glasma is obtained by averaging over many nuclear collisions, each with its distribution of color sources in the two nuclei\footnote{Owing to color confinement at distance scales $1/\Lambda_{\rm QCD}$, one requires $\int_0^{1/\Lambda_{\rm QCD}} d^2 x_\perp \rho_{1,2}^a =0$ for each such configuration.}. This can be expressed as 
\begin{eqnarray}
{d\langle\langle N\rangle\rangle_{\rm LO} \over dY d^2 p_\perp} &=& \int [D\rho_1] [D\rho_2] W_{Y_{\rm beam}-Y}^{\rm MV} [\rho_1] W_{Y_{\rm beam}+Y}^{\rm MV}[\rho_2] \nonumber \\
&\times& {d\langle N\rangle_{\rm LO} \over dY d^2 p_\perp}[\rho_1,\rho_2]\,,
\label{eq:two-nuclei}
\end{eqnarray}
where $Y_{\rm beam} = \log(\sqrt{s}/m_N)$ is the beam rapidity and  $W_{Y_{\rm beam}-Y}^{\rm MV}$ ($W_{Y_{\rm beam}+Y}^{\rm MV}$) are the weight functionals in the MV model in \Eq{eq:Gaussian-weight-functional} and at LO are independent of $Y_{\rm beam}- Y$ ($Y_{\rm beam}+ Y$). 

With the initial conditions in \Eq{eq:KMW}, the YM equations for $\tau=0^+$ can be solved perturbatively to lowest non-trivial order in ${\mathcal O}(\frac{\rho_1}{\nabla_\perp^2}\frac{\rho_2}{\nabla_\perp^2})$;  in this ``dilute-dilute" approximation, one obtains the following for identical nuclei:
\begin{equation}
{d\langle\langle N\rangle\rangle_{\rm LO} \over dY d^2 p_\perp} = {\pi R_A^2}\frac{g^6 \mu_A^4}{(2\pi)^4}\frac{2 N_c(N_c^2-1)}{p_\perp^4} {\cal L}(p_\perp,\Lambda) \, . 
\label{eq:Gunion-Bertsch}
\end{equation}
This result, which agrees with the pQCD bremsstrahlung formula first derived by Gunion and Bertsch~\cite{Gunion:1981qs} is valid for $p_\perp \gg g^2\mu_A$ and ${\cal L}(p_\perp,\Lambda)$ is a logarithmically divergent function, screened at $\Lambda\approx \Lambda_{\rm QCD}$. 

From our dipole model discussion [see \Eq{eq:K-T} and the related discussion], $Q_S^2 \propto G_A(x,p_\perp^2)$, where $p_\perp$ is the momentum conjugate to the dipole size. This suggests that \Eq{eq:Gunion-Bertsch} (employing $Q_S^2\propto\mu_A^2$, as noted in footnote~\ref{ft:MV}) can be 
generalized to a ``$k_\perp$ factorization" form ${d\langle\langle N\rangle\rangle_{\rm LO} \over dY d^2 p_\perp} \propto \alpha_S \int dk_\perp^2 \phi_A(x_1,k_\perp^2) \phi_B(x_2,(k_\perp-p_\perp)^2)$.  Here  
$\frac{\phi_{A,B}(x,k_\perp^2)}{k_\perp^2}$ is the Fourier transform of the dipole scattering amplitude\footnote{This distribution is distinct from the WW-distribution and coincides with it only for large $k_\perp$~\cite{Kharzeev:2003wz,Blaizot:2004wu}.} in each of the hadrons discussed in \Sec{sec:RG-GS}. This $k_\perp$ factorization formula \cite{Gribov:1984tu,Blaizot:1987nc} is widely used in phenomenological studies of hadron-hadron collisions. 

The dilute-dilute analytical approximation for shockwave collisions  can be generalized to compute the inclusive multiplicity to lowest order ${\mathcal O}(\frac{\rho_1}{\nabla_\perp^2})$ in one of the 
sources but to all orders ${\mathcal O}((\frac{\rho_2}{\nabla_\perp^2})^n)$ in the other. In this ``dilute-dense" case as well,  the inclusive gluon multiplicity can be expressed as a $k_\perp$-factorized convolution of the unintegrated gluon distributions in the projectile and target. It is valid for $Q_{S,1}^2(x_1) \ll Q_{S,2}^2(x_2)$, corresponding to the forward (or backward) kinematic regions of the shockwave collision where the parton momentum fractions are  $x_1\gg x_2$ . Alternately, it can be a good approximation in proton-nucleus collisions, where $Q_{S,A}^2\sim A^{1/3} Q_{S,p}^2$~\cite{Kovchegov:1998bi,Dumitru:2001ux}. 
 
\subsection{Non-perturbative evolution of high occupancy fields} 
\label{sec:Glasma-Evolution}

\subsubsection{Real time evolution of boost invariant fields on the lattice}
\label{sec:Lattice}

While analytical results for the inclusive multiplicity are available only in limited kinematic regions, the YM equations for shockwave collisions can be solved numerically to all orders 
${\mathcal O}((\frac{\rho_{1,2}}{\nabla_\perp^2})^n)$~\cite{Krasnitz:1997zj,Krasnitz:1998ns} to obtain the full non-perturbative result for \Eq{eq:two-nuclei}~\cite{Krasnitz:1999wc,Krasnitz:2000gz,Krasnitz:2001qu,Lappi:2003bi,Krasnitz:2003jw}. 
Hamilton's equations are solved in the Fock-Schwinger gauge ${\cal A}^\tau=0$ with the initial conditions at $\tau=0$ specified by \Eq{eq:KMW}. To preserve gauge invariance, lattice gauge theory techniques can be adapted to this problem. The boost invariance of the LO shockwave gauge fields provides a significant simplification whereby the (3+1)-dimensional [(3+1)-D] Kogut-Susskind QCD lattice Hamiltonian~\cite{Kogut:1974ag} can be ``dimensionally reduced" to the (2+1)-D form~\cite{Krasnitz:1998ns}
\begin{align}\label{eq:ham}
  aH &= \sum_{\xt} \left[ \frac{g^2 a}{\tau} {\rm tr}\, E^i E^i + \frac{2 \tau}{g^2 a}\left(N_c-{\rm Re}\,{\rm tr}\, U_{1,2}\right)\right.\nonumber\\
   \left.\right. &~~~~~~~~~~~\left.+\frac{\tau}{a}{\rm tr}\, \pi^2 + \frac{a}{\tau}\sum_i{\rm tr} \left(\Phi - \tilde{\Phi}_i\right)^2\right]\,.
\end{align}
In \Eq{eq:ham} the trace refers to $SU(2)$ color and the sum is over all discretized cells with lattice spacing $a$ in the transverse plane. For clarity, we have omitted the cell index $j$ for all quantities in \Eq{eq:ham}. Further, the   
$E^i$ with $i\in\{1,2\}$ are the components of the transverse electric field living on each site; discretizing the initial conditions gives $E^i=0$ at $\tau=0$.  The spatial plaquette of link 
variables  $U^{i}_{j}$,
\begin{equation}\label{eq:plaq}
 U_{1,2}^j = U^1_j \, U^2_{j+\hat{e}_1} \, U^{1\dag}_{j+\hat{e}_2} \, U^{2\dag}_j\,,
\end{equation}
(where $+\hat{e_i}$ indicates a shift from $j$ by one lattice site in the $i=1,2$ transverse direction) represents the squared longitudinal magnetic fields in the Glasma.  In \Eq{eq:ham}, we  represent  $A_\eta(\tau,x_\perp)$ as an adjoint scalar field $\Phi$ because, as 
a result of boost invariance, it transforms covariantly under $\eta$-dependent gauge transformations:  
\begin{equation}
  \tilde{\Phi}_{i}^j = U_j^i \Phi_{j+\hat{e}_i} U_j^{i\dag}\,. 
  \label{eq:Phi-field}
\end{equation}
Finally, $\pi=E_\eta=\dot{\Phi}/\tau$ in \Eq{eq:ham} represents the longitudinal electric field. 

Details pertaining to the numerical simulations of the real time evolution of gauge fields were given in \cite{Krasnitz:1998ns,Lappi:2003bi}. In the early work, only uniform sheets of nuclei were considered 
with constant ($x$ independent) values of $Q_S$. These were subsequently relaxed to consider finite nuclei~\cite{Krasnitz:2002ng,Krasnitz:2002mn}; more realistic simulations with 
event-by-event simulations of RHIC and LHC collisions were later developed in the IP-Glasma model that we shall discuss shortly~\cite{Schenke:2012wb}.
 
 As anticipated, the numerical results reproduce the perturbative result in \Eq{eq:Gunion-Bertsch} at large $k_\perp \gg Q_S$. However, unlike in that expression, there is 
no logarithmic factor ${\cal L}(k_\perp,\Lambda_{\rm QCD})$. At momenta $k_\perp < Q_S$, the $1/k_\perp^4$ distribution is modified to a form that is well fit by a Bose-Einstein exponential distribution~\cite{Krasnitz:2003jw}. Even more remarkably, the non-linear dynamics generates a plasmon mass\footnote{This plasmon mass is parametrically larger than the confining scale; its properties were investigated recently using a number of approaches~\cite{Dumitru:2014nka,Boguslavski:2019aba}.} that screens the momentum distribution in the infrared~\cite{Krasnitz:2000gz,Lappi:2017ckt}. The 
energy density is therefore well-defined at all proper times without infrared or ultraviolet divergences~\cite{Lappi:2006hq}. 

 \subsubsection{Glasma flux tubes}
 \label{sec:Glasma-tubes}
 
An interesting consequence of the LO Glasma solution is that the Weiz\"{a}cker-Williams plane polarized $E$ and $B$ fields in the colliding CGCs become purely longitudinal immediately after the collision at 
$\tau=0^+$; $E_\eta, B_\eta\neq 0$ and $E_i,B_i=0$. It was pointed out in \cite{Kharzeev:2001ev} that this configuration satisfies the identity 
\begin{equation}
Q_{\rm CS} = \frac{\alpha_S}{2\pi}\int d^4 x\, {\rm Tr} \,E_\eta\cdot B_\eta\,,
\label{eq:CS}
\end{equation}
where the topological charge $Q_{\rm CS}= \frac{\alpha_S}{16 \pi}\int d^3 x\, K^0$ and $K^\mu$ is the Chern-Simons current. A neat interpretation~\cite{Lappi:2006fp,Chen:2015wia} of this result is that the YM equations at $\tau=0^+$ can be 
expressed as $\nabla\cdot E= \rho_{\rm el.}$ and $\nabla\cdot B=\rho_{\rm mag.}$, where $\rho_{\rm el.}$, $\rho_{\rm mag.}$ are respectively electric and magnetic charge densities\footnote{These induced charge densities are proportional to the commutators $\delta^{ij}[A_{1,{\rm cl.}}^i,A_{1,{\rm cl.}}^j]$ and $\epsilon^{ij}[A_{1,{\rm cl.}}^i,A_{1,{\rm cl.}}^j]$ respectively.}
on the gluon shockwaves after the collision. 

\begin{figure}
\centering
\includegraphics[width=0.35\linewidth]{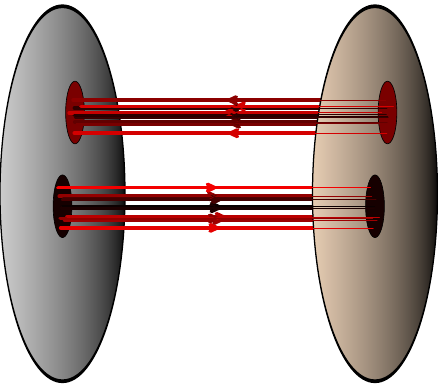}
\caption{Glasma flux tubes: Boost invariant LO Glasma configurations of transverse size $1/Q_S$ at $\tau=0^+$ with parallel $E_\eta$ and $B_\eta$, corresponding to finite Chern-Simons charge. 
Such configurations decay rapidly and are unstable to quantum fluctuations. From \cite{Dumitru:2008wn}.}
\label{fig:fluxtubes}
\end{figure}

As sketched in \Fig{fig:fluxtubes}, the induced electric and magnetic charges generate a ``stringy" Glasma flux tube~\cite{Dumitru:2008wn} of chromo-electromagnetic fields that is uniform in rapidity stretching between the fragmentation regions of the nuclei and are color screened~\cite{Krasnitz:2002mn} on transverse distance scales $\geq 1/Q_S$.

One can straightforwardly compute the energy densities and pressures in the Glasma from the different components of the stress-energy tensor\footnote{Note that 
\begin{equation}
T^{\mu\nu} = -g^{\mu\alpha}g^{\nu\beta}g^{\gamma\delta} F_{\alpha\gamma} F_{\beta\delta} + \frac{1}{4} g^{\mu\nu}g^{\alpha\gamma}g^{\beta\delta} F_{\alpha\beta}F_{\gamma\delta}.
\label{eq:Glasma-stresstensor2}
\end{equation}
}. We obtain $\mathcal E = 2\mathcal P_T + \mathcal P_L$ where 
\begin{eqnarray}
& &\mathcal P_T \equiv  \frac{1}{2}\left(T^{xx} + T^{yy}\right) = {\rm Tr}\left( F_{xy} + E_\eta^2\right)\,\, \nonumber \\
& & \mathcal P_L \equiv \tau^2 T^{\eta\eta} = \frac{1}{\tau^2}{\rm Tr}\left(F_{\eta i}^2 + E_i^2\right)- {\rm Tr}\left( F_{xy} + E_\eta^2\right)\,.
\label{eq:Glasma-stresstensor}
\end{eqnarray}
At the earliest times after the collision $\tau=0^+$, as noted, only the longitudinal $E_\eta$ and $B_\eta= F_{xy}$ fields are non-zero. Equation \eq{eq:Glasma-stresstensor} then immediately gives 
$\mathcal P_T = \mathcal E$ and $\mathcal P_L= -\mathcal E$. Thus, at the earliest times the pressure in the Glasma is purely transverse; after initial transverse dynamics, the longitudinal pressure $\mathcal P_L\rightarrow 0$ from below by $\tau\sim 1/Q_S$. Since the Glasma at LO is conformal, the energy density satisfies $\mathcal E = 2 \mathcal P_T$ at this time. 

Stringy models capture essential features of confining dynamics in QCD~\cite{Bali:2000gf}. In high energy collisions, they have a long history and capture the bulk features of the spectrum of 
multi-particle production~\cite{Andersson:1978vj,Artru:1979ye}; they underlie event generators such as PYTHIA~\cite{Andersson:1983ia}. These models however screen color at distance scales $1/\Lambda_{\rm QCD}$ and  carry only electric flux and no magnetic flux; particle production is assumed to arise from the Schwinger mechanism~\cite{Andersson:1978vj}.  It is remarkable nevertheless to observe that similar  stringy solutions emerge from the more fundamental framework of classical YM equations. 

Motivated by this stringy picture, we expect the number of gluons per unit rapidity to equal the number of flux tubes 
[$S_\perp/(1/Q_S^2)$] times the gluon occupancy in a flux tube ($2(N_c^2-1)/{{\bar \alpha}_S}/(2\pi)^3$) multiplied by a non-perturbative coefficient of ${\mathcal O}(1)$. 
Extracting the number density from the correlator of gauge fields at $\tau\sim 1/Q_S$~\cite{Krasnitz:2000gz}, one indeed finds that\footnote{Here and henceforth for simplicity   
the path integral over gauge fields (moot at LO) and over sources ($\langle \langle \rangle\rangle$) is implicit.} 
\begin{equation}
{d N_{\rm LO} \over dY} = c_N \frac{2\, (N_c^2-1)} {(2\pi)^3}\,\frac{Q_S^2 S_\perp}{{\bar \alpha}_S}\,,
\label{eq:LO-incl-mult}
\end{equation}
where $S_\perp$ is the transverse area of the collision, ${\bar \alpha}_S = \alpha_S N_c/\pi$, and $c_N$ is a gluon liberation coefficient~\cite{Mueller:1999fp} estimated from the numerical simulations to be $c_N=1.1$ with $10\%$ accuracy~\cite{Lappi:2007ku}. 

The YM simulations can also be extended to compute two particle correlations in the Glasma~\cite{Lappi:2009xa}:
\begin{equation}
\frac{d^2 N_{\rm LO}^{\rm conn.}}{dY_1 d^2 p_{\perp} dY_2 d^2 k_\perp} = \frac{\kappa_2}{(N_c^2-1)Q_S^2 S_\perp}\, {dN_{\rm LO} \over dY_1 d^2 p_\perp}\, \,{dN_{\rm LO} \over dY_2 d^2 k_\perp} \,,
\end{equation}
where $\kappa_2$ is a non-perturbative constant\footnote{The results have a weak dependence on the ratio $m/Q_S$, where $m$ is an infrared lattice regulator.}. Again the numerical simulations bear out the Glasma flux tube interpretation: the likelihood that two particles are correlated is suppressed by the number of flux tubes, and non-factorizable color connected graphs are suppressed by ${\mathcal O}(1/N_c^2)$. Perturbative arguments suggest that this picture can be extended to $n$-particle cumulants and that the $n$-particle multiplicity distribution that generates these cumulants is a negative binomial distribution~\cite{Gelis:2009wh}. For $n$-particle multiplicities, this expectation is confirmed by non-perturbative numerical simulations~\cite{Schenke:2012hg}. 

\subsubsection{The IP-Glasma model\label{sec:ipglasma}}
In the discussion thus far, color charge fluctuations on the scale $1/Q_S$ provide the only structure in the colliding gluon shockwaves. However, nucleon distributions in nuclei are not uniformly smooth and can fluctuate from event to event. These fluctuations in nucleon positions are extremely important for understanding 
key features of the data such as the azimuthal moments $v_n$ of the flow distributions at low momenta~\cite{Alver:2010gr,Alver:2010dn}. 
Another important ingredient in the realistic modeling of heavy-ion collisions is the dependence of the saturation scale in the nuclei on 
$x$ (or, equivalently, $\sqrt{s}$), which describes the variations of particle multiplicites in energy and rapidity at RHIC and the LHC. 

We will outline here the IP-Glasma model~\cite{Schenke:2012wb,Schenke:2012hg,Schenke:2013dpa,Schenke:2014tga}, and improvements thereof, which incorporates 
the fluctuations in the nucleon positions to construct event-by-event lumpy color charge distributions and  corresponding gluon field configurations in the LO Glasma framework. As we will also discuss, the energy dependence of these configurations at a given $Y$ or $\sqrt{s}$ is determined by the saturation scales in the two nuclei. 

An essential input is the dipole cross-section of the proton. We consider here the IP-Sat model \cite{Bartels:2002cj,Kowalski:2003hm} which, as discussed in Sec.~\ref{sec:DIS-dipole}, is an impact parameter dependent generalization of the MV model. As noted, high precision combined data from the H1 and ZEUS collaborations~\cite{Aaron:2009aa,Abramowicz:1900rp} are used to constrain the parameters of the model and excellent fits are obtained~\cite{Rezaeian:2012ji}. 

The dipole cross-section for each nucleus at a given $x$ is constructed by taking the product of the S-matrices corresponding to the dipole cross-sections of overlapping nucleons at a given spatial location 
${\mathbf{x}_\perp}$. It can be expressed as~\cite{Kowalski:2007rw}
\begin{equation}
\frac{1}{2}\frac{\mathrm{d}\sigma^{\textrm{A}}_{\textrm{dip}}} {{\mathrm{d}}^2 {\mathbf{x}_\perp}}  
=1-e^{-\frac{\pi^2}{2N_{c}} {\mathbf{r}_\perp}^2 \alpha_{S}(Q^2) x G(x,Q^2)\sum_{i=1}^A T_p({\mathbf{x}_\perp}-{\mathbf{x}_T}^i)}\,,
\label{eq:nuc-dipole}
\end{equation}
where $T_p$ stands for the Gaussian thickness function for each of the $A$ nucleons in each nucleus and $Q^2=4/\mathbf{r}_\perp^2+Q_0^2$, with $Q_0$ fixed by the HERA inclusive data. 
The gluon distribution $x G(x,Q^{2})$ is parametrized at the initial scale $Q_0^2$ and then evolved up to the scale $Q^2$ using LO DGLAP-evolution. We define the nuclear saturation scale $Q_S=1/\sqrt{\mathbf{r}_{\perp,s}^2}$ at  $r_\perp = r_{\perp,s}$ for which the argument of the exponential in \Eq{eq:nuc-dipole} equals one-half. To obtain the spatial dependence of $Q_S$, one self-consistently solves $x = 0.5\,Q_S({\mathbf{x}_\perp},x)/\sqrt{s}$ for every ${\mathbf{x}_\perp}$. 

\begin{figure}
\centering						
 \includegraphics[width=0.25\textwidth]{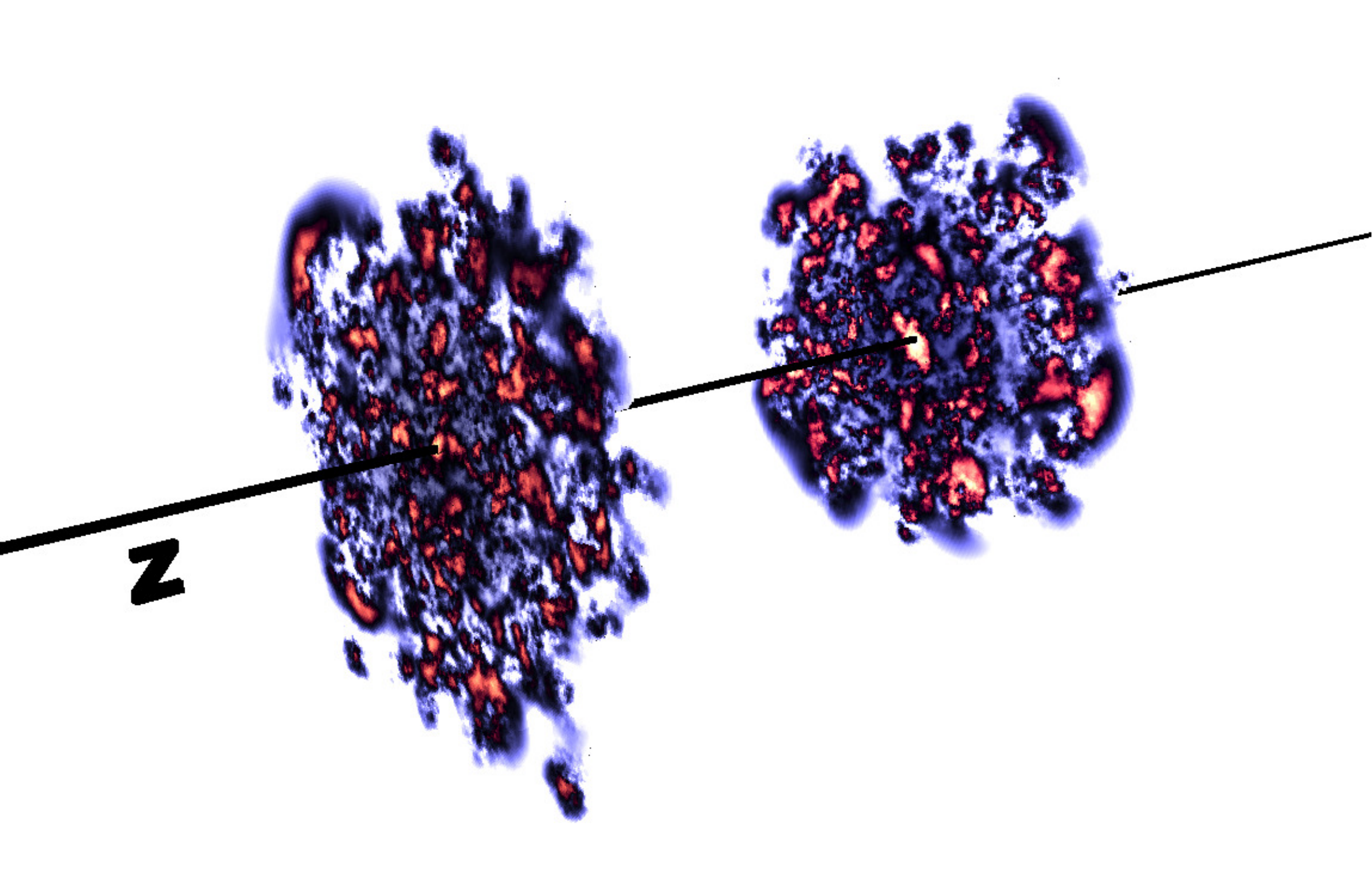}
 \vskip -0.2in
 \includegraphics[width=0.3\textwidth]{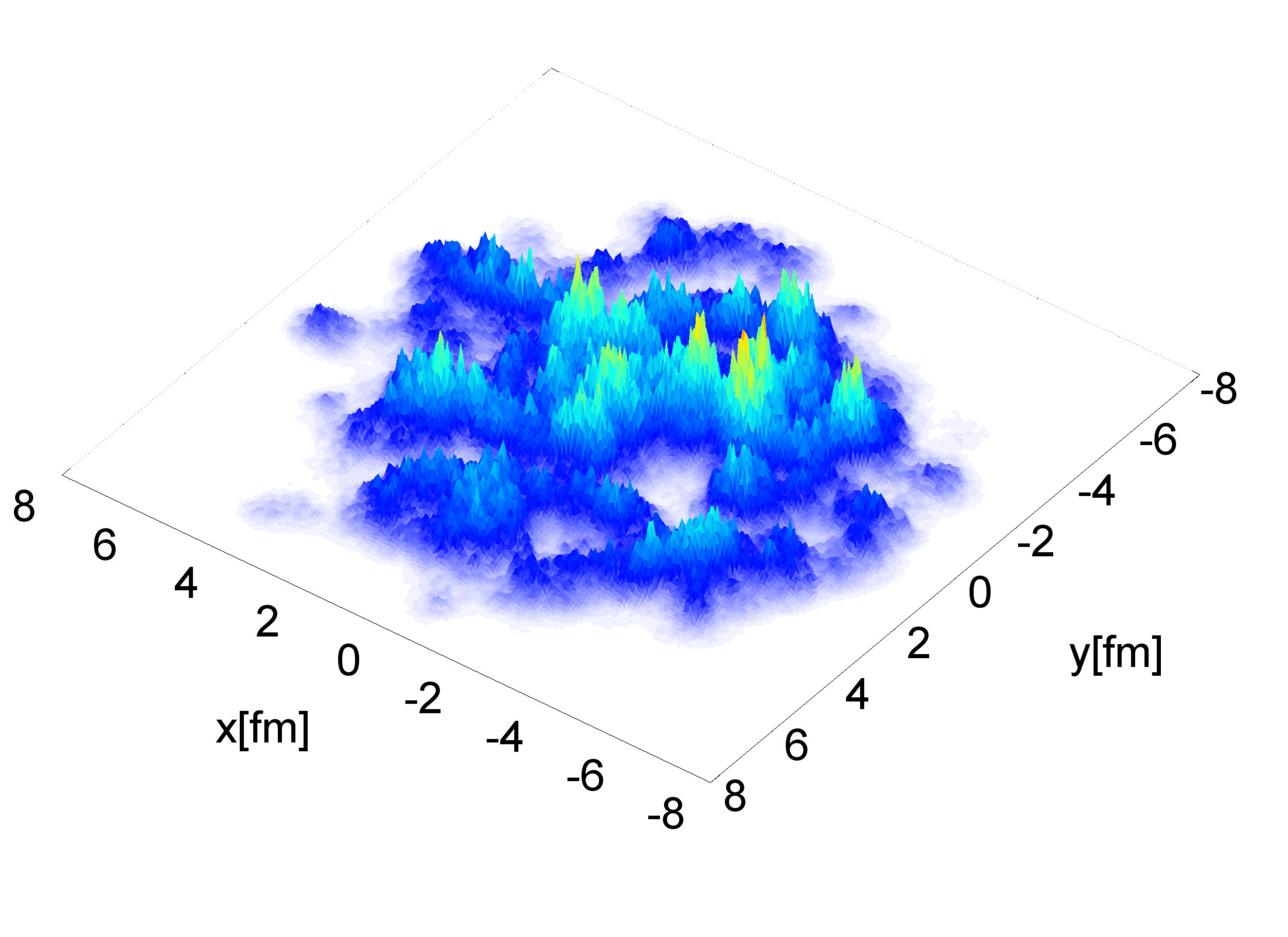}
  \caption{Top panel: collisions of nuclei with sub-nucleon color charge fluctuations determined by the IP-Sat model. Bottom panel: the LO energy density in the Glasma at $\tau= 1/Q_S.$  From 
  \cite{Schenke:2012hg}.}
  \label{fig:IP-Glasma}
\end{figure}

The result of this procedure is a lumpy distribution of $Q_S^2({\mathbf{x}_\perp}, x)$  denoting the sub-nucleon structure of the nucleus. Since the IP-Sat model is a simple generalization of the MV model,
one can extract the variance of the color charge density $g^2\mu_A^2({\mathbf{x}_\perp})$ at each $x$ from $Q_S^2({\mathbf{x}_\perp}, x)$~\cite{Lappi:2007ku}. One then samples random color charges 
$\rho^a({\mathbf{x}_\perp})$ on a transverse lattice:
\begin{equation}
  \langle \rho_k^a({\mathbf{x}_\perp})\rho_l^b({\mathbf{y}_\perp})\rangle = \delta^{ab}\delta^{kl}\delta^2({\mathbf{x}_\perp}-{\mathbf{y}_\perp})\frac{g^2\mu_A^2({\mathbf{x}_\perp})}{N_y}\,,
\end{equation}
where the indices $k,l=1,2,\dots,N_y$ label the $N_y$ points of representing the width of the nucleus in $x^-$. The path ordered Wilson line in the dipole model S-matrix (see \eq{eq:dip-corr}) is discretized as 
\begin{equation}
  V_{A (B)}({\mathbf{x}_\perp}) = \prod_{k=1}^{N_y}\exp\left(-ig\frac{\rho_k^{A (B)}({\mathbf{x}_\perp})}{\boldsymbol{\nabla}_T^2 - m^2}\right)\,,
\end{equation}
where $m$ is a infrared cut-off and $A$ and $B$ distinguish the color charge distributions in the two colliding nuclei. The corresponding dipole distributions in each of the incoming nuclei for a particular configuration of color sources is shown in the top panel of \Fig{fig:IP-Glasma}.

To each lattice site $j$, one then assigns two $SU(N_c)$ matrices $V_{(A),j}$ and $V_{(B),j}$, each of which defines a pure gauge configuration with the link variables
$U^{i}_{(A,B),j} = V_{(A,B),j}V^\dag_{(A,B),j+\hat{e_i}}$, where $+\hat{e_i}$ indicates a shift from $j$ by one lattice site in the $i=1,2$ transverse direction. The link variables in the future lightcone $U_{j}^{i}$ that are input into \Eqs{eq:plaq} and \eq{eq:Phi-field} are determined~\cite{Krasnitz:1998ns} from solutions of the lattice classical Yang-Mills equations at $\tau=0$,
\begin{align}
  &{\rm tr} \left\{ t^a \left[\left(U^{i}_{(A)}+U^{i}_{(B)}\right)(1+U^{i\dag})\right.\right.\nonumber\\
   &  ~~~~~~~~~ \left.\left.-(1+U^{i})\left(U^{i\dag}_{(A)}+U^{i\dag}_{(B)}\right)\right]\right\}=0\,,\label{eq:initU}
\end{align}
where $t^a$ are the generators of $SU(N_c)$ in the fundamental representation. (The cell index $j$ is omitted here.) The $N_c^2-1$ equations in \Eq{eq:initU} are highly non-linear and are solved iteratively for $N_c=3$. With these initial conditions,  Hamilton's equations corresponding to \Eq{eq:ham}, are solved to compute inclusive quantities in the LO Glasma. The lower panel of \Fig{fig:IP-Glasma} shows the result for the energy density in the transverse plane at $\tau= 1/Q_S$

The IP-Glasma model gives a good description of bulk features of distributions at RHIC and the LHC~\cite{Schenke:2013dpa,Schenke:2014tga}. In particular, when matched with the 
MUSIC relativistic viscous hydrodynamic code~\cite{Schenke:2010rr}, the IP-Glasma+MUSIC model provides an excellent description of the multiplicity distributions, the inclusive centrality, and the $p_\perp$ distributions, as well as the $v_n$ distributions in heavy-ion collisions putting strong constraints on the extracted transport coefficients of the quark-gluon plasma~\cite{Gale:2012rq,Ryu:2015vwa}. 

There have been several developments since. First, the model has been extended to include JIMWLK evolution of the sources $\rho(x_\perp)\rightarrow \rho(x_\perp,x^\mp)$ for nuclei with large $P^\pm$, enabling one to study rapidity correlations of produced gluons~\cite{Dusling:2009ni,Schenke:2016ksl} and 3D evolution of the LO Glasma fields~\cite{Schenke:2016ksl,Muller:2019bwd,McDonald:2020oyf}. Further, the extension of the IP-Glasma+MUSIC model to hadron-hadron and hadron-nucleus collisions~\cite{Bzdak:2013zma} indicates that sub-nucleon  shape fluctuations in the Glasma are essential to understanding final state contributions to two and multi-particle cumulants of azimuthal anisotropies for high multiplicity events in small systems~\cite{Schenke:2014zha}, the so-called ``ridge" correlations~\cite{Dusling:2015gta}. 

Data on incoherent diffraction from HERA are sensitive to such  non-perturbative ``shape" fluctuations~\cite{Mantysaari:2016ykx,Mantysaari:2016jaz,Mantysaari:2020axf}; the framework developed here allows one to constrain the latter with HERA data and in the future likely more precisely with the EIC. Numerical simulations suggest that long range two particle correlations in the Glasma~\cite{Lappi:2015vta} when combined with hydrodynamic flow can explain the systematics of high multiplicity azimuthal moments in small systems~\cite{Schenke:2016lrs,Schenke:2019pmk}. 
 
\subsection{The Glasma at NLO}
\label{sec:NLO-Glasma}
Thus far we have focused on the leading order dynamics of classical fields ${\cal A} \equiv {\mathcal O}(1/g)$ in the Glasma. As we shall discuss now,  quantum fluctuations that 
are parametrically ${\mathcal O}(1)$ and that contribute to $\Gamma_{+-}$ in \Eq{eq:inclusive-mult} play a large role both before ($p^\eta=0$ modes) and after ($p^\eta\neq 0$ modes) the collision\footnote{$p^\eta$ is the Fourier conjugate of the spacetime rapidity $\eta$.}. 
We discussed the former previously in the context of the small $x$ evolution of the hadron wavefunctions. We will discuss here the role of these modes after the collision.  The $p^\eta\neq 0$ modes  appear only after the collision; as we shall  subsequently discuss, they play a fundamental role in the thermalization of the Glasma. 

\subsubsection{Dynamics of $p^\eta=0$ modes: QCD factorization and energy evolution}
\label{sec:Factorization}
At NLO [${\mathcal O}(1)$ relative to the leading ${\mathcal O}(1/\alpha_S)$ contribution] for the inclusive multiplicity in \Eq{eq:inclusive-mult}, one of the two terms is 
the amputated small fluctuations propagator $\Gamma_{+-}$ and the other is a one loop correction to $\Gamma_\pm$ (or equivalently, the classical field). The $p^\eta=0$ modes  
lie close to the beam rapidities $\pm Y_{\rm beam}$; before the collision, they can be visualized as the fur of wee gluon modes accompanying the valence partons moving along the light cone. 

After the collision, the valence partons are stripped of the small $x$ wee gluon modes that then populate $p^\eta \neq 0$. The surviving $p^\eta=0$ modes are valence modes and the quasi-static cloud of large $x$ partons than accompany them into the fragmentation 
region of the nuclear collision. Thus $p^\eta=0$ modes after the collision are likely not very interesting from the perspective of thermalization at central rapidities. 

Before the collision, all one has are the $p^\eta=0$ modes. These modes are further  separated into sources and fields with the latter dynamically absorbed into the former via the ``small $x$" evolution of the weight functionals $W_{\rm Y_{\rm beam}\pm Y} [\rho_{1,2}]$ of each of the comoving nuclei. This, however, requires a factorization of the quantum fluctuations of each of the two nuclei from each other. 

The resulting factorized form of  \Eq{eq:two-nuclei} can be proven to leading logarithmic accuracy in $x$~\cite{Gelis:2008rw,Gelis:2008ad}. An important ingredient in the proof is the structure of 
the cut propagator $G_{+-}(u,v) \propto \int  \frac{d^2 k_\perp dk^+}{k^+} e^{ik^+(u^- - v^-) + i\frac{k_\perp^2}{2k^+}(u^+ - v^+)}$. If the spacetime points $u$ and $v$ reside on one of the 
nuclei, say, moving along $x^+$, then $u^-\approx v^-$ and  one of the phases vanishes. The other phase oscillates rapidly when $k^+\rightarrow 0$, giving a convergent contribution. However, for $k^+\rightarrow \infty$ it converges to unity, and one obtains a logarithmic divergence $dk^+/k^+$ that is the source of the large logarithms resummed in the small $x$ evolution of the nucleus. 

In the case where quantum fluctuations in the two nuclei  could ``talk" to each other before the collision, the spacetime points $u$ and $v$ reside respectively on the lightcones of the incoming nuclei corresponding to $u^\pm - v^\pm \neq 0$. The phases therefore oscillate rapidly when $k^\pm\rightarrow \infty$ and there are no logarithmic divergences from such contributions. 
The only possible region in which such fluctuations may contribute is where the nuclei overlap. The area of this region is $x^+x^- =\frac{1}{P^+P^-}\sim \frac{1}{s}$; such contributions are therefore suppressed by the squared c.m. energy. 

Thus, the factorized form in \Eq{eq:two-nuclei} at LLx is satisfied to high accuracy, and one can replace $W_{\rm Y_{\rm beam}\pm Y}^{\rm MV} [\rho_{1,2}]$ with $ W_{\rm Y_{\rm beam}\pm Y} [\rho_{1,2}]$, where the latter satisfies the JIMWLK equation in \Eq{eq:JIMWLK-RG}. This allows one to go beyond the boost invariant MV expression and  treat the dynamical evolution in $Y$ of the weight functionals in the two nuclei. While our arguments suggest that the factorization theorem can be extended to NLLx, a formal proof is lacking. 

As $Y_{\rm beam}$ increases with increasing energy, the $W$'s in \Eq{eq:two-nuclei} describe the energy evolution of the  inclusive multiplicity\footnote{This LLx result is implicitly assumed in the 
3+1-D IP-Glasma simulations~\cite{Schenke:2016ksl}.}. Running coupling corrections, which are part of the NLLx contributions, improve the accuracy of the computations significantly. In the future, one may anticipate using the NLLx JIMWLK Hamiltonian as a systematic improvement in describing  energy evolution and rapidity correlations in heavy-ion collisions. 

Details of the factorization of the $W$'s, and their energy evolution, are crucial to phenomenology because they dictate concretely the dependence of final state observables (such as the energy density and the correlators thereof) on the saturation scales in the wavefunctions of the colliding nuclei.

\subsubsection{Dynamics of $p^\eta\neq 0$ modes: plasma instabilities and the classical-statistical approximation}
\label{sec:Weibel}

The $p^\eta\neq 0$ modes are generated right after the collision when the sources become time dependent and produce gluon modes away from the rapidities of the beams. At NLO, their contribution to the gluon spectrum for a fixed distribution of color sources can be written as~\cite{Gelis:2007kn}
\begin{eqnarray}
&&\frac{dN_{\rm NLO}}{dY d^2 p_\perp} = \frac{1}{16\pi^3} \int d^4x\, d^4y\, e^{ip\cdot (x-y)}\partial_x^2\partial_y^2 \sum_\lambda \epsilon_\mu^\lambda \epsilon_\nu^\lambda\nonumber \\
&\times& \left[{\cal A}^\mu(x) \delta {\cal A}^\nu(y) + \delta {\cal A}^\mu(x) {\cal A}^\nu(y) + G_{+-}(x,y)\right]\,,
\label{eq:NLO-peta}
\end{eqnarray}
where $\epsilon_\mu^\lambda$ is a gluon polarization vector of helicity $\lambda$. The first two terms in \Eq{eq:NLO-peta} represent the NLO contribution to $\Gamma_{+}(x) \Gamma_{-}(y)$ in \Eq{eq:inclusive-mult}, with $\delta {\cal A}$ the one-loop correction to the classical field 
${\cal A}\equiv {\cal A}[\rho_1,\rho_2]$ and the last term representing $\Gamma_{+-}$, which first appears at NLO. 

We first consider the cut propagator term $G_{+-}$ in \Eq{eq:NLO-peta}. 
Its contribution to the NLO multiplicity can be written as 
\begin{equation}
\sum_{\lambda,\lambda^\prime} \int \frac{d^3 k}{(2\pi)^3 2 E_k}\,\left|\int_{x^0\rightarrow\infty}\!\!\!\!\!\!\!\! d^3 x\, e^{ip\cdot x}\, (\partial_x^0-iE_q)\, \epsilon_\mu^\lambda \,a_{\lambda^\prime k}^\mu\right|^2\,,
\label{eq:small_fluct-cut-prop.}
\end{equation}
where $a_{\lambda^\prime a k}^\mu(x)$ is a small fluctuation field of ${\mathcal O}(1)$ about ${\cal A}^\mu$ with the plane wave initial condition 
$e_{\lambda^\prime}^\mu T^a e^{ik\cdot x}$, where $T^a$ are the $SU(3)$ generators in the adjoint representation\footnote{For compactness, we will suppress color indices henceforth.}. 
Note that the previous structure is analogous to that of \Eq{eq:classical-evolution} except that the classical field is replaced by the small fluctuation field. The latter obeys the small fluctuation equations of 
motion, and its solution can be expressed as 
\begin{equation}
a^\mu(x) = \int_{\tau=0^+} d^3 u \left[a(y)\cdot {\bf T}_y\right]\, {\cal A}^\mu(x) \,,
\label{eq:formal-small_fluct}
\end{equation}
where $ {\bf T}_y$ is a linear operator that corresponds to a shift of the initial data on the classical fields and their derivatives~\cite{Gelis:2008rw,Dusling:2011rz}, 
\begin{equation}
a(y)\cdot {\bf T}_y = a^\mu(y)\frac{\delta}{\delta {\cal A}^\mu(y)} + (\partial^\nu a^\mu(y))\frac{\delta}{\delta (\partial^\nu {\cal A}^\mu)} \,,
\end{equation}
on the initial spacelike surface at $\tau=0^+$. 

The key insight provided by \Eq{eq:formal-small_fluct} is that, to  compute the small fluctuation field at a spacetime point $x$ in the forward lightcone, 
it is sufficient to know the small fluctuation field at $\tau=0^+$ rather than solve the small fluctuation equations on a time-dependent background. We will return to this point shortly. 

Plugging \Eq{eq:formal-small_fluct} into \Eq{eq:small_fluct-cut-prop.} and then into \Eq{eq:NLO-peta}, one obtains 
\begin{eqnarray}
\frac{dN_{\rm NLO}}{dY d^2 p_\perp} &=& \left[ \int_{\Sigma_y} [ \delta {\cal A}(y) \cdot {\bf T}_y] + \int_{\Sigma_y,\Sigma_z} [\Gamma_2(y,z)\cdot {\bf T}_y {\bf T}_z]\right]_{\tau=0^+}\nonumber \\
&\times& \frac{dN_{\rm LO}}{dY d^2 p_\perp}\,,
\label{eq:NLO-Glasma}
\end{eqnarray}
where $\Sigma_y=\int d^3 y$ denotes the initial spacelike surface  $\tau=0^+$ and 
\begin{equation}
\Gamma_2(y,z) = \sum_{\lambda} \int  \frac{d^3 k}{(2\pi)^3 2 E_k} a_{+k\lambda}(y) a_{-k\lambda}(z) \,,
\label{eq:small-fluctuations-propagator}
\end{equation}
is the small fluctuation propagator evaluated on this surface\footnote{Discussions of the computation of this propagator at $\tau=0^+$ can be found in \cite{Fukushima:2006ax,Dusling:2011rz,Epelbaum:2013waa}.}.

\begin{figure}[t!]
\centering						
 \includegraphics[width=0.4\textwidth]{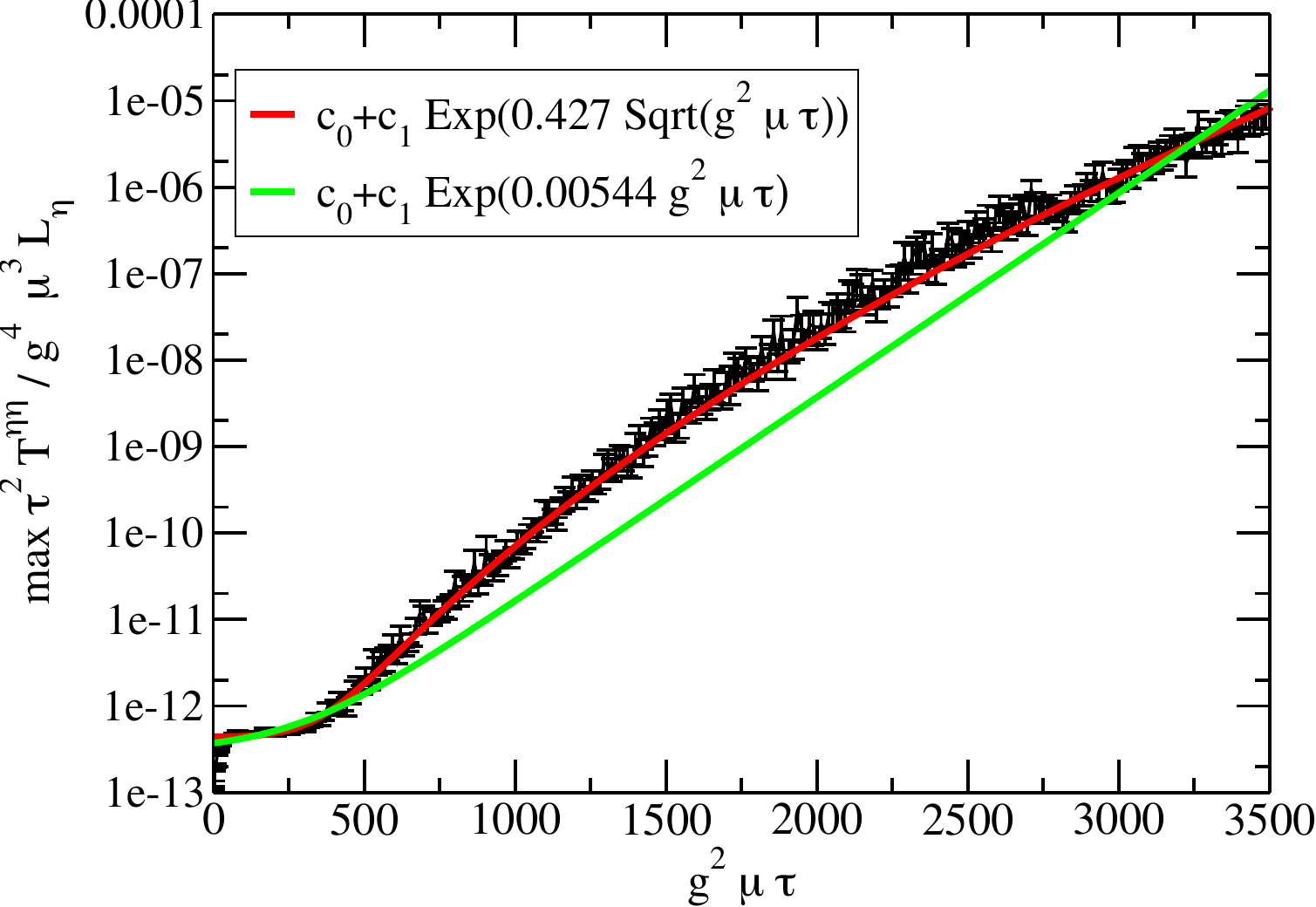}
 \caption{Growth of the maximally unstable Fourier mode of the longitudinal pressure $P_L=\tau^2 T^{\eta\eta}$. Note that since $g^2\mu\propto Q_S$, the results are in units of 
 $Q_S^3/g^2$, with $g\sim 10^{-5}$ and $L_\eta=1.6$. From \cite{Romatschke:2005pm}.}
  \label{fig:Weibel}
\end{figure}

This NLO result, however, is not suppressed parametrically by ${\mathcal O}(\alpha_S)$ relative to the LO result because the LO Glasma is very unstable to small fluctuations:
\begin{equation}
{\bf T}_y {\cal A}(x) \sim \frac{\delta{\cal A}(x)}{\delta {\cal A}(y)}\sim g e^{\sqrt{\gamma_{\rm inst.} \tau}}\,,
\end{equation}
where $\gamma_{\rm inst.}$, parametrically of the order of $Q_S$, denotes the growth rate of the instability. This exponential growth of small fluctuations in \Eq{eq:formal-small_fluct} with $\sqrt{\tau}$ is clearly demonstrated in \Fig{fig:Weibel} using (3+1)-D numerical simulations of the YM equations for an $\eta$-dependent fluctuation $a(\eta)$ on top of the boost invariant Glasma background~\cite{Romatschke:2005pm,Romatschke:2006nk}. The very small values of $g$ in the plot\footnote{At  RHIC (LHC) energies, $g^2\mu\propto Q_S
\sim 1-2$ GeV on the x-axis of \Fig{fig:Weibel}. With these values, $\tau \gg 10 $ fm, the typical life time of such a collision. However, for $g\sim 10^{-5}$, from QCD running, $g^2\mu$ is larger than the Planck scale. The takeaway message from \Fig{fig:Weibel} is the functional form of the fit and not the absolute values.} are chosen to ensure that the classical-statistical approximation is satisfied in the numerical simulations. This point is discussed further in Section~\ref{sec:classicalstatistical}.

The existence of such instabilities was previously predicted~\cite{Mrowczynski:1993qm} and studied with the context of a finite temperature hard thermal loop effective field theory~\cite{Rebhan:2004ur,Attems:2012js}. They are understood to be analogous to the Weibel instabilities familiar in plasma physics~\cite{Arnold:2003rq}; for a recent review, we refer the reader to \cite{Mrowczynski:2016etf}.

As a result of the instability, the exponentially growing small fluctuations can become of the order of the LO classical field for $\tau\sim \frac{1}{\gamma_{\rm inst.}} \log^2\frac{1}{\alpha_S}$. In a 
so-called classical-statistical approximation~\cite{Aarts:2001yn}, these leading instabilities can be resummed to all orders, modifying \Eq{eq:NLO-Glasma} as 
\begin{equation}
\frac{dN_{\rm resum}}{dY d^2 p_\perp} = \int [Da] F[a] \frac{dN_{\rm LO}}{dY d^2 p_\perp}[{\cal A}+a]\,,
\end{equation}
where $F[a] \sim \exp\left[ - \int_{\Sigma_y\Sigma_z} a(y) \Gamma_2^{-1}(y,z) a(z)\right]$. 

To conclude our discussion of the classical-statistical approximation, as a final step we need to perform the average of the color sources to obtain the inclusive multiplicity distribution at early times in the Glasma:
\begin{eqnarray}
\frac{\langle\langle dN\rangle\rangle}{dY d^2 p_\perp} &=&  \int [D\rho_1] [D\rho_2] W_{Y_{\rm beam}-Y} [\rho_1] W_{Y_{\rm beam}+Y}[\rho_2] \nonumber \\
&\times& \int [Da] \, F[a] \, {d N_{\rm LO} \over dY d^2 p_\perp}[{\cal A} +a]\,.
\end{eqnarray}
This result of course applies to other inclusive quantities, such as the components of  the stress-energy tensor given in Eq.~(\ref{eq:Glasma-stresstensor2}). 

In the classical-statistical approximation, the one loop correction to the classical field ($\delta {\cal A}$) is suppressed at early times relative to the $G_{+-}$ term that we consider here. In general, the classical-statistical approximation does not account for the full quantum evolution of the Glasma fields. In  Section~\ref{sec:classicalstatistical}, we will discuss the dynamical power counting of quantum fields within the framework of the two-particle irreducible (2PI) effective action that specifies the range of validity of the classical-statistical approximation and the nature of the corrections beyond, as well as numerical results from the implementation of this approximation and the consequences thereof.

\section{Far-from-equilibrium gluon and quark production: From plasma instabilities to non-thermal attractors\label{sec:classicalstatistical}}

We have seen in the section~\ref{sec:Glasma} that the overoccupied Glasma is unstable with respect to small quantum fluctuations that break longitudinal boost invariance. As noted, the growth of fluctuations is caused by primary (Weibel-like~\cite{Mrowczynski:2016etf}) instabilities \cite{Romatschke:2005pm,Romatschke:2006nk,Fukushima:2011nq}. However, there are also secondary instabilities that arise due to the nonlinear interactions of unstable modes~\cite{Berges:2012cj}. The fluctuations that are initially small grow with time and an over-occupied plasma emerges on a time scale $Q_S \tau \sim \log^{2}(\alpha_{S}^{-1})$. 

At this stage, details about the initial spectrum of fluctuations are effectively lost as a consequence of the strongly nonlinear evolution. The apparent loss of information at such an early stage gives rise to decoherence toward a more isotropic equation of state in this prethermalization regime~\cite{Berges:2004ce,Arnold:2004ti,Dusling:2010rm}. Subsequently, a universal scaling behavior emerges far from equilibrium with increasing anisotropy~\cite{Berges:2014bba}, which is described in terms of non-thermal attractor solutions~\cite{Berges:2013eia,Berges:2013fga}, representing the first stage of the ``bottom-up'' thermalization scenario~\cite{Baier:2000sb,Bodeker:2005nv}.

In the following, we will describe how this nonlinear behavior emerges, starting with the underlying quantum field theory, formulated as an initial value problem in time. Essential aspects of the far-from-equilibrium quantum evolution can be approximated by a controlled weak-coupling expansion around the full (non-perturbative) classical-statistical theory, which was first pointed out in the context of scalar field theories~\cite{Son:1996zs,Khlebnikov:1996mc,Aarts:2001yn} and then extended to include fermions~\cite{Aarts:1998td,Borsanyi:2008eu,Saffin:2011kc,Berges:2010zv,Kasper:2014uaa}. 

In strong field  QCD, this corresponds to an expansion in $\alpha_S \equiv g^2/(4\pi)$, where the leading order contribution includes the full classical-statistical theory of gluons described in Sec.~\ref{sec:Glasma}. The next-to-leading order contributions take into account the back-reaction of the quarks onto the gluons and encode important quantum effects such as anomalies. The non-equilibrium time evolution of gluons with dynamical quarks was been studied numerically on the lattice in Refs.~\cite{Gelis:2005pb,Gelfand:2016prm,Tanji:2017xiw}.

Such an expansion around the full classical-statistical field theory breaks down on the time scale $Q_S \tau \sim \alpha_{S}^{-3/2}$~\cite{Baier:2000sb,Berges:2013eia}, where typical gluon occupancies become of the order of unity. To continue further and capture the late-time evolution toward local thermal equilibrium, one employs a resummed perturbative description of quantum field theory in an on-shell approximation. This also underlies the effective kinetic theory that we will discuss in Sec.~\ref{sec:kinetictheory}. 

The range of validity of both approximation schemes, the expansion around the classical-statistical theory at early times and effective kinetic theory employed at late times with their common overlap at intermediate times~\cite{Mueller:2002gd,Jeon:2004dh}, can be efficiently discussed using the two-particle irreducible (2PI) quantum effective action~\cite{Baym:1962sx,Cornwall:1974vz} on the closed time path~\cite{Calzetta:1986cq,Berges:2004yj}. 

\subsection{Non-equilibrium time evolution equations from the quantum effective action}\label{sec:1PI}

Quantum evolution equations can be formulated in terms of expectation values of field operators, such as the macroscopic field ${\cal A}(x)$ and the connected two-point correlation function or propagator $G(x,y)$ on the closed time contour ${\cal C}$ introduced in Sec.~\ref{sec:Glasma}. In practice, the space-time evolution of the one-point, two-point or higher-point correlation functions cannot be computed for the full quantum theory without approximations. However, one can formally write exact evolution equations, which provides an efficient starting point justifying the applicability of systematic expansion schemes. 

Writing for simplicity only the gauge field part, the evolution equations for connected one and two-point correlation functions follow from the stationarity of the 2PI effective action~\cite{Baym:1962sx,Cornwall:1974vz}
\begin{eqnarray}
\Gamma[{\cal A},G]=S[\cal{A}]&+&\frac{i}{2}\text{tr}\big(\text{ln}\, G^{-1}\big)+\frac{i}{2}\text{tr}\big[G_0^{-1}({\cal A})\,G\big] \nonumber \\
&+&\Gamma_2[{\cal A} ,G]+\text{const}\;,
\end{eqnarray}
where $i G_{0;ab}^{-1,\mu\nu}(x,y;{\cal A}) \equiv \delta^2S[{\cal A}]/\delta {\cal A}^a_\mu(x)\delta {\cal A}^b_\nu(y)$ is the inverse propagator with Lorentz indices $\mu,\nu$ and color indices $a$, $b = 1, \ldots, N_c^2-1$ for $SU(N_c)$ gauge theories with classical action $S[{\cal A}]$. Here  $\Gamma_2[{\cal A} ,G]$ contains all two-particle irreducible contributions, which leads to the self-energy $\Pi^{\mu\nu}_{ab}(x,y) \equiv 2i \delta \Gamma_2[{\cal A} ,G]/\delta G_{\mu\nu}^{ab}(x,y)$. Higher $n$-point correlation functions can be obtained from $\Gamma[{\cal A},G]$ by functional differentiation with respect to the fields once the solutions for ${\cal A}$ and $G$ are known.

\subsubsection{Macroscopic field, spectral and statistical functions}

The full quantum evolution equation for the macroscopic field is obtained from the stationarity of $\Gamma[{\cal A},G]$ with respect to variations in ${\cal A}(x)$ and is given by
\begin{eqnarray}
\label{eq:AEOM}
\frac{\delta S[{\cal A}]}{\delta {\cal A}_{\mu}^a(x)}=-J^{\mu}_a(x)-\frac{i}{2}\text{tr}\left[\frac{\delta G_0^{-1}({\cal A})}{\delta {\cal A}_{\mu}^a(x)}G\right]-\frac{\delta \Gamma_2[{\cal A},G]}{\delta {\cal A}_{\mu}^a(x)} \,.\nonumber \\
\end{eqnarray}
For our discussion of the evolution equations for two-point functions, it is convenient to introduce spectral and statistical components by 
\begin{eqnarray}
G_{\mu\nu}^{ab}(x,y) \equiv F_{\mu\nu}^{ab}(x,y)-\frac{i}{2}\rho_{\mu\nu}^{ab}(x,y)\, \text{sgn}_{\cal C}(x^0-y^0)\,\,
\end{eqnarray}
where the spectral function $\rho(x,y)$ is associated with the expectation value of the commutator of two fields and the statistical function $F(x,y) $ by the anti-commutator for bosons\footnote{In terms of the Keldysh components of the propagator employed in Sec.~\ref{sec:Glasma}, this reads\\ $G_{++}(x,y)=F(x,y)-i \rho(x,y) \text{sgn}(x^0-y^0)/2$,\\ $G_{--}(x,y)=F(x,y)+i \rho(x,y) \text{sgn}(x^0-y^0)/2$,\\ $G_{+-}(x,y)=F(x,y)+i \rho(x,y)/2$,\\ and $G_{-+}(x,y)=F(x,y)-i \rho(x,y)/2$.}~\cite{Berges:2004yj}. A similar decomposition can be done for the self-energy, $\Pi(x,y) \equiv -i \Pi^{(0)}(x) \delta(x-y)  + \Pi^{(F)}(x,y) -i \Pi^{(\rho)}(x,y)\text{sgn}_{\cal C}(x^0-y^0)/2$, where $\Pi^{(0)}$ describes a local contribution to the self-energy. With this notation, the equations for spectral and statistical two-point correlation functions, which follow from the stationarity of $\Gamma[{\cal A},G]$ with respect to variations in $G$, can be written as \cite{Berges:2004yj}
\begin{widetext}
\begin{eqnarray}
\left[iG_{0,ac}^{-1,\mu\gamma}(x;{\cal A})+\Pi^{(0),\mu\gamma}_{ac}(x)\right]\rho_{\gamma\nu}^{cb}(x,y)=&-&\int_{y^0}^{x^0}dz~\Pi^{(\rho),\mu\gamma}_{ac}(x,z)\rho_{\gamma\nu}^{cb}(z,y)\;,
\nonumber\\
\label{eq:FEOM}
\left[iG_{0,ac}^{-1,\mu\gamma}(x;{\cal A})+\Pi^{(0),\mu\gamma}_{ac}(x)\right]F_{\gamma\nu}^{cb}(x,y)=&-&\int_{t_0}^{x^0}dz~\Pi^{(\rho),\mu\gamma}_{ac}(x,z)F_{\gamma\nu}^{cb}(z,y)+\int_{t_0}^{y^0}dz~\Pi^{(F),\mu\gamma}_{ac}(x,z)\rho_{\gamma\nu}^{cb}(z,y)\;.
\end{eqnarray}
\end{widetext}
In \Eq{eq:FEOM} we denote $\int_a^b dz\equiv \int_a^b dz^0 \int d^3z \sqrt{-g(z)}$ with given initial time $t_0$ and $g$ as the determinant of the metric. The inverse propagator enters \Eq{eq:FEOM} as
\begin{align}
&&iG_{0,ab}^{-1,\mu\nu}(x;{\cal A}) = \left(-g\right)^{-\frac{1}{2}} D^{ac}_\gamma({\cal A})\left(-g\right)^{\frac{1}{2}} g^{\gamma\alpha} g^{\mu\nu} D^{cb}_\alpha({\cal A})
\nonumber\\
&&- \left(-g\right)^{-\frac{1}{2}} D^{ac}_\gamma({\cal A})\left(-g\right)^{\frac{1}{2}} g^{\gamma\nu} g^{\mu\alpha} D^{cb}_\alpha({\cal A}) - g f^{abc} {\cal F}_c^{\mu\nu}({\cal A})
\nonumber
\end{align}
with the covariant derivative $D^{ab}_\mu({\cal A}) = \delta^{ab}\partial_\mu - g f^{abc}{\cal A}^c_\mu$ and ${\cal F}^a_{\mu\nu}({\cal A}) = \partial_\mu {\cal A}_\nu^a - \partial_\nu {\cal A}_\mu^a + g f^{abc} {\cal A}^b_\mu {\cal A}^c_\nu$ as the field strength tensor.

The non-zero spectral and statistical parts of the self-energy $\Pi^{(\rho/F)}(A,F,\rho)$ on the r.h.s and the space-time local part  $\Pi^{(0)}(F)$ on the l.h.s of this coupled set of equations make the evolution equations nonlinear in the fluctuations. In general, they contain contributions from the interaction vertices of QCD, where in addition to the standard three- and four-vertices there is a three-gluon vertex associated with the presence of a non-vanishing field expectation value. The explicit expressions for the derivatives on the r.h.s of \Eq{eq:AEOM} and the self-energy contributions entering \Eq{eq:FEOM} were given to three loop order ($g^6$) in Ref.~\cite{Berges:2004pu}, and the corresponding expressions in co-moving $(\tau,\eta)$ coordinates can be found in Ref.~\cite{Hatta:2011ky}. The inclusion of quark degrees of freedom follows along the same lines and can also be found in Ref.~\cite{Berges:2004pu}.

The non-equilibrium initial conditions for the coupled evolution equations \eq{eq:AEOM} and \eq{eq:FEOM} 
can be formulated in $(\tau,\eta)$ coordinates (and Fock-Schwinger gauge ${\cal A}_{\tau}=0$) for the Glasma initial conditions discussed in Sec.~\ref{sec:Glasma}. The gauge field expectation values in \Eq{eq:KMW} correspond to the Glasma background fields, while the spectral and statistical two-point functions describe the fluctuations. At all times the former satisfy the equal-time commutation relations 
\begin{eqnarray}
\label{eq:commrel}
\left.\rho_{\mu\nu}^{ab}(x,y)\right|_{x^0=y^0}&=&0\;, \nonumber \\
\left.\partial_{x^0}\rho_{\mu\nu}^{ab}(x,y)\right|_{x^0=y^0}&=&-\delta^{ab}\frac{g_{\mu\nu}}
{\sqrt{-g(x)}} \delta(\vec{x}-\vec{y}) \;, \nonumber \\
\left.\partial_{x^0}\partial_{y^0}\rho_{\mu\nu}^{ab}(x,y)\right|_{x^0=y^0}&=&0 \;.
\end{eqnarray}

\subsubsection{Resummed evolution equations to leading order}

To isolate the leading contributions one has to take into account the strong external currents $J \sim\mathcal{O}(1/g)$ in the Glasma, which induce non-perturbatively large background fields ${\cal A}\sim\mathcal{O}(1/g)$. In contrast, the statistical fluctuations $F$ originate from the vacuum and are therefore initially $\mathcal{O}(1)$. The spectral function $\rho$ encodes the equal-time commutation relations and is therefore parametrically $\mathcal{O}(1)$ at any time. 

Considering only the leading contributions in a weak coupling expansion, the evolution equation \eq{eq:AEOM} reduces to the classical Yang-Mills equation for the classical Glasma field ${\cal A}$, and the equations for the spectral and statistical two-point correlation functions read 
\begin{eqnarray}
iG_{0,ac}^{-1,\mu\gamma}(x;{\cal A})~\rho_{\gamma\nu}^{cb}(x,y)&=&0 \;, \nonumber\\
 iG_{0,ac}^{-1,\mu\gamma}(x;{\cal A})~F_{\gamma\nu}^{cb}(x,y)&=&0\;.
\label{eq:linF}
\end{eqnarray}
In \Eq{eq:linF} sub-leading contributions are suppressed by at least a factor of $g^2$ relative to the leading contribution. 

At this order the evolution of the Glasma background fields decouples from that of the fluctuations.  
The evolution of vacuum fluctuations of the initial state is taken into account by \Eq{eq:linF} to linear order in the fluctuations. This was an important assumption in the derivation in \Sec{sec:Weibel} and was exploited in Ref.~\cite{Dusling:2011rz,Epelbaum:2013waa} to obtain the spectrum of initial fluctuations right after the collision. These approximations are therefore valid only for evolution times short enough that the fluctuations have parametrically small values. 

In general, it is difficult to find suitable approximation schemes for the 2PI effective action in gauge theories beyond the linear regime~\cite{Arrizabalaga:2002hn}. However, it provides a formal justification of a resummed coupling expansion of the quantum field theory around the full classical-statistical solution; as we will soon discuss, this scheme can be implemented numerically  on a lattice to describe dynamics that are far from equilibrium.

Furthermore, as we shall also later discuss, the different dynamical stages of the Glasma undergoing a non-equilibrium instability at early times can be conveniently understood analytically from power counting in the 2PI effective action beyond the linear regime~\cite{Berges:2012cj}. Not least, the 2PI effective action approach allows for efficient on-shell approximations employing a gradient expansion; these lead to effective kinetic equations describing non-equilibrium evolution at later times~\cite{Blaizot:2001nr}. We will discuss these equations and their numerical solutions in \Sec{sec:kinetictheory}.

\subsection{Nonlinear evolution of plasma instabilities}
\label{sec:plasmainstabilities}

In \Sec{sec:Weibel}, we demonstrated that the highly anisotropic state of the Glasma is unstable with respect to small quantum fluctuations. In the language of \Eq{eq:linF}, these correspond to the quasi-exponential growth of the statistical function, \cite{Romatschke:2005pm,Romatschke:2006nk,Fukushima:2011nq,Berges:2014yta,Fukushima:2007ja}
\begin{eqnarray}
\label{eq:expgrowth}
F_{\mu\nu}^{ab}(\tau,\tau,x_T,y_T,\nu) \sim \exp\left[\Gamma(\nu) \sqrt{g^2\mu\tau}\right]\;, 
\end{eqnarray}
where we recall that $g^2 \mu\propto Q_S $ and $\Gamma(\nu)$ is a function of the order of unity for characteristic modes $\nu$ that are Fourier coefficients with respect to the relative rapidity\footnote{Here $\nu$ is equivalent to the 
momentum $p^\eta$ in the $(\tau,\eta)$ coordinate system.}
\begin{eqnarray}
F_{\mu\nu}^{ab}(x,y)=\int \frac{d \nu}{2\pi} F_{\mu\nu}^{ab}(x,y,\nu) e^{i\nu(\eta_x-\eta_y)}\;.
\end{eqnarray}

\subsubsection{Dynamical power counting}

The behavior of the quantum evolution beyond the linear regime is captured by a dynamical power counting scheme~\cite{Berges:2002cz,Berges:2012iw,Berges:2007re,Berges:2008zt}. Self-energy corrections are classified according to powers of the coupling constant $g$, of the background field ${\cal A}$, and of the statistical fluctuations $F$. Thus, a  generic self-energy contribution is of the order of $g^n F^m{\cal A}^{l}\rho^k$ and 
contains the suppression factor from powers of the coupling constant ($n$),  as well as the enhancement due to a parametrically large background field ($l$) and large fluctuations ($m$). The ``weight'' of the spectral function ($k$) remains parametrically of order one at all times as encoded in the equal-time commutation relations, see \Eq{eq:commrel}. 

For the strong macroscopic fields ${\cal A}\sim1/g$ in the Glasma, sizable self-energy corrections occur once fluctuations  grow as large as $F \sim 1/g^{(n-l)/m}$ for characteristic modes. This yields a hierarchy of time scales, where diagrammatic contributions with smaller values of $r=(n-l)/m$ become important at earlier times (since $g\ll 1$)  than contributions with larger values of $r$. 

The quasi-exponential growth stops when fluctuations become $\mathcal{O}(1/g^2)$, where they saturate. At $\mathcal{O}(1/g^2)$ the fluctuations lead to sizable contributions from every given loop-order and the perturbative power-counting scheme breaks down. The corresponding time scale may be estimated from the one-loop correction  
\begin{center}
  \includegraphics[width=0.12\textwidth]{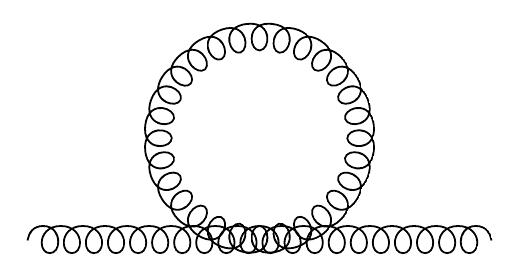} \, ,
\end{center}
which has $r=2$ ($n=2$, $l=0$, $m=1$). Using the quasi-exponential growth behavior [\Eq{eq:expgrowth}] the factor of $\sim g^2$ from the vertex is compensated for by the propagator line $F \sim \mathcal{O}(1/g^2)$ at time
\begin{eqnarray}
\label{eq:tSecParam}
\tau_\text{occ}\stackrel{g\ll1}{\sim}~\frac{1}{Q_S}\,\log^2\left(g^{-2}\right) \;,
\end{eqnarray}
which denotes the characteristic time for the end of the instability regime. 

The earliest time for nonlinear amplification to set in can be inferred from the diagram with the lowest value of $r$. For our problem,
this is realized by the one-loop contribution with $r=1$ ($n=2$, $l=0$, $m=2$),
\begin{center}
  \includegraphics[width=0.12\textwidth]{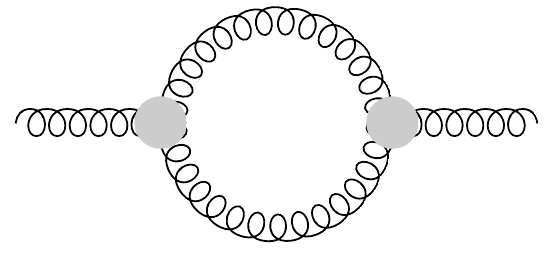} 
\end{center}
which already becomes sizable when $F \sim \mathcal{O}(1/g)$, where the two propagator lines compensate for the 2 powers of the coupling. Again using the quasi-exponential growth behavior [\Eq{eq:expgrowth}] of the primary unstable modes, we find that the time at which this $\mathcal{O}(1/g)$ correction becomes important  relative to the $\mathcal{O}(1/g^2)$ in \Eq{eq:tSecParam} is $\sim \tau_\text{occ}/4$ in the weak-coupling limit. This is followed by a series of higher-loop corrections, all leading to a fast broadening of the primary unstable range in rapidity wave number $\nu$~\cite{Berges:2012cj}.   

\subsubsection{Classical-statistical field theory limit}

\begin{figure*}
\includegraphics[width=0.7\textwidth]{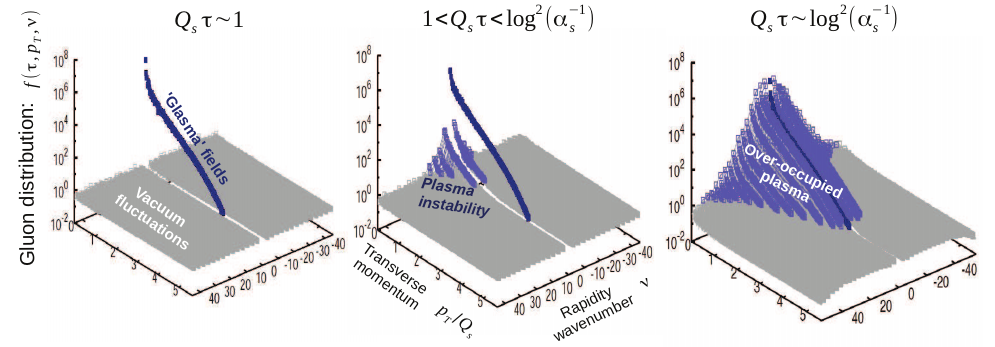}
\caption{\label{fig:InstOverview} Time evolution of the gluon distribution at early times $0 \lesssim Q_S \tau \lesssim \log^{2}(\alpha_{S}^{-1})$ from next-to-leading order CGC initial conditions~\cite{Epelbaum:2013waa} at very weak coupling ($\alpha_S\sim 10^{-6}$). From Ref.~\cite{Berges:2014yta}.}
\end{figure*}

The evolution of the Glasma to later times than $\tau_\text{occ}$ is non-perturbative. While  there are different ways to address this in scalar quantum field theories, with an example being large-$N$ resummation techniques \cite{Berges:2001fi,Aarts:2002dj}, for gauge theories the most frequently employed approach is the classical-statistical approximation. The latter can be understood starting with the full quantum 2PI effective action by a set of well-defined approximations. 

One first notes that a given propagator line of a diagram may be associated with either the statistical ($F$) or the spectral ($\rho$) correlation function. The set of diagrams included in the classical-statistical approximation can be identified as those corrections that contain the most powers of the statistical function relative to powers of the spectral function for each type of diagram \cite{Aarts:2001yn}. This corresponds to resumming the leading effects of the instability to all orders in the coupling constant \cite{Dusling:2011rz,Epelbaum:2013waa}. 

Therefore, in contrast to expansions at fixed loop-orders, the classical-statistical approach provides a controlled approximation scheme that is particularly well suited to problems involving large statistical fluctuations. Specifically, for the large $F\sim \mathcal{O}(1/g^2)$ values encountered at the end of the plasma instability regime, neglecting powers of $\rho \sim \mathcal{O}(1)$ compared to those of $F$ represents a systematic weak-coupling approximation of a system that is strongly correlated because of the large fluctuations. 

While leading order in this expansion corresponds to the full non-equilibrium classical-statistical field theory for the gauge fields, genuine quantum corrections for the dynamics arise. As we will soon discuss, the dynamical evolution of quarks and anti-quarks represent a class of such genuine quantum corrections~\cite{Tanji:2017xiw}.

We can conclude from this discussion that for the far-from-equilibrium overoccupied Glasma there is a well-controlled mapping of the weak-coupling quantum dynamics for correlation functions onto a classical-statistical field theory.  The latter can be simulated numerically on a lattice. In principle, starting with large field amplitudes, the mapping involves two steps: (I) The field is separated into a large coherent part and a small fluctuation part in which one linearizes the field evolution equations. The set of linearized equations is given by \Eq{eq:linF}. (II) Although small initially, the fluctuations grow because of plasma instabilities. Once the fluctuations become sizable, the time evolution of the linearized equations is stopped and the results are used as input for a subsequent classical-statistical simulation that is fully non-linear. 

A virtue of the two-step procedure of mapping the original quantum theory to the classical description is that it has a well-defined continuum limit, enabling one to recover the full physical results for certain quantities in the weak-coupling limit~\cite{Aarts:1997kp}. In scalar field theories, this is well tested by comparisons to fully quantum calculations using 2PI effective action techniques~\cite{Aarts:2001yn} and likewise when scalar fields are coupled to  fermions~\cite{Berges:2013oba}. The mapping was first applied in cosmology in the context of post-inflationary scalar preheating dynamics~\cite{Son:1996zs,Khlebnikov:1996mc}. 

The two-step procedure is in practice replaced by a simplified description whereby one already starts with the fully non-linear classical-statistical description from the initial time in the strong-field regime. This can be well controlled, for a given regularization with lattice spacing $a$ in the weak coupling limit, by ensuring that vacuum fluctuations from modes with momenta near the cutoff $\sim 1/a$ do not dominate the dynamics. Several studies have investigated the range of validity of this simplified ``one-step" mapping of the original quantum theory onto the classical-statistical description--see for instance Ref.~\cite{Epelbaum:2014yja}; the limitations of the classical-statistical approximation were studied in detail in Ref.~\cite{Berges:2013lsa} for scalar field theories. 

Figure \ref{fig:InstOverview} provides snapshots of the time evolution of the gluon distribution for an analytically computed initial spectrum of fluctuations given in Ref.~\cite{Epelbaum:2013waa}, already employing the fully non-linear classical-statistical description from the initial time in the strong-field regime. The non-equilibrium evolution is computed numerically  using the Wilson formulation of lattice gauge theory in real time~\cite{Berges:2014yta}. In addition to gauge invariant quantities, Coulomb type gauge fixed distribution functions can be extracted for comparison to effective descriptions such as kinetic theory. The definition of the distribution function shown in \Fig{fig:InstOverview} employs the two-point correlation function of the gauge field following Ref.~\cite{Berges:2013eia}.
While the gluon distribution as a function of transverse momentum $p_T$ and rapidity wave number $\nu$ is dominated by the boost-invariant $(\nu=0)$ background at early times $Q_S \tau \sim 1$, an over-occupied plasma emerges on a time scale $Q_S \tau \sim \log^{2}(\alpha_{S}^{-1})$. 

\begin{figure}[b]
\includegraphics[width=0.45\textwidth]{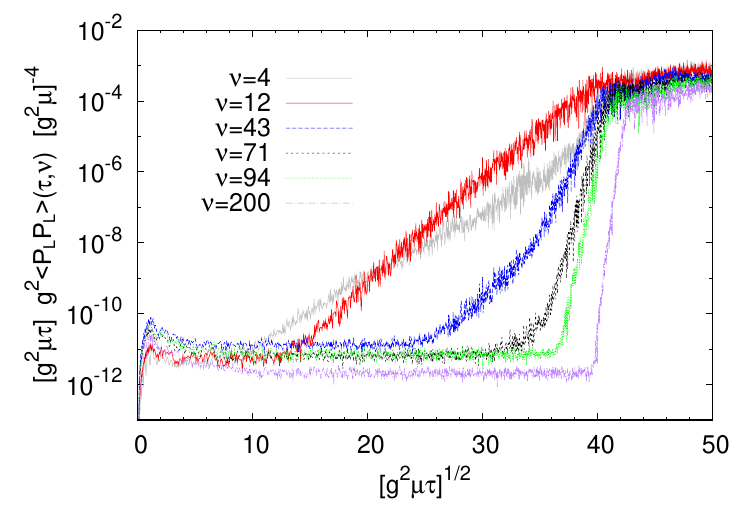}
\caption{\label{fig:Modes} Time evolution of the longitudinal pressure-pressure correlator for different rapidity wave numbers $\nu$ with the parameters described in Ref.~\cite{Berges:2012cj}. Once the initial fluctuations have grown larger, one observes the emergence of secondary instabilities at larger $\nu$ with enhanced growth rates.}
\end{figure}

A corresponding evolution is found irrespective of the details of the fluctuations in the initial conditions. Figure~\ref{fig:Modes} shows the example of the gauge-invariant longitudinal pressure-pressure correlation function for different rapidity wave numbers $\nu$, averaged over transverse coordinates, as a function of time~\cite{Berges:2012cj}.
The evolution starts at initial conditions with simplified initial fluctuations taken as an additive contribution to the strong background gauge fields. While primary unstable modes at non-zero rapidity wave number exhibit quasi-exponential amplification first, secondary instabilities with enhanced growth rates set in with a delay for higher momentum modes due to the previously described nonlinear processes. Subsequently the instability propagates toward higher momenta until saturation occurs and the system exhibits a much slower dynamics~\cite{Romatschke:2006nk,Berges:2012cj}. This behavior is similar to that observed in non-expanding gauge theories \cite{Berges:2007re,Berges:2008zt} and cosmological models for scalar field evolution~\cite{Berges:2002cz}.  

\subsection{Non-thermal attractor \label{sec.nonthatt}}

The plasma instabilities lead to a far-from-equilibrium state at time $Q_S \tau_\text{occ} \sim \log^{2}(\alpha_{S}^{-1})$, which exhibits an over-occupied gluon distribution whose characteristic properties may be parametrized as
\begin{equation}
\label{eq:TurbIC}
f(p_T,p_z,\tau_\text{occ}) = \frac{n_0}{2g^2}\, \Theta\!\left( Q - \sqrt{p_T^2+(\xi_0 p_z)^2}\right)\;.
\end{equation}  
In \Eq{eq:TurbIC} $n_0$ denotes the magnitude of the initial over-occupancy of the plasma, averaged over spin and color degrees of freedom up to the momentum $Q$. The momentum scale $Q$ is of comparable magnitude, albeit non-trivially related, to the saturation scale $Q_S$. 
The degree of anisotropy of the gluon distribution in momentum space is described by the parameter $\xi_0$. 

While \Eq{eq:TurbIC} does not capture all details of the state at $\tau_{occ}$, a precise matching to the Glasma appears to be inessential because of the existence of an attractor solution for the subsequent dynamics. In fact, variation of the parameters of \Eq{eq:TurbIC} can be used to visualize attractor properties. 

\begin{figure}[t]
\includegraphics[width=0.4\textwidth]{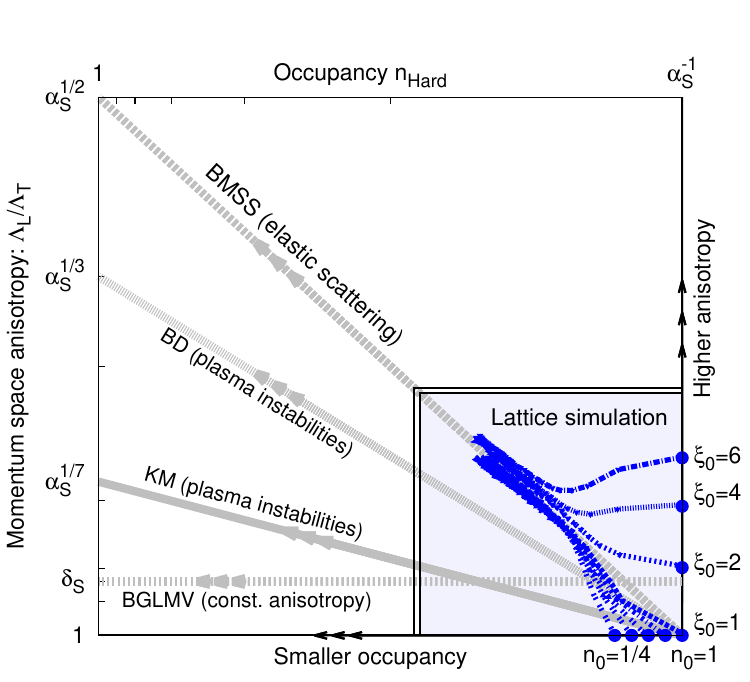}
\caption{\label{fig:Attractor} Evolution in the occupancy--anisotropy plane, from Ref.~\cite{Berges:2013eia}. Indicated are the thermalization scenarios proposed in (BMSS)~\cite{Baier:2000sb}, (BD)~\cite{Bodeker:2005nv}, (KM)~\cite{Kurkela:2011ub} and (BGLMV)~\cite{Blaizot:2011xf}. The blue lines show the results of classical-statistical simulations for different initial conditions~\cite{Berges:2013eia,Berges:2014yta}.}
\end{figure}

Figure~\ref{fig:Attractor}~illustrates the evolution of the plasma in the occupancy-anisotropy plane, which was introduced in Refs.~\cite{Kurkela:2011ti,Kurkela:2011ub}. The horizontal axis shows the characteristic ``hard scale" occupancy $n_{\text{Hard}}(\tau)=f(p_\perp\simeq Q,p_z=0,\tau)$, while the vertical axis shows the momentum-space anisotropy, which can be characterized in terms of the ratio of typical longitudinal momenta ($\Lambda_L$) to the typical transverse momenta ($\Lambda_{T}$). These typical longitudinal and transverse momentum scales are gauge invariant quantities expressed as ratios of the product of covariant derivatives of the field strength tensor normalized by the energy density~\cite{Berges:2013fga}. In a weak coupling limit, these are proportional to $\langle p_\perp\rangle$ and $\langle p_z \rangle$ for a single particle distribution $f(p_\perp,p_z,\tau)$.

The blue lines in \Fig{fig:Attractor} show a projection of lattice simulation results onto the anisotropy-occupancy plane. The different initial conditions are indicated by blue dots. After some time all curves exhibit a similar evolution along the diagonal, thereby illustrating the presence of a non-thermal attractor independent of the initial conditions.
The attractor has a number of interesting properties associated with non-thermal fixed points that we shall discuss in Secs.~\ref{sec:turbulent-attractor}--\ref{sec:scalar-NTA}.

\subsubsection{Far-from-equilibrium universal scaling}
\label{sec:turbulent-attractor}

\begin{figure}[t!]
\centering						
 \includegraphics[width=0.4\textwidth]{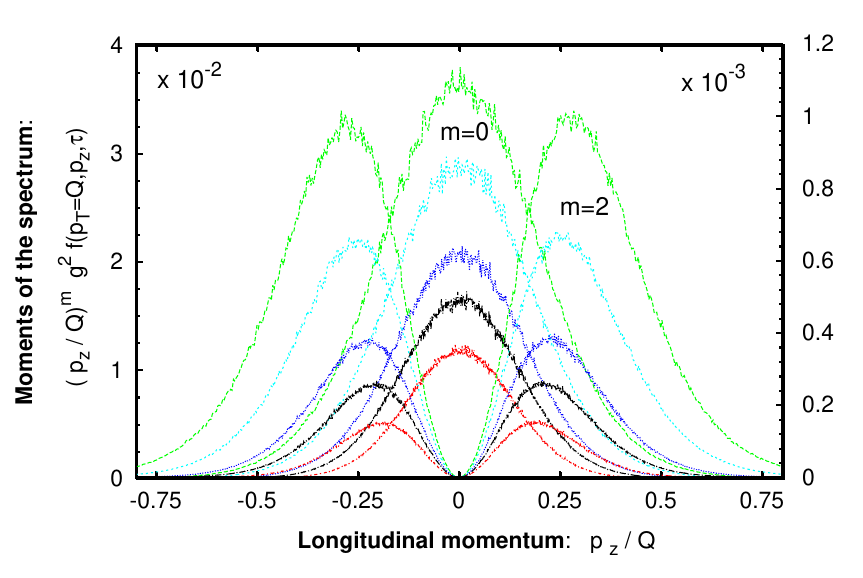}
 \includegraphics[width=0.4\textwidth]{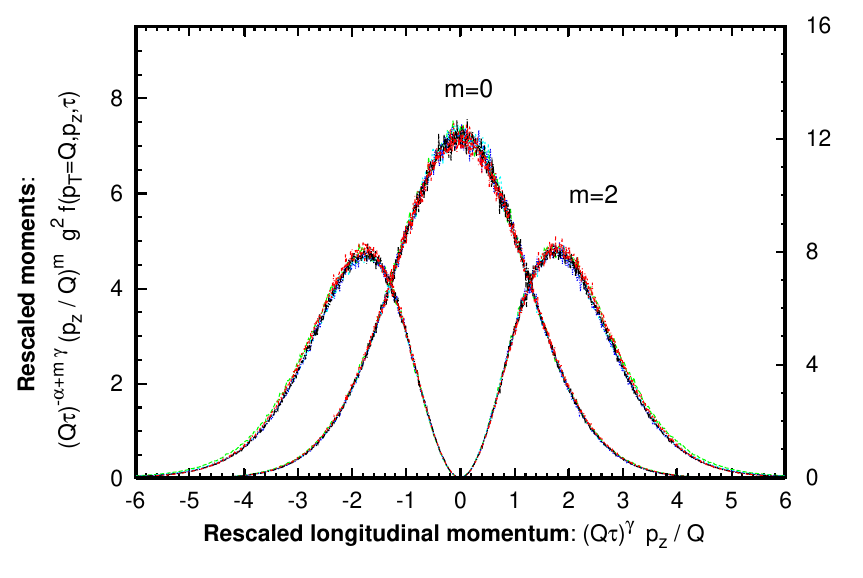}
  \caption{\label{fig:self-similar} Left panel: moments of the single particle distribution function as a function of longitudinal momenta. The longitudinal spectra are evaluated at transverse momentum $p \simeq Q$. The different curves correspond to different times of the evolution: $Q\tau=750,~1000,~1500,~2000,~3000$ (from top to bottom). Right panel: the rescaled moments of the distribution function are found to collapse onto a single curve when plotted as a function of the rescaled longitudinal momentum variable. From Ref.~\cite{Berges:2013eia}.}
\end{figure}

In addition to an insensitivity to details of the initial conditions, the Glasma's evolution exhibits a universal scaling behavior such that the dynamics in the vicinity of the attractor becomes self-similar. In the weak coupling limit, the gluon distribution can be expressed in terms of a time independent scaling function $f_S$~\cite{Berges:2013eia}:
\begin{eqnarray}
f(\tau,p_T,p_z)=\frac{(Q\tau)^{\alpha}}{\alpha_S} f_{S}\Big((Q\tau)^{\beta} p_T,(Q\tau)^{\gamma} p_z\Big)\;.\label{eq:scalingsol} 
\end{eqnarray}
This scaling behavior is characteristic of the phenomenon of wave turbulence and has been observed in a variety of systems that are far from equilibrium \cite{Micha:2004bv,Berges:2015ixa}. As shown in the left panel of \Fig{fig:self-similar}, the moments of the longitudinal momentum distribution at different times in the evolution collapse into  universal curves for each moment $m$ of the single particle distribution. One observes a corresponding behavior for moments of the transverse momentum distribution. This self-similar behavior of the distribution allows one to extract numerically the values of the scaling exponents in \Eq{eq:scalingsol} as $\alpha\simeq -2/3$, $\beta \simeq 0$ and $\gamma \simeq 1/3$  \cite{Berges:2013eia}. 

These values are consistent with those obtained analytically from small-angle elastic scattering as the dominant process and confirm the onset of the ``bottom-up'' thermalization scenario \cite{Baier:2000sb}. The competition between longitudinal momentum broadening via small-angle scattering and the red-shift due to the longitudinal expansion leads to a decrease of the typical longitudinal momenta as $p_z/Q \sim (Q\tau)^{-1/3}$,  while the typical transverse momenta remain approximately constant, $p_T/Q\sim {\rm const}$. At the same time, the gluon occupancy decreases as $f(\tau,p_T\sim Q) \sim {\alpha_S}^{-1}(Q\tau)^{-2/3}$ and becomes of the order of unity on a time scale $Q\tau_\text{quant}\sim \alpha_S^{-3/2}$ when quantum effects can no longer be neglected. Beyond $\tau_\text{quant}$, the classical-statistical framework becomes inapplicable and one may resort to an effective kinetic description as will be discussed in Sec.~\ref{sec:kinetictheory}.

\subsubsection{Identifying the weak-coupling thermalization scenario\label{sec:identifyingEKT}}

In \Fig{fig:Attractor}, we showed the predictions for various thermalization scenarios for the momentum anisotropy with decreasing occupancy. These thermalization scenarios are based on estimates in effective kinetic theory and differ primarily in how infrared momentum modes are treated. Clearly, these differences lead to very different paths in the thermalization process.  As the system evolves with decreasing occupancy from the initial $f\sim {\alpha_S}^{-1}$, classical-statistical field theory simulations accurately capture the physics of the infrared regime. This may be used to distinguish whether a particular thermalization scenario is indeed realized, especially since lattice simulations and effective kinetic theory have an overlapping regime of validity when $1< f < {\alpha_S}^{-1}$. 

The gray lines in Fig.~\ref{fig:Attractor} indicate the different thermalization scenarios put forward in Refs.\ (BMSS)~\cite{Baier:2000sb}, (BD)~\cite{Bodeker:2005nv}, (KM)~\cite{Kurkela:2011ub} and (BGLMV)~\cite{Blaizot:2011xf}. Unlike the BMSS scenario, which is consistent with the lattice simulation results and is discussed in detail in Sec.~\ref{sec:kinetictheory}, the BD scenario considers the possibility that plasma instabilities lead to an overpopulation $f\sim 1/\alpha_S$ of modes with $|\p|\lesssim m_D$. The coherent interaction of hard excitations with the soft sector then causes an additional momentum broadening such that the longitudinal momenta of hard excitations fall at a slower rate. A possible variant of the impact of plasma instabilities for the subsequent quantum evolution also underlies the KM scenario. In the BGLMV scenario, elastic scattering is argued to be highly efficient in reducing the anisotropy of the system. This would generate an attractor with a fixed anisotropy such that $\Lambda_L/\Lambda_T$ remains constant in time. 

The selection of the appropriate effective kinetic theory using lattice simulation data represents the state of the art and is the basis for the thermalization discussion in Sec.~\ref{sec:kinetictheory}. A justification of the kinetic description solely based on perturbation theory in its range of validity raises important open questions on how to incorporate the effects of infrared modes. 

\subsubsection{Non-thermal attractors in scalar field theories
\label{sec:scalar-NTA}}

Non-thermal attractors in overoccupied weakly coupled field theories were studied earlier in the context of cosmological (p)reheating and thermalization after inflation in the early Universe~\cite{Micha:2002ey,Micha:2004bv,Berges:2008wm}. A large class of inflationary models employ scalar field theories, where an initially coherent inflaton field decays due to non-equilibrium instabilities. These may originate from tachyonic or spinodal dynamics or parametric resonance~\cite{Traschen:1990sw,Kofman:1994rk,Berges:2002cz}. The instabilities lead to overoccupied excitations, whose transient dynamics can exhibit self-similar evolution. 

The dynamics is in general spatially isotropic on large scales, in contrast to the longituinal expansion relevant to heavy-ion collisions. To compare the two, if we impose  the isotropic case of no expansion with overoccupied initial conditions for gauge fields, the gluon distribution function in the self-similar regime obeys $f(t,p) = t^{-4/7} f_S(t^{-1/7} p)$ in three spatial dimensions. This is characteristic of an energy cascade toward higher momentum scale due to weak wave  
turbulence~\cite{Kurkela:2011ti,Schlichting:2012es,Kurkela:2012hp}. 

In the fixed box case for a relativistic real scalar field theory in the self-similar 
regime, 
the distribution function obeys $f^\phi(t,p) = t^{-(d+1)/(2l-1)} f^\phi_S(t^{-1/(2l-1)} p)$ for $l$-vertex scattering processes~\cite{Micha:2004bv}. For quartic ($l=4$) self-interactions, the exponents are identical to the gauge theory with the same geometry. However, in the presence of spontaneous symmetry breaking the non-zero field expectation value leads to effective three-vertex scattering processes off the macroscopic field. These analytical estimates were numerically verified using 2PI effective action techniques in Refs.~\cite{Berges:2016nru,Shen:2019jhl} for an $N$-component scalar field theory with quartic self-interactions. In classical-statistical simulations, which construct the ensemble averages from individual runs with a non-zero field value, the observed scaling exponents are consistent with the estimates in the presence of an effective three-vertex~\cite{Micha:2004bv}.   

In Ref.~\cite{Berges:2014bba} longitudinally expanding $N$-component scalar field theories are analyzed starting from over-occupied initial conditions. In the vicinity of the non-thermal attractor, scaling behavior very similar to that in a non-Abelian gauge theory is observed. The universal scaling exponents and shape of the scaling function agree well with those obtained for the early stage of the bottom-up thermalization process for gauge theories for not too late times. 

\begin{figure}[t]
\includegraphics[width=0.35\textwidth]{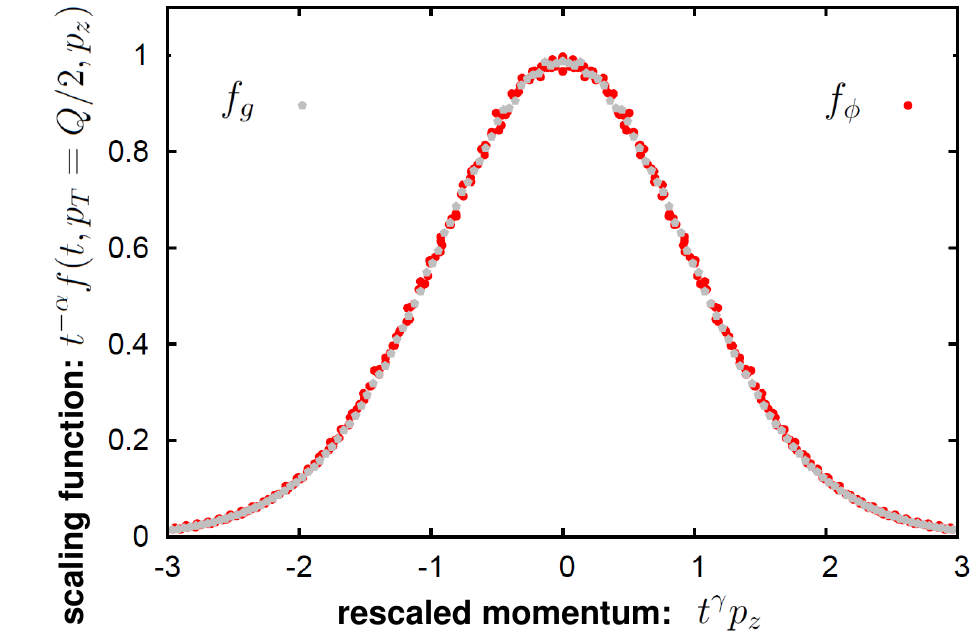}
\caption{\label{fig:UniversalScaling} The normalized distribution for the scalar theory $f_\phi$ as a function of the rescaled longitudinal momentum at different times in the self-similar regime compared to the gauge theory $f_g$. From Ref.~\cite{Berges:2014bba}.}
\end{figure}

As an example, Fig.~\ref{fig:UniversalScaling} shows results for the $N=4$ component scalar theory for intermediate transverse momentum $p_T \sim Q/2$, where the normalized scaling distribution as a function of the rescaled longitudinal momentum is given. All data curves at different times in the scaling regime collapse onto a single curve using the scaling exponents $\alpha=-2/3$,  and $\gamma = 1/3$. This scaling curve is seen to be indistinguishable from the corresponding scaling curve for non-Abelian gauge theory, which shares the same scaling exponents. The results provide a striking manifestation of universality far from equilibrium.

\subsection{Far-from-equilibrium separation of scales and ultrasoft scale dynamics}
\label{sec:IR}

The weakly coupled QCD plasma exhibits a hierarchy of scales in thermal equilibrium at high temperature $T$, with the separation of hard momenta $\sim T$ dominating the system's energy density, soft (electric screening or Debye) momenta $\sim gT$, and ultrasoft (magnetic) momenta $\sim g^2 T$ for $g^2 = 4\pi \alpha_S \ll 1$. A similar separation of scales exists far from equilibrium in the vicinity of the non-thermal attractor, where for comparison we will consider the spatially isotropic case without longitudinal expansion. 

Starting from over-occupied initial conditions, in this fixed-box case the gluon distribution function in the self-similar regime obeys $f(t,p) = t^{-4/7} f_s(t^{-1/7} p)$ in three spatial dimensions~\cite{Kurkela:2011ti,Schlichting:2012es,Kurkela:2012hp}. Accordingly, the time-dependent hard momentum scale dominating the energy density is given by $\Lambda(t) \sim t^{1/7}$. The Debye scale $m_D(t) \sim g \sqrt{\int d^3p\,f(t,p)/p} \sim t^{-1/7}$ decreases with time~\cite{Kurkela:2012hp,Berges:2013fga,Lappi:2016ato,Mace:2016svc,Boguslavski:2018beu}. 

At even lower scales, the dynamics becomes non-perturbative for momenta $K(t)$ where the occupancy reaches $\sim 1/\alpha_S$, and the perturbative notion of a gluon distribution function becomes problematic in this ultrasoft regime. As suggested in Ref.~\cite{Kurkela:2012hp}, the evolution of the ultrasoft scale may be estimated as $K(t) \sim t^{-2/7}$ using the power law form of the occupation number distribution extracted in the perturbative regime. While all characteristic momentum scales are initially of the same order $Q_S$, this suggests that during the self-similar evolution a dynamical separation of these scales $K(t) \ll m_D(t) \ll \Lambda(t)$ occurs as time proceeds.

\subsubsection{Non-equilibrium evolution of the spatial Wilson loop}

\begin{figure}[t]
\includegraphics[width=0.43\textwidth]{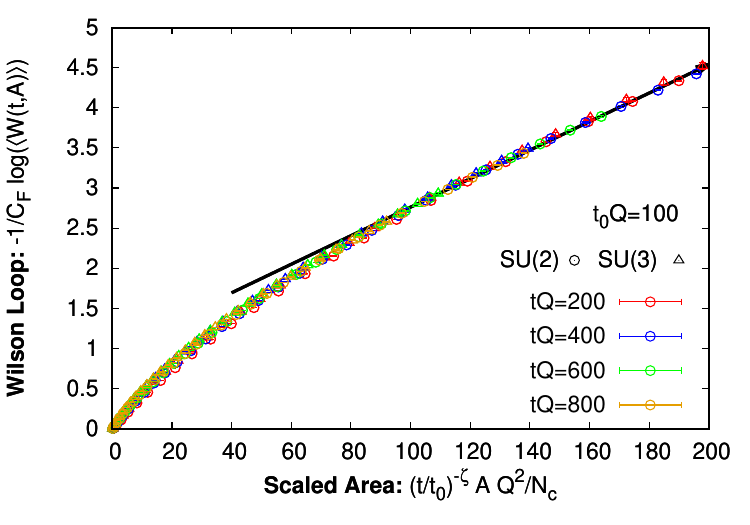}
\caption{\label{fig:WilsonLoopScaling} Self-similar behavior of the spatial Wilson loop as a function of the time-rescaled area $\sim t^{-\zeta} A$ with universal scaling exponent $\zeta$ for gauge groups $SU(N_c)$ with $N_c =2$ (circles) and $N_c=3$ (triangles)~\cite{Berges:2017igc}.}
\end{figure}

A proper description of the non-perturbative low momentum regime can be based on gauge-invariant quantities. This should take into account that the infrared excitations of non-Abelian gauge theories are extended objects, which can be computed from Wilson loops~\cite{Berges:2007re,Dumitru:2014nka,Mace:2016svc,Berges:2017igc,Berges:2019oun}. At the magnetic scale, spatial Wilson loops capture the long-distance behavior of gauge fields
$\mathcal{A}$, which is defined as
\begin{align}\label{eq:Wilson}
  W = \frac{1}{N_c} \mathrm{Tr} \, {\mathcal{P}} 
  e^{-i\, g \int_{\mathcal{C}} \mathcal{A}_i(\mathbf z,t)\, dz_i} \,, 
\end{align}
where the index $i$ labels spatial components~\cite{Montvay:1994cy}. 
Here ${\mathcal{P}}$ denotes path ordering along a closed line $\mathcal{C}$, 
and the trace is in the fundamental representation of $SU(N_c)$. 

The behavior of the spatial Wilson loop for large areas $A \gg 1/Q_S^2$ enclosed by the line $\mathcal{C}$ reflects the long-distance or infrared properties of the strongly correlated system. Like the large-distance behavior of the spatial Wilson loop in a high-temperature equilibrium plasma, the spatial Wilson loop exhibits an area law in the overoccupied regime of the non-equilibrium plasma, i.e.,~$-\log \langle W \rangle \sim A$~\cite{Berges:2007re,Dumitru:2014nka,Mace:2016svc}. 

However, here the area-law behavior occurs in the self-similar regime of the non-equilibrium evolution. This is demonstrated in Fig.~\ref{fig:WilsonLoopScaling}, which shows the logarithm of the Wilson loop as a function of the time-rescaled area \mbox{$\sim t^{-\zeta} A$} with universal scaling exponent $\zeta$~\cite{Berges:2017igc,Berges:2019oun}. Results for the $SU(2)$ and $SU(3)$ gauge groups are both displayed. After we take into account the Casimir color factors, normalizing the data points with
$C_F = (N_c^2-1)/(2N_c)$ discloses  very similar behaviors for
$N_c = 2$ and $N_c = 3$~\cite{Berges:2017igc}. The scaling exponent  $\zeta = 0.54 \pm 0.04\; \text{(stat.)} \pm 0.05\; \text{(sys.)}$ agrees for both gauge groups to very good accuracy~\cite{Berges:2019oun}. This value of the scaling exponent for the ultra-soft scale $\sqrt{\sigma}$ obtained from lattice simulations and the perturbatively motivated result for the scaling of $K(t)$~\cite{Kurkela:2012hp} are rather close, corroborating $\sqrt{\sigma} \sim K$. 

The positive value for $\zeta$ signals evolution toward larger length scales, with a growing characteristic area $A(t) \sim t^{\zeta}$.
For large $A/t^{\zeta}$ one observes from Fig.~\ref{fig:WilsonLoopScaling} the generalized area-law behavior~\cite{Berges:2017igc,Berges:2019oun} 
\begin{equation}\label{eq:asymptoticomega}
  - \log \langle W \rangle 
  \sim  A/t^{\zeta} \, .
\end{equation} 
This implies a time-dependent string tension scale $\sigma(t) = -\partial \log \langle W\rangle / \partial A \sim t^{-\zeta}$. 
   
In Ref.~\cite{Mace:2016svc}, this behavior is related to the rate of topological transitions, the so-called sphaleron transition rate: 
\begin{equation}
\label{eq:sphaleron}
\Gamma_{\rm sphaleron} = C\, \sigma^2\,,
\end{equation}
where $C$ is a number of the order of unity. The picture that emerges is that the rate of topological transitions is large at early times ($\Gamma_{\rm sphaleron}\sim Q_S^4$) but subsequently decreases with time at a rate dictated by the universal scaling exponent $\zeta$. One expects this rate to converge from above to the thermal rate for sphaleron transitions in a high-temperature plasma~\cite{Moore:2010jd}. We will return to the implications of these results for the evolution of anomalous currents in \Sec{sec:Diracfermions}.

\begin{figure}[t]
\includegraphics[width=0.4\textwidth]{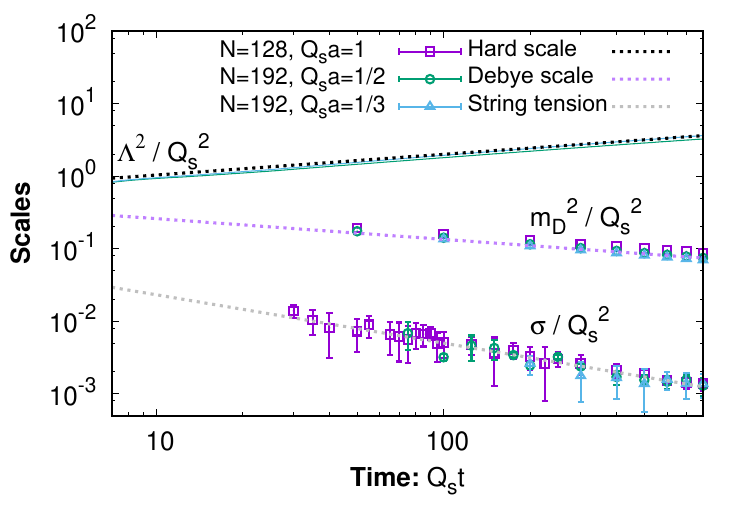}
\caption{\label{fig:StringTension} Time evolution of the hard scale ($\Lambda^2$), the electric screening scale ($m^2_D$), and the spatial string tension ($\sigma$). Symbols for $m^2_D$ and $\sigma$ denote lattice results for the different lattice spacings employed, while the solid green line represents a continuum extrapolation for the hard scale. The dotted lines represent perturbative estimates for the hard and Debye scales, and a fit to the lattice results for the non-perturbative string tension scale. From Ref.~\cite{Mace:2016svc}.}
\end{figure}

Figure~\ref{fig:StringTension} summarizes the behavior of the different characteristic scales in the self-similar regime far from equilibrium. Apart from the perturbative behavior of the hard scale, classical-statistical lattice simulations results are given for the Debye and the non-perturbative string tension scale~\cite{Mace:2016svc}. The results clearly demonstrate the dynamical separation of scales as a function of time.   

\subsubsection{Effective condensate dynamics}
\label{sec:ECD}

The traced Wilson loop in \Eq{eq:Wilson} may be directly related to correlation functions of a gauge-invariant scalar field~\cite{Gasenzer:2013era,Ford:1998bt,Mitreuter:1996ze}. In thermal equilibrium, this scalar field serves as an order parameter for the confinement-deconfinement phase transition of the underlying gauge theory~\cite{Braun:2007bx,Fister:2013bh}. In the self-similar scaling regime of the non-thermal attractor, the dynamical evolution of the scalar order-parameter field modes toward the infrared bears many similarities~\cite{Berges:2019oun} to the dynamics of Bose condensation in non-relativistic field theories far from   equilibrium~\cite{Berges:2012us,Orioli:2015dxa,Chantesana:2018qsb}. Even quantitatively, the values for the infrared scaling exponents in the different theories agree well within errors~\cite{Berges:2019oun}. 

The non-equilibrium infrared dynamics for scalars starting from over-occupation
has been studied in great detail~\cite{Berges:2008wm,Scheppach:2009wu,Berges:2010ez,
  Nowak:2010tm,Nowak:2011sk,Berges:2012us,Berges:2014bba,
  Orioli:2015dxa,Moore:2015adu,Walz:2017ffj,Chantesana:2018qsb,Deng:2018xsk,PineiroOrioli:2018hst,Shen:2019jhl,Boguslavski:2019ecc}. The emergence of self-similar scaling behavior is closely related to the existence of non-thermal fixed points~\cite{Berges:2008wm,Berges:2008sr,Berges:2012ty,Corell:2019jxh}. For scalar $N$-component theories, the behavior can be approximately described by a large-$N$ effective kinetic theory to next-to-leading order, which describes the perturbative higher momentum regime as well as the non-perturbative infrared dynamics~\cite{Walz:2017ffj}. 
  
  Both relativistic and non-relativistic scalar theories can show the same infrared scaling and condensation properties~\cite{Orioli:2015dxa}. This is true even for the anisotropic dynamics of relativistic scalars with longitudinal expansion along the $z$-direction; the latter geometry is relevant in the context of heavy-ion collisions, and scalar and gauge theories show  very similar behaviors for higher momenta in this case~\cite{Berges:2015ixa}. Because of the strong enhancement in the overoccupied infrared regime, the low momentum modes exhibit essentially isotropic properties despite longitudinal expansion.
  
A remarkable development in this regard is that table-top experiments with ultracold quantum gases have discovered universal transport processes toward the infrared starting from initial overoccupation of bosonic excitations of trapped atoms~\cite{Prufer:2018hto,Erne:2018gmz}, which is similar to the case discussed here. This is discussed further in Sec.~\ref{sec:interdisciplinary}.

\subsection{Early-time fermion production and quantum anomalies\label{sec:Diracfermions}}

In the high energy limit, strong gauge fields dominate the earliest stages of the plasma's space-time evolution. However, the Bose enhancement from over-occupied gluons can lead to a rapid production of quarks with important phenomenological consequences for heavy-ion collisions, such as direct photon production from the electrically charged quarks~\cite{Chatterjee:2009rs} or the breaking of classical symmetries due to anomalies, with a prominent example being the chiral magnetic effect~\cite{Kharzeev:2015znc,Skokov:2016yrj}. At early times these processes occur far from equilibrium and require suitable techniques for their computation. We will now discuss these techniques and their consequences for the production and evolution of fermions off-equilibrium.

\subsubsection{Real-time simulations for fermions and gauge fields beyond the classical-statistical approximation}
\label{sec:2PI-fermions}

Since identical fermions cannot occupy the same state, their quantum nature is in general highly relevant and a consistent quantum treatment of their dynamics is  crucial. In the QCD Lagrangian in \Eq{eq:QCD-Lagrangian}, quarks appear to be bilinear fields.  Their real-time quantum dynamics may therefore be computed by numerically solving the operator Dirac equation coupled to the gluon fields. 

This can be achieved in an approximation where the gauge fields are treated using classical-statistical field theory and by employing a mode function analysis of the operator Dirac equation for quarks with available lattice simulation techniques~\cite{Aarts:1998td,Borsanyi:2008eu,Saffin:2011kc,Berges:2010zv,Kasper:2014uaa}. For strong gauge fields $\sim 1/g$, this approximate description amounts to a systematic expansion of the quantum dynamics in $\alpha_S \equiv g^2/(4\pi)$, where the leading order includes the full (non-perturbative) classical-statistical theory of gluons, and the next-to-leading order takes into account back-action of the quarks onto the gluons, which is controlled by $\sim \alpha_S N_f$ for $N_f$ quark flavors.  

This can be also directly understood from the path integral formulation of the quantum theory as described in detail in Ref.~\cite{Kasper:2014uaa} for Abelian and non-Abelian gauge theories with fermions on a lattice. Performing the Gaussian integration for the quark fields in QCD analytically yields a path integral for the gauge fields $A^\pm$ on the forward ($+$) and backward ($-$) part of the closed time contour (see Sec.~\ref{sec:Glasma}) with an effective action 
\begin{equation}
    S_{\mathrm{eff}}[A^+,A^-] = {\mathrm{Tr}} \log \Delta^{-1}[A^+,A^-] + i S_{\mathrm{YM}}[A^+,A^-]\, .
\end{equation}
The term ${\mathrm{Tr}} \log \Delta^{-1}[A^+,A^-]$ arises from the Gaussian integral over the quarks, where $i\Delta^{-1}[A^+,A^-]$ denotes the inverse fermion propagator in the presence of the gauge fields. Here $S_{\mathrm{YM}}[A^+,A^-]$ is the Yang-Mills action of the pure gauge theory evaluated on the upper and lower branches of the closed time contour.

\begin{figure}[t]
\includegraphics[width=0.3\textwidth]{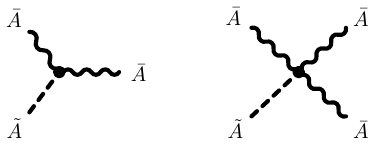}
\caption{\label{fig:classvert} Illustration of rescaled ``classical'' three- and four-vertices, that are independent of the coupling. From Ref.~\cite{Kasper:2014uaa}.}
\end{figure}

The power counting for strong gauge fields is most efficiently done by a rotation of the $\pm$-basis for the gauge fields, splitting the gauge fields into a ``classical'' part $\bar{A}$ and a ``quantum'' one $\tilde{A}$, according to
\begin{eqnarray}
\label{eq:retadv}
A^+ = \frac{1}{g}\bar{A} + \frac{g}{2}\tilde{A} \quad , \quad 
A^- = \frac{1}{g}\bar{A} - \frac{g}{2}\tilde{A} \, .
\end{eqnarray}
Expressed thus in terms of $\bar{A}$ and $\tilde{A}$, the interaction terms of $S_{\mathrm{YM}}$ can be similarly decomposed into classical and quantum parts. 

\begin{figure}[t]
\includegraphics[width=0.3\textwidth]{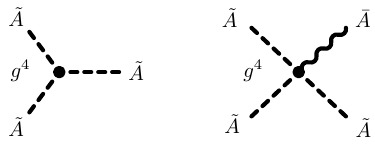}
\caption{\label{fig:quantvert} Illustration of rescaled ``quantum'' three- and four-vertices, that are $\sim g^4$. From Ref.~\cite{Kasper:2014uaa}.}
\end{figure}

This is illustrated in Fig.~\ref{fig:classvert}, which indicates the classical three-vertex $\sim \bar{A}^2\tilde{A}$ and four-vertex $\sim \bar{A}^3\tilde{A}$ parts of $S_{\mathrm{YM}}$, which are linear in the quantum field $\tilde{A}$. Figure~\ref{fig:quantvert} gives the corresponding quantum three-vertex $\sim g^4 \tilde{A}^3$ and four-vertex $\sim g^4 \bar{A}\tilde{A}^3$ parts of $S_{\mathrm{YM}}$, which are cubic in the quantum field $\tilde{A}$ and suppressed by 2 powers of $\alpha_S$ relative to their classical counterparts.

A similar analysis can be done for the ${\mathrm{Tr}} \log \Delta^{-1}[\bar{A},\tilde{A}]$ contribution coming from the quark fluctuations. Expanding this contribution in powers of the quantum field $\tilde{A}$ yields~\cite{Kasper:2014uaa} 
\begin{equation}
\label{eq:fermcont}
    {\mathrm{Tr}} \log \Delta^{-1}[\bar{A},\tilde{A}] \sim g^2\, {\mathrm{Tr}} \left( j_q[\bar{A}] {\tilde{A}} \right) + g^4\, {\cal O}(\tilde{A}^3) \, .
\end{equation}
The linear term in $\tilde{A}$ is proportional to the quark vector-current in the presence of the classical gauge field $j_q[\bar{A}]$~\cite{Aarts:1998td,Borsanyi:2008eu,Saffin:2011kc,Berges:2010zv,Kasper:2014uaa}.  

Correspondingly, in this formulation the limit $g=0$ represents the classical-statistical field theory limit of pure Yang-Mills theory. In fact, the rescalings with the gauge coupling employed in \Eq{eq:retadv} reflect the fact that for classical-statistical field theory the coupling can always be scaled out by suitable field re-definitions, while this not possible in the presence of quantum corrections. Since fermions are genuinely quantum, one cannot scale out the coupling from their contributions, as seen in \Eq{eq:fermcont}, which starts at the order of $\alpha_S$. 

According to the previous analysis, genuine quantum corrections to the dynamics in pure Yang-Mills theory enter only at the order of $\alpha_S^2$. Both the classical-statistical field contribution for the Yang-Mills part, and the lowest contribution from quark fluctuations to $S_{\mathrm{eff}}$, are linear in $\tilde{A}$. When we neglect higher-order corrections coming from terms with higher powers of $\tilde{A}$, the stationarity condition $\delta S_{\mathrm{eff}}[\bar{A},\tilde{A}]/\delta \tilde{A} = 0$ yields the classical Yang-Mills evolution equation for $\bar{A}$ with the quark current as a source term. This can be efficiently implemented numerically with sampling techniques using the Wilson plaquette formulation on a lattice~\cite{Aarts:1998td,Borsanyi:2008eu,Saffin:2011kc,Berges:2010zv,Kasper:2014uaa}.

Numerical solutions of the non-equilibrium time evolution of gluons with dynamical quarks were obtained in Ref.~\cite{Gelis:2005pb} from (2+1) dimensional boost invariant simulations,  in Ref.~\cite{Gelfand:2016prm} in 3+1 space-time dimensions for a non-expanding system, and in Ref.~\cite{Tanji:2017xiw} for a realistic case with longitudinal expansion. The calculations provide important first principles results on early quark production and the approach toward chemical equilibrium. The results for the gluon sector are in line with earlier simulations without quarks as expected at weak couplings, including self-similar scaling characteristic of the first stage of the bottom-up thermalization scenario~\cite{Baier:2000sb,Berges:2013eia}. Several properties of the quark number distributions are carried over from the gluon distributions, such as longitudinal momentum broadening~\cite{Tanji:2017suk,Tanji:2017xiw}. 

We also note recent work on the real-time propagation of heavy quarks in the Glasma that are important for a first-principles understanding of quarkonium production in heavy-ion collisions~\cite{Lehmann:2020kjg}. 

Classical-statistical lattice simulations cannot correctly describe the late-time thermalization dynamics, when typical gluon occupancies become of the order of unity. The evolution may then be continued with effective kinetic descriptions, as reported in Sec.~\ref{sec:ektchemical}.

\subsubsection{Real-time off-equilibrium dynamics of quantum anomalies}
\label{sec:quantum-anomalies}

The pair production of quarks and antiquarks lead to macroscopic manifestations of quantum anomalies, corresponding to the breaking of classical symmetries by quantum effects. These may be observable in heavy-ion collisions in the form of a Chiral Magnetic Effect (CME) whereby topological transitions in the very strong electromagnetic $B$ fields at early times generate a vector current in the direction of the $B$ field ~\cite{Kharzeev:2007jp,Fukushima:2008xe}. The prospects for the discovery of this and related phenomena were reviewed in Refs.~\cite{Kharzeev:2015znc,Skokov:2016yrj}.

The key idea is that transitions between different topological sectors of the non-Abelian gauge theory can induce a net axial charge asymmetry $j^0_a$ of light quarks, which can fluctuate on an event-by-event basis. In off-central heavy-ion collisions where strong electromagnetic $\vec{B}$-fields are present, this axial charge asymmetry can be converted into an electric current $\vec{j} \sim j^0_a \vec{B}$ that is potentially observable. Since the large ``magnetar strength" $B$ fields die off very quickly after the collision~\cite{Skokov:2009qp}, the CME is most pronounced at the earliest times after the collision. 

The non-equilibrium dynamics of topological transitions in a highly occupied, albeit non-expanding, Glasma was studied in Ref.~\cite{Mace:2016svc} by performing classical-statistical simulations and employing a cooling technique to isolate infrared dominated topological transitions. Since gluon saturation generates a large scale $Q_S\gg \Lambda_{\rm QCD}$, so-called sphaleron transitions generate real-time transitions between configurations characterized by integer valued topological charge that may be separated by an energy barrier. 

Interestingly, the boost invariant Glasma configurations discussed in \Sec{sec:Glasma-tubes} do not correspond to integer valued configurations of topological charge~\cite{Kharzeev:2001ev}; sphaleron transitions therefore go hand in hand with the explosive growth of plasma instabilities that break boost invariance, a phenomenon named ``exploding sphalerons"~\cite{Shuryak:2002qz}. 
As noted in \Eq{eq:sphaleron}, the sphaleron transition rate is controlled by the spatial string tension in the Glasma.

While off-equilibrium topological transitions are an essential ingredient, the CME in heavy-ion collisions is mediated by the transport of quarks in this topological background and in the presence of external $B$ fields. To address this problem of anomaly transport in such backgrounds, real-time lattice simulations were performed with dynamical fermions for $3+1$ dimensional Abelian and non-Abelian gauge theories in Refs.~\cite{Mueller:2016ven,Mace:2016shq} for given background gauge fields. In addition, transient anomalous charge production in strong-field QCD was studied in Refs.~\cite{Tanji:2016dka,Tanji:2018qws}.

Anomalies have been investigated for Abelian theories off-equilibrium for the fully dynamical situation, including the back-reaction of the fermions onto the gauge fields, in one~\cite{Zache:2018cqq,Kharzeev:2020kgc}, two~\cite{Ott:2019dkn} and three~\cite{Mueller:2016aao,Mace:2019cqo} spatial dimensions. In Refs.~\cite{Zache:2018cqq,Kharzeev:2020kgc} dynamical topological transitions in the massive Schwinger Model with a $\theta$-term as a prototype model for CP-violation, are studied. A dynamical order parameter for quantum phase transitions between different topological sectors is established, that can be accessed through fermion two-point correlators. Using exact diagonalization techniques,  the topological transitions have been shown to persist beyond the weak-coupling regime~\cite{Zache:2018cqq}.

Quantum fluctuations lead to an anomalous violation of parity symmetry in quantum electrodynamics for an even number of spatial dimensions, that was studied in Ref.~\cite{Ott:2019dkn} using the previously described lattice simulation techniques. While the leading parity-odd electric current vanishes in vacuum, a non-cancellation of the anomaly for strong electric fields off-equilibrium is observed with distinct macroscopic signatures.

The non-linear dynamics of the CME in QED was computed in Ref.~\cite{Mueller:2016aao} using real-time lattice simulations. For field strengths exceeding the Schwinger limit for pair production, one encounters a highly absorptive medium with anomaly-induced dynamical refractive properties. An intriguing tracking behavior is found, in which the system spends the longest time near collinear field configurations with maximum anomalous current. 

An interesting phenomenon observed in such simulations of off-equilibrium QED plasmas is that of chiral instabilities proceeding through the primary and secondary instabilities that we discussed previously culminating in a self-similar turbulent magnetic helicity transfer to macroscopic length scales~\cite{Mace:2019cqo}; see also Ref.~\cite{Buividovich:2015jfa}.

\section{Equilibration in QCD kinetic theory\label{sec:kinetictheory}}

\subsection{The quasi-particle description of QCD plasmas}
\label{sec:quasiparticle}

To solve the quantum equations of motion [\Eq{eq:FEOM}] for the late time evolution toward local thermal equilibrium, an effective description with a well defined range of validity at certain long time and distance scales is needed.
A well known example is kinetic theory, which describes the state of the system in terms of  phase space distributions of particles. 
Such an effective kinetic description of the plasma may be obtained from $n$-particle irreducible quantum effective action techniques by following along the lines of Refs.~\cite{Blaizot:2001nr,Berges:2004pu,Carrington:2009xf}. 

The derivation of kinetic theory from the underlying quantum field theory involves a series of approximations. 
First, for the notion of particles with a well defined position and momentum between collisions to be valid, the de Broglie wavelength of the \mbox{(quasi-)}particles must be small compared to the mean free path between collisions  Likewise, quantum interference effects between successive scattering events should not spoil a description in terms of independent scatterings. For the weakly coupled QCD plasma at high temperature these questions were addressed in a series of works culminating in the kinetic theory formulation by Arnold, Moore and Yaffe~\cite{Arnold:2002zm}.

The phase space distribution functions employed in kinetic descriptions are derived from two-point correlation functions of the underlying quantum field theory. In local thermal equilibrium, the system is locally homogeneous and time independent. Therefore
all two-point functions can depend only on the relative coordinate $s^\mu=x^\mu-y^\mu$. For slow variations in space and time
of the central coordinates [$X^\mu= (x^\mu+y^\mu)/2$], one considers the  
evolution in $X$ given by a gradient expansion of
\Eq{eq:FEOM} for the spectral function $\rho$ and the statistical function $F$. To the lowest order in gradients,
the evolution equation for $\rho$ is not dynamical, and a quasi-particle description emerges from an on-shell spectral function $\rho$ in the weak coupling limit~\cite{Berges:2004pu}. 

Here we consider the temperature $T$ of the QCD plasma to be the single dominant energy scale in the problem.  At leading order in the coupling, the self-energy already receives contributions from
 an infinite number of perturbative loop diagrams with hard $\mathcal O(T)$ internal momentum: Hard Thermal Loops (HTLs)~\cite{Braaten:1989mz}.
 This results in quasi-particles acquiring a screening mass $m\sim gT$.
 
The equation of motion for the statistical function is solved by generalizing the 
 Kubo-Martin-Schwinger (KMS) relation
to introduce a non-equilibrium distribution function $f(X,p)$
\begin{equation}
    F(X,p) =-i\left[\frac{1}{2}\pm f(X,p)\right]\rho(X,p),
\end{equation}
where ``$+$" is for bosons and ``$-$" is for fermions and the quasi-particle momentum $p^\mu$ is
the Fourier conjugate to the relative coordinate $s^\mu$. In general, there can be separate distributions for different color, spin and polarization components of the two-point correlation function.

From the equation of motion for the statistical function one obtains the kinetic Boltzmann equation for the distribution function, which  is  written as\footnote{Keeping interactions with strong background gauge fields leads to more general equations 
~\cite{Mrowczynski:2016etf}.}
\begin{equation}
  p^\mu\partial_\mu f(X,p)  = -C[f]\label{eq:Boltz}
.\end{equation}
The leading order collision  term $C[f]$ is obtained using a systematic power counting in the coupling constant;  this computation is non-trivial and various diagrammatic approaches have been employed to derive the relevant collision processes.
For a systematic derivation of kinetic theory from the underlying field theory see \cite{Jeon:1994if, Jeon:1995zm,Calzetta:1999ps} for the scalar case and \cite{Aarts:2005vc, Gagnon:2007qt} for Abelian field theories.

For non-Abelian gauge theories at high temperatures, the leading order collision kernel appears at $g^4$ order. However, in addition to elastic scattering processes, there are collinear splitting processes that contribute at the same order. The importance of the latter was recognized only later~\cite{Aurenche:1998nw,Arnold:2001ba}.
The corresponding vertex corrections for the underlying quantum field theory can be formulated using higher $n$PI effective actions~\cite{Berges:2004pu,Carrington:2009xf}.

Once relevant physics processes are accounted for at the given order,
\Eq{eq:Boltz} describes the non-equilibrium  evolution of QCD plasmas with
the coupling constant $g$ as the only free parameter at high temperature (with the possible exception of heavy quark masses). In particular, one
can use linearized kinetic theory to compute various transport properties of
the plasma around thermal equilibrium: shear and bulk viscosities, conductivity,
diffusion and higher order transport coefficients~\cite{Arnold:2003zc, Arnold:2006fz, York:2008rr}.
 For a recent comprehensive review on  perturbative thermal QCD techniques in kinetic theory and beyond, see Ref.~\cite{Ghiglieri:2020dpq}. As we will later discuss in detail, the
QCD kinetic theory also provides a phenomenologically successful picture of QCD thermalization  in heavy-ion collisions~\cite{York:2014wja, Kurkela:2014tea, Kurkela:2015qoa, Keegan:2016cpi}. For a complementary review, see \cite{Schlichting:2019abc}.

\subsubsection{Chiral kinetic theory} 
\label{sec:chiral-KE}
 
In the rest of Section~\ref{sec:kinetictheory}, we will discuss in detail the equilibration of QCD in the framework of spin and color averaged kinetic theory. Spin and color dependent kinetic descriptions require extensions of phase space distributions~\cite{Berezin:1976eg,Mueller:2019gjj}. Such theories must include a relativistic covariant description of Berry curvature and of the dynamics of the chiral anomaly for spinning and colored particles in external background fields~\cite{Son:2012zy,Stephanov:2012ki,Chen:2015gta}. 
 
 Recent work in this direction includes a Wigner function approach~\cite{Weickgenannt:2019dks,Hattori:2019ahi,Gao:2019znl,Yang:2020hri,Sheng:2020oqs}, chiral effective field theory~\cite{Carignano:2019zsh}, and a worldline formalism~\cite{Mueller:2017arw}. An important question to resolve in this context is the relation of the dynamics of Berry's phase to that of the chiral anomaly~\cite{Fujikawa:2005cn,Mueller:2017lzw,Yee:2019rot}. A common goal of these approaches is a consistent framework to describe anomalous transport in QCD that can be matched to an anomalous relativistic hydrodynamic description at late times~\cite{Inghirami:2019mkc}. These studies have strong interdisciplinary connections to chiral transport across energy scales ranging from Weyl and Dirac semi-metals to astrophysical phenomena~\cite{Landsteiner:2016led}.

\subsection{Leading order kinetic theory\label{sec:ekt}}

We will briefly recap here the main ingredients of QCD effective kinetic theory at leading order in the coupling constant~\cite{Arnold:2002zm}. We will consider the time evolution of the color and spin/polarization averaged distribution function $f_s$ 
with effective $2\leftrightarrow 2$ scatterings and $1\leftrightarrow 2$ collinear radiation terms.  For a transversely homogeneous and boost invariant system (applicable at early times in central heavy-ion collisions), the phase space distribution $f^s(\tau,\p)\equiv f^s_\p$ is  a function only of Bjorken time $\tau=\sqrt{t^2-z^2}$ and momentum. The resulting Boltzmann equation is
 \begin{align}
\left(\partial_\tau  - \frac{p^z}{\tau} \frac{\partial}{\partial p^z}\right)f^s_\p &= -{C}^s_{2\leftrightarrow 2}[f](\p)-{C}^s_{1\leftrightarrow 2}[f](\p)
\label{eq:bolz}
 \end{align}
 with the massless\footnote{At leading order, we can neglect the thermal mass correction $m_s\sim gT$ to the dispersion relation $p=\sqrt{|\p|^2+m_s^2}$ for hard momenta $|\p|\sim T$ on external legs.}  dispersion relation, $p^0 = |\p |=p$. Consequently, this kinetic theory describes a conformal system
 with temperature $T$ as the only dimensionful scale.
The index $s$ refers to different particle species in the theory such as quarks and gluons in $SU(N_c)$ gauge theory with $N_f$ fermion flavors. The second term on the left hand side is due to the longitudinal gradients in a boost invariant expansion~\cite{Baym:1984np}.  The expansion redshifts the distribution in the $p^z$ direction, thereby making it more anisotropic along the longitudinal direction. 
Different stages of the thermalization process are defined by the competition between the expansion that drives the system
away from equilibrium and the collision terms that isotropize and equilibrate the system.

\subsubsection{Elastic two-body scattering\label{sec:elastic}}

 The $2\leftrightarrow 2$ collision term for particle species $s=a$ is
\begin{align}
&\label{eq:2to2}
{C}^a_{2\leftrightarrow 2}[f](\p) = \frac{1}{4 p \nu_a}
 \sum_{bcd}\int \frac{d^3\k d^3\p' d^3\k'}{(2\pi)^9 2 k2p'2k'}\nonumber \\
&\times \left\{ f^a_\p f^b_\k (1\pm f^c_{\p'})(1\pm f^d_{\k'})-f^c_{\p'} 
f^d_{\k'} (1\pm f^a_{\p})(1 \pm f^b_{\k}) \right\}\nonumber\\
&\times
 \left|\mathcal{M}^{ab}_{cd} 
\right |^2(2\pi)^4 
 \delta^{(4)}(p^\mu+k^\mu-p'^\mu-k'^\mu),
\end{align}
where  $\sum_{bcd}$ is the sum over all particle and antiparticle species. The second line represents the phase space loss and gain terms. $|\mathcal{M}^{ab}_{cd}|^2$ is the $2\leftrightarrow2$ scattering amplitude squared and summed over spin and color degrees of freedom of the external legs, with $\nu_g=2 (N_c^2-1)$ for gluons and $\nu_q=2N_c$ for quarks.

The scattering matrix element 
$|\mathcal{M}^{ab}_{cd}|^2$  in \Eq{eq:2to2} should be calculated using in-medium corrected propagators and vertices from the HTL effective Lagrangian~\cite{Ghiglieri:2020dpq}. At leading order in the coupling constant and for hard $p\sim T$ external legs,  the scattering matrix element coincides with the tree level vacuum matrix element; for instance, 
in the case of two gluon scattering,
\begin{align}
|\mathcal{M}^{gg}_{gg}|^2 = 8 \nu_g  N_c^2 g^4 \left( 3 - \frac{st}{{u}^2}- 
\frac{su}{{t}^2}- \frac{tu}{s^2}\right),\label{c2to2}
\end{align}
where $s,t$ and $u$ are the Mandelstam variables.
In-medium corrections become relevant when $-t,-u\sim (gT)^2$ is small but $s$ is large, as is the case for the small angle scattering of hard particles.
When the exchanged gluon or quark  
is soft, so that $q = |\p' - \p|\ll T$  in the $t$\nobreakdash-channel (and likewise in the $u$\nobreakdash-channel),
the vacuum collision matrix suffers from a soft Coulomb divergence $|\mathcal M|^2\propto 1/(q^2)^2$.
Therefore the problematic scattering matrix elements in this region need to be reevaluated
using the non-equilibrium propagators for internal lines, which regulate the divergence~\cite{Arnold:2002zm}.

For isotropic distributions and hard $p\sim T$ external legs, the soft self-energy (which cuts off the long range Coulomb interactions) is proportional to the in-medium effective masses of hard gluons and quarks~\cite{Arnold:2002zm}. For gluons, it is given by (assuming that $f^q_\p=f^{\bar q}_\p$)
\begin{align}
&m_{g}^2 =  2g^2 \int \!\frac{d^3 \p}{  (2\pi)^3 p}\big[  N_cf^g_\p
+ N_f f^q_\p\big],
\end{align}
However, for anisotropic distributions the HTL
resummed gluon propagator\footnote{Note that there are no unstable fermionic
modes in anisotropic plasmas~\cite{Mrowczynski:2001az, Schenke:2006fz}.}
develops poles at imaginary frequency indicating the presence of a soft gauge
instability~\cite{Mrowczynski:1996vh,Mrowczynski:2016etf}. Formally,
this restricts the applicability of kinetic theory to parametrically
small anisotropies~\cite{Arnold:2002zm}. 

The rich physics of plasma 
instabilities has been studied extensively~\cite{Mrowczynski:2016etf}.
While such instabilities are of fundamental importance at early times, 
remarkably, classical-statistical simulations of the non-equilibrium field dynamics of the Glasma (discussed in
\Sec{sec.nonthatt}) show that such instabilities do not play a significant role {\it at late times} in expanding (3+1) dimensional non-Abelian plasmas. 
Motivated by these findings,
phenomenological approaches in kinetic theory simulations for anisotropic distributions use an isotropic screening prescription~\cite{York:2014wja, Kurkela:2015qoa}.

\subsubsection{Fokker-Planck limit of elastic scatterings}

For isotropic distributions, the elastic collision kernel for soft momentum exchange
can be rewritten as a drag and diffusion process in momentum space~\cite{Moore:2004tg,Hong:2010at,Ghiglieri:2015zma,Ghiglieri:2015ala,Ghiglieri:2018dib, Schlichting:2019abc,Blaizot:2014jna}.
First, one needs to separate the full collision kernel into a diffusion term for soft momentum transfers $q<\mu$ and large-angle scatterings $q> \mu$, where the cutoff scale $\mu$ satisfies $g T \ll\mu\ll T$:
\begin{equation}
  {C}^g_{2\leftrightarrow 2}[f](\p) =  \left. {C}^g_{\text{diff}}[f](\mu)\right|_{q<\mu}  +   \left.{C}^g_{2\leftrightarrow 2}[f](\p) \right|_{q>\mu} \label{eq:diff}
.\end{equation}
The physics of the diffusion term is that of hard particles being kicked around by the fluctuating soft gauge fields generated by other particles.  For an isotropic non-equilibrium plasma,
the expectation value of such gauge field fluctuations can be related to equilibrium fluctuations with the help of an effective 
temperature $T_*$ (taking $f^q_\p=f^{\bar q}_\p$)
\begin{equation}
T_* \equiv \frac{g^2}{m_g^2} \int \!\!
\frac{d^3\p}{(2\pi)^3} [N_c f^g_\p(1+ f^g_\p) +  N_f f^q_\p(1- f^q_\p)].
\end{equation}
Note that although $T_*=T$ in equilibrium,
$T_*$ is distinct from the effective temperature defined by the energy density and used in Secs.~\ref{sec:extrastrong} and \ref{sec:strongcoupling}.
Evaluating the collision kernel in the limit of soft momentum transfer and isotropic distributions, results in a Fokker-Planck
type collision term
\begin{align}
  {C}^g_{\text{diff}}[f](\mu)  &= \eta_D(p)\hat{p}^i\frac{\partial }{\partial p^i} \left[ f_\p^g (1+f^g_\p) \right]+\frac{1}{2} q^{ij}\frac{\partial^2 f^g_\p}{\partial p^i \partial p^j}\label{eq:fokkerplanck}
,\end{align}
where $\eta_D$ is the drag coefficient, $q^{ij} = \hat{q}_L \hat{p}^i \hat{p}^j + \frac{1}{2}\hat{q}\left( \delta^{ij} - \hat{p}^i \hat{p}^j \right) $ is the diffusion tensor, and  $\hat{p}^i = p^i/p$ is the unit vector. 

The transport coefficients $\hat{q}$ and $\hat{q}_L$ can be evaluated using the resumed HTL propagators, while $\eta_D$ is constrained by the  Einstein relation and the requirement
that \Eq{eq:fokkerplanck} vanish in equilibrium~\cite{Moore:2004tg,Arnold:1999va,Arnold:1999uza,Ghiglieri:2015ala}. The leading order result for $\hat q$ is
\begin{align}
  \begin{split}
  \hat{q}(\mu) &= \frac{g^2 N_c T_* m^2_g}{2\pi} \log \frac{\mu^2}{2 m^2_g}.
\end{split}\label{eq:qhat}
\end{align}
The UV divergence in the diffusion term is canceled by the corresponding IR divergence in the large-angle scattering term in \Eq{eq:diff}. 

We can now specify the isotropic screening prescription
for regulating the elastic collision kernel for anisotropic distributions:
 for a soft gluon exchange in the $t$\nobreakdash-channel  (likewise for the $u$-channel),
the divergent term is replaced by the IR regulated term
$t \rightarrow t(q^2 + \xi_g^2 m^2_g)/{q^2}$,
where $\xi_g = e^{5/6}/2 $ is a numerical constant fixed such 
that the new matrix element 
reproduces the full HTL result for the 
drag and momentum diffusion properties of soft gluon scattering~\cite{York:2014wja}.

Similarly, one can regulate divergent soft fermion exchanges 
to reproduce gluon to quark conversion $gg\rightarrow q\bar q$ at leading order for isotropic distributions~\cite{Ghiglieri:2015ala, Kurkela:2018oqw}.
Formally, this regulated collision kernel is accurate for small couplings and for near-isotropic systems. However, in practice numerical simulations for phenomenological applications are often performed
for stronger couplings $g\approx 1$ and anisotropic systems.

\subsubsection{Effective collinear one-to-two processes\label{sec:coll}}

\def\n{\hat{\mathbf{n}}}
\def\h{{\mathbf{h}}}
\def\q{{\mathbf{q}}}
In addition to the momentum diffusion of hard particles, 
 soft gluon exchange can also take the particle slightly off shell 
and make it kinematically possible for it to split into two nearly collinear
hard particles. Naively, such a $2\to 3$ process has an additional vertex relative to elastic $2\leftrightarrow 2$
scattering making it subleading in the coupling constant. However 
both the soft gluon exchange and the perturbed off-shell hard particle have
$\sim 1/(g^2 T^2)$ enhancements from the propagators. These compensate for the additional vertex insertion and the nearly-collinear emission phase space~\cite{Arnold:2001ba}.
For the same reason, multiple soft scatterings $N+1 \to N+2$ also have to be summed over. 

Physically, this means that the nearly on-shell hard particle lives
long enough before splitting to receive multiple kicks from the plasma that destructively interfere, leading to the suppression of emissions from very energetic particles. This phenomenon is known as the Landau-Pomeranchuk-Migdal (LPM) effect~\cite{Landau:1953gr, Landau:1953um,Migdal:1956tc,Migdal:1955nv}.
Collectively these processes are described as an effective $1\leftrightarrow 2$ matrix element.
In \Eq{eq:bolz} it is denoted by ${C}^{1\leftrightarrow 2}[f](\p)$
and has the explicit form,
\begin{align}
\label{1to2}
&{C}^a_{1\leftrightarrow 2}[f](\p)= 
\frac{(2\pi)^3}{2 \nu_a p^2}\sum_{bc}\int_0^{\infty}dp' dk' \, \Big [\\
&\quad
\gamma^{a}_{bc}(p;p',k')\delta(p-p' 
-k')\nonumber \\
&
\times
\big\{f_{p\hat{\p}}^a[1\pm f^b_{p'\hat{\p}}][1\pm f^c_{k'\hat{\p}}] - f^b_{p'\hat{\p}}f^c_{k'\hat{\p}}[1\pm f^a_{p\hat{\p}}]\big\}\nonumber\\
&- 2\gamma^{b}_{ac}(p';p,k')\delta(p'-p 
-k')\nonumber \\
&
\times
\big\{f_{p'\hat{\p}}^b[1\pm f^a_{p\hat{\p}}][1\pm f^c_{k'\hat{\p}}] - f^a_{p\hat{\p}}f^c_{k'\hat{\p}}[1\pm f^b_{p'\hat{\p}}]\big\}\Big],\nonumber
\end{align}
where the unit vector $\hat{\p}=\p/|\p|$ defines the splitting direction and $\gamma^a_{bc}(p; p', k')$ is the effective collinear splitting rate. 

As required by detailed balance, 
\Eq{1to2} describes both particle splitting  $p\leftrightarrow p'+k'$ and fusion
$p+k' \leftrightarrow p'$ processes.
Factoring out the kinematic splitting function $P_{g\to g}\left(z=\frac{p'}{p}\right)= N_c\frac{1 + z^4 + (1-z)^4}{z(1-z)} $ for the gluonic process $g \to gg$, this rate is given by 
\begin{align}
\gamma^g_{gg}(p;p',k')=&P_{g\to g}(z)\frac{\nu_g g^2}{4 \pi}\!\int\! \frac{d^2 
h}{(2\pi)^2} \frac{2 \h \cdot \text{Re}\, \mathbf{F}_g(\h; p, p', k')}{4\left( 2\pi \right)^3 p p'^2 k'^2}\,,
\end{align}
where the integral has mass dimension 2 and is proportional to the virtuality acquired
by the hard particle due to interactions with the soft gauge field.
The complex two-dimensional function $\mathbf{F}_g(\h; p, p', k')$ (with mass dimension 1)
solves the integral equation
~\cite{Arnold:2001ba, Arnold:2002zm,Arnold:2002ja}
\begin{align}
 &2 \h = i \delta E(\h) \mathbf{F}_g(\h) + g^2 \frac{N_c}{2} T_* \int 
 \frac{d^2 \q_\perp}{(2\pi)^2} \mathcal A(\q_\perp)  \label{diffeq}\\
 &\left\{ 
3\mathbf{F}_g(\h) - \mathbf{F}_g(\h-k' \q_\perp)  - \mathbf{F}_g(\h-p' \q_\perp)  - \mathbf{F}_g(\h+p  \q_\perp)\right\},\nonumber
\end{align}
 where the energy difference between the incoming and the outgoing states is 
\begin{align}
\delta E(\h; p,p',k')&\equiv \frac{m_{g}^2}{2 k'}+\frac{m_{g}^2}{2 p'} 
- \frac{m_{g}^2}{2 p}+ \frac{\h^2}{2 p k' p'}\,,
\end{align}
and $\h = (\p'\times \k')\times \hat{\p}$ quantifies the transverse momentum in the near collinear splitting. 

The second term on the r.h.s.\ of \Eq{diffeq} can be interpreted as a linearized collision integral
with loss and gain terms describing the probability of a particle to scatter in and out 
of transverse momentum $\h/p$. The scattering rate $\mathcal A(\q_\perp)$ is proportional to the
mean square fluctuation of soft gauge fields; for  isotropic distributions 
 it is given by~\cite{Aurenche:2002pd},
\begin{equation}
\mathcal A(\q_\perp)= \frac{1}{\q_\perp^2}- 
\frac{1}{\q_\perp^2+2m_g^2}.\label{eq:Aisotropic}
\end{equation}
Even with this isotropic approximation, \Eq{diffeq} is highly non-trivial. Various numerical schemes have been proposed for solving it~\cite{Aurenche:2002wq,Ghiglieri:2014kma, Ghisoiu:2014mha}.

\subsubsection{Bethe-Heitler and LPM limits of collinear radiation}

We will now discuss two limiting cases of the soft gluon radiation $z=\frac{p'}{p}\ll1$.
In the first case, the so-called Bethe-Heitler (BH) limit, the interference between successive scattering
events can be neglected. This corresponds to the first (decoherence) term in \Eq{diffeq} being much
larger than the scattering integral ($pz g^2 T_*/m_g^2\ll1$). In this case, the equation can
be solved iteratively. One obtains~\cite{Ghiglieri:2018dib}
\begin{align}
\left.\gamma^g_{gg}(p;p',k')\right|^{z\ll1}_\text{BH}=&P_{g\to g}(z) \frac{\nu_g \alpha_S}{(2\pi)^4} \left.\frac{\hat{q}(\mu)p }{ m_g^2}\right|_{\mu=e m_g},\label{eq:BH}
\end{align}
where $\hat{q}(\mu)$ is as given in \Eq{eq:qhat}.
In the opposite limit $z p T_*/m_g^2\gg 1$  (but still $z\ll 1$), 
the successive scatterings by the medium interfere destructively, reducing
the emission rate to 
\begin{align}
  \left.\gamma^g_{gg}(p;p',k')\right|_\text{LPM}^{z\ll1} =&P_{g\to g}(z)\frac{\nu_g \alpha_S}{(2 \pi)^4} \left(   \frac{\hat q(\mu)p }{z}\right)^{1/2},\label{eq:LPM}
\end{align}
where at next-to-leading-logarithmic order $\mu$ solves
$  \mu^2 =2 \sqrt{2} e^{2-\gamma_\text{E}+\pi/4} \sqrt{\hat{q}(\mu) p z}$~\cite{Arnold:2008zu}.

Owing to soft gauge field instabilities, 
collinear radiation in anisotropic plasmas contains unstable modes~\cite{Hauksson:2017udm,Hauksson:2020etn}. In phenomenological applications these unstable modes are neglected and the isotropic approximation in \Eq{eq:Aisotropic} is employed instead.

\subsection{Bottom-up thermalization\label{subsec:bottomup}}

\subsubsection{Initial conditions}
BMSS~\cite{Baier:2000sb} spelled out a bottom-up scenario for thermalization beginning with the overoccupied Glasma discussed  in  Sections \ref{sec:hadrons}, \ref{sec:Glasma} and \ref{sec:classicalstatistical}.
In this framework, the momentum  modes $p\sim Q_S$ can be interpreted as quasi-particles with a well-defined anisotropic distribution after time\footnote{Plasma instabilities that are operational over shorter time-scales are well described by classical-statistical simulations; see \Sec{sec:plasmainstabilities}.} $\tau \,Q_S\geq \log^2 \alpha_S^{-1}$.
The initial gluon distribution in kinetic simulations of this scenario is parametrized at  $Q_S\tau_0=1$ as~\cite{Kurkela:2015qoa}
\begin{align}
f^g_\p= \frac{2 A}{g^2 N_c} \frac{\left<p_T\right>_c}{\sqrt{p_\perp^2+p_z^2\xi_0^2}} e^{-\frac{2}{3} \frac{p_\perp^2+\xi_0^2 p_z^2}{\left<p_T\right>_c^2}}.\label{eq:focc}
\end{align}
The normalization $A$ is chosen to reproduce the co-moving energy density $\tau \e = \left<p_T\right>_c {dN_g}/{d^2\x_\perp dY}$. In this expression the gluon number density at a fixed initial rapidity is determined at LO by numerical simulations of the Glasma
and the result 
can be read off  \Eq{eq:LO-incl-mult}. To obtain the first principles input for the initial gluon production as a function of rapidity, one further needs to solve the JIMWLK equations described previously in \Sec{sec:RG-EFT}. Likewise, one can determine the average transverse momentum $\left< p_T\right>_c\approx 1.8 \,Q_S$~\cite{Lappi:2011ju}. Finally, the anisotropy parameter $\xi_0$ 
is varied to quantify our ignorance of the longitudinal momentum distribution.

For the evolution of the overoccupied and highly anisotropic initial state, specified at its initial time by  \Eq{eq:focc},
the typical gluon occupancy and the deviation from isotropy can be monitored by computing the following ratios: 
\begin{equation}
\frac{\left<pf \right>}{\left<p\right>} = \frac{\int \frac{d^3p}{(2\pi)^3} p f^g_\p f^g_\p}{\int \frac{d^3p}{(2\pi)^3} p f_\p^g},\quad
\label{eq.PToPL}
\frac{\mathcal P_T}{\mathcal P_L} = \frac{\frac{1}{2}\int \frac{d^3p}{(2\pi)^3p} p_\perp^2 f^g_\p}{\int \frac{d^3p}{(2\pi)^3p} p_z^2 f^g_\p}.
\end{equation}

\begin{figure*}
\centering
\includegraphics[width=0.7\linewidth]{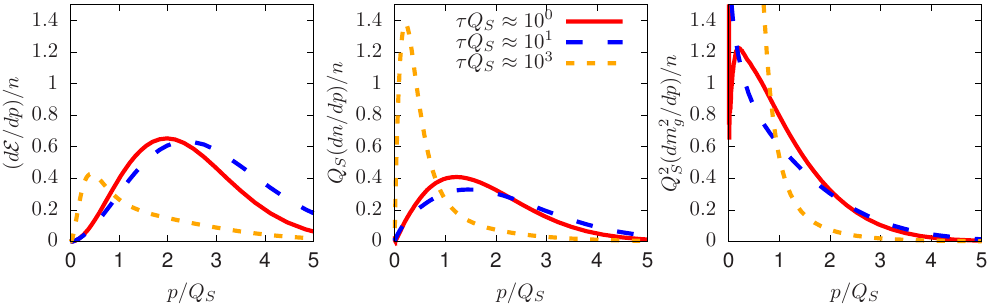}
\caption{Momentum differentiated plots  of the (a) energy density (b) number density and (c) screening-mass at different stages of  bottom-up thermalization. All curves are normalized by the instantaneous number density $n$
and lines correspond to different times ($\tau Q_S = 10^0, 10^1,  10^3$).}
\label{fig:plotcombkzprlfig2}
\end{figure*}
\subsubsection{Stage one: collisional broadening}
The solution of the collisionless Boltzmann equation in the boost invariant expansion is a simple rescaling of initial longitudinal momentum
that does not change the typical occupancy but increases the anisotropy quadratically in time. However in the presence of elastic collisions, gluons scatter into the longitudinal momentum direction thus broadening the distribution. The longitudinal momentum diffusion for anisotropic distributions can be estimated
from the Fokker-Planck equation \eq{eq:fokkerplanck}:
 \begin{align}
\left(\partial_\tau  - \frac{p^z}{\tau} \frac{\partial}{\partial p^z}\right)f^g_\p &= \frac{\hat{q}}{4} \frac{\partial^2 f_\p^g}{\partial p_z^2},
\label{eq:fokerplanck}
 \end{align}
where we kept the dominant term on the right hand side.
Note that for a highly occupied anisotropic system $\hat q\sim \int_{\p}(f^g_\p)^2$, \Eq{eq:fokerplanck} admits the scaling solution \Eq{eq:scalingsol}; as discussed in \Sec{sec:identifyingEKT}, this solution is singled out in the classical-statistical simulations.

The physical picture is that the longitudinal momentum diffuses as $\left<p_z^2\right>\sim \hat{q} \tau$,
where $\hat{q}\sim \alpha_S^2 {n_g^2}/({Q_S^2\sqrt{\left<p_z^2\right>} })$ and the hard
gluon number density per rapidity  is constant ($\alpha_S n_g \tau Q_S^{-2}\sim 1$). From this, it follows that the longitudinal momentum decreases as
\begin{equation}
  \left<p_z^2\right> \sim Q_S^2 (Q_S \tau)^{-2/3}
.\end{equation}
This clearly shows that the increase in anisotropy is milder than in the free streaming case. One obtains
 ${\mathcal P_T}/{\mathcal P_L}\propto \left({\tau}/{\tau_0}\right)^{2/3}$ and $ {\left<pf \right>}/{\left<p\right>}\propto\left({\tau}/{\tau_0}\right)^{-2/3}$, which are in agreement with the scaling behavior of the non-thermal attractor of \Sec{sec:turbulent-attractor}.
The typical occupancy becomes $\mathcal O(1)$ at the time 
\begin{equation}
\tau Q_S \geq \alpha_S^{-3/2}\,.
\end{equation}
This is the first stage of bottom-up thermalization. As  previously discussed, this corresponds to a ``quantum breaking" time where the classical-statistical approximation breaks down definitively. 
After this time, hard gluons with $p_T\sim Q_S$ are no longer overoccupied, although they still carry most of the energy and particle number.

\subsubsection{Stage two: collinear cascade}
Once the typical hard parton occupancy becomes $\mathcal{O}(1)$, the diffusion coefficient scales as
$\hat{q}\sim \alpha_S^2 n_g$, where we still have $\alpha_S n_g \tau Q_S^{-2}\sim 1$. At this time, the longitudinal momentum diffusion rate and the expansion rate are comparable, with the result that  
the longitudinal momentum reaches the constant value
\begin{equation}
  \left<p_z^2\right> \sim  \alpha_S \,Q_S^2 \,.
\end{equation}
This ensures that the momentum anisotropy remains constant in the second 
bottom-up stage.

In this stage, in addition to elastic scatterings,  medium induced collinear radiation becomes important, 
as it rapidly increases the population of soft gluons. 

The soft gluon multiplicity can be estimated using the Bethe-Heitler formula [\Eq{eq:BH}]; 
 integrating over  soft momentum  $m_D\!<p\!<\!\sqrt{\left<p_z^2\right>}$ and neglecting logarithmic factors, one obtains
$n^{\text{soft}}_g \sim \tau  \frac{\alpha_S^3 }{m_g^2}\left(n^{\text{hard}}_g\right)^2$.
The screening mass is now dominated by soft isotropic gluons 
($m_g^2\sim \alpha_S n_g^{\text{soft}}/\sqrt{\left<p_z^2\right>}$). Using the previous expression for the longitudinal momentum, we can show that the soft and hard gluon multiplicities are of the same order at times
\begin{equation}
Q_S \tau \geq \alpha_S^{-5/2}.
\end{equation}
At this time, the soft gluons have thermalized among themselves, forming a bath with an effective temperature. This marks the end of the second stage of bottom-up thermalization.

\subsubsection{Stage three: mini-jet quenching}
Even though the soft gluons have thermalized, the hard gluons 
still dominate the energy density. They are, however, highly diluted ($\left<fp\right>/\left<p\right> \ll 1$); the non-equilibrium modes are now underoccupied, as opposed to being overoccupied in the first bottom-up stage. 
Although soft gluon emission is very efficient in populating the infrared, the  successive
$z\sim 1/2$ branching of modes is more efficient for energy transfer.
Such branching suffers from the LPM suppression.
The hard gluons are finally absorbed by 
the thermal bath in a ``mini-jet" quenching that is formally identical to the jet quenching formalism that is typically applied when describing much harder modes.

The system finally thermalizes when the energy in soft and hard components becomes comparable.
 This happens at the time
\begin{equation}
\tau_{\rm thermal} =  C_1 \,Q_S^{-1}\,\alpha_S^{-13/5} \, ,
\label{eq:thermalization-time}
\end{equation}
with the thermalization temperature $T = C_2 \alpha_S^{2/5} Q_S$. Here $C_1$ and $C_2$ are $\mathcal O(1)$ constants~\cite{Baier:2002bt,Baier:2011ap}.  
This time scale is parametrically  $\alpha_S^{-1/10}$ longer than when  stage two ends and therefore only cleanly distinguishable at asymptotically small values of the coupling.

The bottom-up thermalization scenario provides an intuitive picture of equilibration at weak coupling. It is remarkable, given the complexity of the thermalization process in QCD, that this scenario allows one to relate the final thermalization time and temperature to the scale for gluon saturation in the nuclear wavefunctions.

Asymptotic freedom tells us that the coupling constant must run with $Q_S$, which is  the relevant hard scale in the problem. 
Therefore an interesting consequence of \Eq{eq:thermalization-time} is that $\tau_{\rm thermal} \sim \log^{13/5}({Q_S})/Q_S \rightarrow 0$ as $Q_S\rightarrow \infty$.
{\it Thus, contrary to naive expectations, the bottom-up thermalization scenario predicts that thermalization in the asymptotic Regge limit of QCD will occur nearly instantaneously relative to the size of the system. }

\subsubsection{Numerical realization of bottom-up thermalization}

The thermalization time scales in the previous discussion were only parametric estimates. We will now discuss the results of a numerical implementation of the bottom-up kinetic  evolution from the overoccupied initial phase space distribution in \Eq{eq:focc} to the Bose-Einstein distribution~\cite{Kurkela:2015qoa}.

For 't~Hooft coupling $\lambda=N_c g^2=1$ and  initial anisotropy  $\xi_0=10$,  we show in  \Fig{fig:plotcombkzprlfig2}(a-c) the evolution of the gluon distribution (integrated over the spherical angle) with different momentum weights.
The three panels correspond, respectively, to the distribution of the gluon energy density $d\e/dp$, the number density $dn/dp$, and the screening mass $dm_g^2/dp$ as a function of gluon momentum. To factor out the dilution due to expansion, all of these quantities are 
normalized by the total gluon number
density $n$. The lines correspond to different times $\tau \,Q_S \approx 1,10,10^3$.

We see 
that at early times $\tau \,Q_S \approx 1-10$ 
the hard $p>Q_S$ modes dominate both the energy and particle number, and even have
significant contributions to the screening mass. At very late times ($\tau Q_S\approx 10^3$), the particle number and the screening mass are completely dominated by the soft sector, but there is still a noticeable contribution to the  energy density  from the modes with $p>Q_S$. 

\begin{figure}
	\centering
	\includegraphics[width=0.9\linewidth]{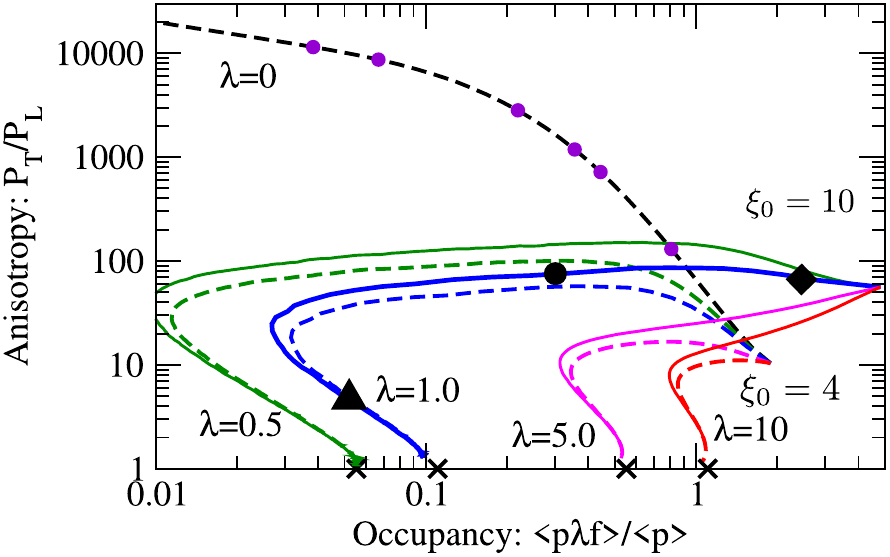}
  \caption{Gluon kinetic theory equilibration in anisotropy-occupancy plane for initial anisotropy $\xi_0=10$ and different values of the coupling constant.
  Times corresponding to $\tau Q_S = 10^0, 10^1, 10^3$ are indicated by black symbols. Simulations with the smaller initial anisotropy $\xi_0=4$ are shown
as dashed curves.
Adapted from Ref.~\cite{Kurkela:2015qoa}.}
	\label{fig:kurkelazhu}\label{fig:plotcombkzprlfig1}
\end{figure}

It is interesting to compare the momentum distributions in  \Fig{fig:plotcombkzprlfig2} to the anisotropy and occupancy evolution in \Fig{fig:plotcombkzprlfig1} (which is a kinetic theory extension of the lattice computation in \Fig{fig:Attractor}).
We mark the times $\tau \,Q_S\approx 1,10,10^3$ with a diamond, a circle, and a triangle, respectively, on the $\lambda=1, \xi_0=10$ simulation trajectory (blue solid line).  We observe that typical occupancies drop rather quickly below unity and see a slight increase of anisotropy as it happens.
However the slope of the anisotropy increase is different than the naive expectation in 
the first stage of bottom-up thermalization and is dependent on the choice
of initial conditions.

The anisotropy plateau of the second stage
is already reached at $\tau \,Q_S\approx 10$,  somewhat quicker than the parametric estimates suggest.
Finally, because the soft sector is more isotropic than that of hard gluons,  we observe that as the gluon number
shifts toward lower momentum (see \Fig{fig:plotcombkzprlfig2}(b)), the anisotropy starts falling sharply in  \Fig{fig:plotcombkzprlfig1}. This marks the onset of the third stage of bottom-up thermalization. 
Although dilute hard modes
still contribute significantly to the energy density,
the balance shifts toward more densely populated soft modes whose  
occupancy steadily increases as the system isotropizes.

The bottom-up process  finally ends when the system isotropizes. In 
practice, the third stage of bottom-up equilibration
is significantly longer than the second stage, in contrast to 
the $\alpha_S^{-1/10}$ difference in the parametric time scales.

For an initial distribution with different initial anisotropy values [$\xi_0=4,10$ dashed and solid lines in \Fig{fig:plotcombkzprlfig1}], the evolution follows a qualitatively similar
path. Although we expect all initial 
conditions to converge at thermal equilibrium,
it is remarkable that different initializations   already merge at rather large values of the anisotropies $\mathcal P_T/\mathcal P_L\approx 10$, when the system is still far away from  local thermal equilibrium.
This precocious collapse to a universal curve, independent of the initial conditions, is termed a ``hydrodynamic attractor." This phenomenon is discussed further in Secs.~\ref{sec:extrastrong} and \ref{sec.strongattractors}.

Finally, \Fig{fig:kurkelazhu} also shows kinetic equilibration with an increasing coupling constant (and decreasing shear viscosity $\upeta/s$).
For $\lambda\geq 5$, corresponding to small values of $\upeta/s \lesssim 2$ (and for which the initial occupancy is already below unity),
the system starts to isotropize almost immediately and the distinct stages of the bottom-up scenario are no longer clearly discernible.

\subsection{Self-similar evolution in the high-occupancy regime\label{sec:scaling}}

\subsubsection{Self-similar scaling}

 When characteristic field occupancies are sufficiently large for
the classical-statistical approximation to be valid, but small enough for the perturbative kinetic expansion to
apply, there is an overlapping regime where both approximations to the 
dynamics of the system are valid~\cite{Mueller:2002gd,
Aarts:1997kp,Jeon:2004dh}.

As discussed in Secs.~\ref{sec.nonthatt}  and \ref{subsec:bottomup}, the non-equilibrium dynamics of the overoccupied plasma
undergoes a remarkable simplification in complexity by exhibiting  self-similar evolution.
In kinetic theory language, the self-similar  behavior refers to the situation in which the particle distributions 
at different times can be related by  rescaling  the momentum arguments and the overall normalization; see \Eq{eq:scalingsol},
where $\alpha, \beta$, and $\gamma$ denote the universal scaling exponents. The relations between the exponents are
constrained by  conservation laws and the Boltzmann equation \eq{eq:Boltz}, for which \Eq{eq:scalingsol} provides a solution.

Longitudinally expanding systems are anisotropic and subject to soft gauge instabilities. Therefore from a perturbative viewpoint it is very surprising that
plasma instabilities do not seem to affect the late time evolution of
the classical-statistical real time simulations, as shown in \Fig{fig:Attractor}. The self-similar evolution near the non-thermal attractor is consistent with the bottom-up thermalization scenario
and numerical QCD kinetic theory simulations~\cite{Kurkela:2015qoa},  which explicitly neglect plasma instabilities. How to consistently solve the effective kinetic theory in anisotropic plasmas is an important open question~\cite{Mrowczynski:2016etf}.

Finally, as mentioned in \Sec{sec:identifyingEKT}, in the case of the non-expanding isotropic systems, the self-similar direct energy cascade plays an important role in equilibration of overoccupied bosons.
The same scaling exponents and the scaling function are also
reproduced in kinetic theory simulations~\cite{York:2014wja, Kurkela:2014tea}. Fermions are never overoccupied and chemical equilibration takes place over longer timescales than the
direct energy cascade~\cite{Kurkela:2018oqw}.

\subsubsection{Pre-scaling phenomenon}

\begin{figure}
  \centering
  \includegraphics[width=0.8\linewidth]{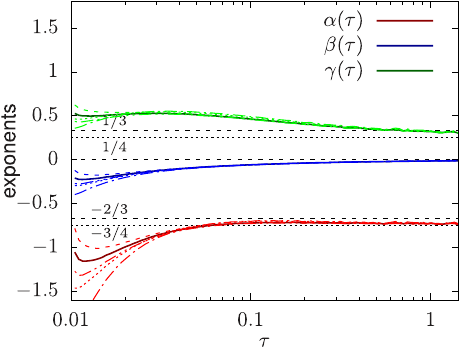}
  \caption{\label{fig:prescaling} Time evolution of instantaneous scaling exponents extracted from different
  sets of integral moments of the distribution. Horizontal lines indicate possible asymptotic values.  From Ref.~\cite{Mazeliauskas:2018yef}.}
\end{figure}

In Ref.~\cite{Mazeliauskas:2018yef} it was found that the far-from-equilibrium QGP
already exhibits a self-similar behavior before the scaling exponents attain their
constant values $\alpha=-2/3$, $\beta=0$ and $\gamma=1/3$.
The pre-scaling phenomenon is realized
through the time dependent rescaling of the distribution function and its arguments
(cf.\ \cite{Micha:2004bv}),
\begin{equation}
  f^g_\p\stackrel{\mathrm{prescaling}}{=}  \frac{(Q\tau)^{\alpha(\tau)}}{\alpha_S} f_S\left( (Q\tau)^{\beta(\tau)} p_\perp, (Q\tau)^{\gamma(\tau)} p_z\right), 
\label{eq:prescaling}
\end{equation}
where $\alpha(\tau)$, $\beta(\tau)$ and $\gamma(\tau)$ are generic time dependent functions.

Figure \ref{fig:prescaling} shows the evolution
of time dependent scaling exponents in QCD kinetic theory at very small couplings and overoccupied initial conditions~\cite{Mazeliauskas:2018yef}. 
The value of the exponents is calculated from 
the time dependence of various  moments of the distribution: 
\begin{equation}
  n_{m,n}(\tau) = \int\!\frac{d^{3}p}{(2\pi)^{3}}\,p_T^m |p_z|^n f^g_\p \label{eq:moments}\,.
  \end{equation}
Different lines of the same color in \Fig{fig:prescaling} correspond to integrals with different powers of the momentum. It is important to note that the rescaling in \Eq{eq:prescaling}
is implicitly assumed to be valid in a certain physically relevant momentum range. Therefore a finite
set of moments of \Eq{eq:moments} contains all the physically relevant information in the distribution. As shown in \Fig{fig:prescaling},
different extractions rapidly collapse onto each other and a unique set of scaling exponents
emerge that govern the time evolution of all probed  moments.

The time dependent scaling exponents provide a more differential picture of how self-similar behavior and information loss
emerge near the non-thermal attractor. Here the scaling exponents act as effective degrees of freedom whose
slowly varying evolution constitutes a hydrodynamic description of the system around the non-thermal attractor. In particular, the time dependent exponents could be well suited to studying the evolution away from the attractor in equilibrating systems even if the non-thermal attractor is never fully reached, for instance, at larger values of the coupling. For related studies in scalar field theory, see also Ref.~\cite{Schmied:2018upn}.

\subsection{Extrapolation to stronger couplings\label{sec:extrastrong}}

Thus far we have discussed a non-equilibrium QCD evolution scenario that is strictly valid only for $g\ll 1$. 
However the coupling constant is not parametrically small even at the $Z$ boson mass scale, where $\alpha_S(M_Z^2)\approx 0.1179\pm0.0010$ ($g=\sqrt{4\pi\alpha_S}\approx 1.2$)~\cite{Tanabashi:2018oca}.
In the case of finite temperature perturbation theory, the expansion parameter is $\sim \alpha_S T/m_D\sim g$ -- the convergence is therefore very slow~\cite{Blaizot:2003iq}. In this section, we will therefore discuss phenomenological extrapolations of the QCD kinetic theory to ``realistic" couplings.

\begin{figure}
	\centering
	\includegraphics[width=0.7\linewidth]{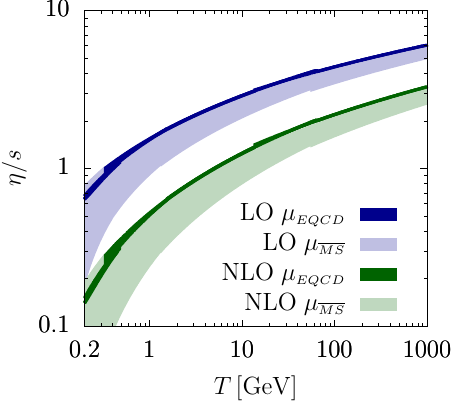}
  \caption{The shear viscosity over entropy ratio as a function of temperature at leading (LO) and (nearly) next to leading order (NLO) thermal QCD. The bands correspond to the scale variation of running coupling prescriptions.  Figure taken from \cite{Ghiglieri:2018dib}.}
	\label{fig:etavst-crop}
\end{figure}
The first calculation at next-to-leading order for QGP transport properties was performed for heavy quark diffusion and the corrections were found to be large~\cite{CaronHuot:2008uh}. On the other hand, the NLO contributions to the photon emission nearly cancel and the overall contribution is only $\sim 20\%$~\cite{Ghiglieri:2013gia}. 
Recently computations of the shear viscosity, quark diffusion
and second order transport coefficients have been extended to include higher order contributions (named ``almost NLO" in  \cite{Ghiglieri:2018dib, Ghiglieri:2018dgf}) thanks to the breakthrough idea of evaluating HTL correlations on the  lightcone~\cite{CaronHuot:2008ni}.
In \Fig{fig:etavst-crop}, we see that NLO results for the specific shear viscosity $\upeta/s$ can be a factor of 5 smaller than the leading order result at the accessible QGP temperatures $T\lesssim 1\, \text{GeV}$.
It is conceivable that a better reorganization of the perturbative expansion would result in
an improved convergence at NLO~\cite{Ghiglieri:2018dib}.

Nevertheless, for phenomenological applications in heavy-ion collisions, the strong coupling constant value $\alpha_S\approx0.3$ ($g\approx 2$) is commonly used in leading order calculations. Examples of these include 
 thermal photon emission~\cite{Paquet:2015lta}, heavy quark transport~\cite{Yao:2020xzw}, and parton energy
loss~\cite{Burke:2013yra}.
At this point, it is fair to admit that the leading order kinetic theory applications to equilibration processes in the QGP  do not provide a  controlled expansion at realistic energies and therefore have large theoretical uncertainties. 

On the other hand, QCD kinetic theory does contain the necessary physical processes, such as elastic and inelastic scatterings, to describe QCD thermalization at weak coupling. Therefore in the absence of real time non-perturbative QCD computations, extrapolating the weak coupling results to larger couplings provides a useful baseline that can be systematically improved upon.

As we will later discuss, the dependence on the coupling constant 
is better replaced by the 
 value of shear viscosity $\upeta/s$, a physical property of the QGP. 
The relaxation to equilibrium is naturally controlled by the
strength of the dissipative processes. Therefore 
rescaling weakly coupled kinetic theory dynamics to small values of $\upeta/s$ (favored by hydrodynamic modeling of QGP)
can be fairly compared to heavy-ion phenomenology and other microscopic models. This includes the genuinely strongly coupled systems discussed in \Sec{sec:strongcoupling}.
 An indication that lessons learned from QGP equilibration in leading order kinetic theory are more robust than the LO expansion itself.

There have been a number of phenomenological applications of kinetic theory to the study of thermalization in QCD. Early notable examples include Refs.~\cite{Hwa:1985tv, Biro:1993qt,Geiger:1991nj}.
Numerical implementations of classical kinetic theory including elastic $gg\leftrightarrow g g$ and inelastic $gg\leftrightarrow ggg$ gluon
scatterings were pioneered in Ref.~\cite{Xu:2004mz,El:2007vg}.
We will now focus on the results from the numerical implementations of quantum kinetic theory, including all of the leading order processes that were discussed in \Sec{sec:ekt}.

\subsubsection{Hydrodynamic attractors in QCD kinetic theory \label{sec.hydroattractorkin}}

The  universal macroscopic effective theory  close to local thermal equilibrium is given by fluid dynamics consisting of the conservation laws and  constitutive equations~\cite{LandauFluids}
\begin{equation}
  \partial_\mu T^{\mu \nu}=0,\quad T^{\mu\nu}=T^{\mu\nu}_\text{hydro}(\e, u^\mu,\ldots)\label{eq:EOM}
.\end{equation}
The only surviving information is contained in the macroscopic fluid variables, the  energy density~$\e$ and fluid velocity $u^\mu$; all other
information about the initial conditions has been lost.

The surprising phenomenological success of viscous hydrodynamics in describing many soft hadronic observables in heavy-ion collisions leads one to consider the possibility of whether a fluid dynamic description is applicable  to systems with significant deviations from local thermal equilibrium. This topic was first investigated in strongly coupled holographic models, and subsequently in
the relaxation
time approximation (RTA) kinetic theory and hydrodynamic models; see the reviews~\cite{Romatschke:2017ejr,Florkowski:2017olj} and \Sec{sec:strongcoupling}.

In the QCD kinetic theory simulations of boost invariant expansion of homogeneous plasmas~\cite{Keegan:2015avk,Heller:2016rtz,Kurkela:2018vqr}, it was observed
that the energy-momentum tensor quickly becomes  a sole function of time measured in units of the characteristic kinetic relaxation time\footnote{The effective temperature can be defined as a function of the energy density that would play the role of the temperature in equilibrium. In conformal models it is given by the fourth root of the energy density $T= [\e/( \nu_\text{eff} \pi^2/30)]^{1/4}$. For an ideal gas of quarks and gluons, $\nu_\text{eff}=47.5$ and $\nu_\text{eff}=16$ for gluons only.\label{foot.effT}} $\tau_R \sim \upeta/(sT)$, i.e.,
\begin{equation}
\wtilde \equiv \frac{\tau T}{4 \pi \upeta/s}.\label{eq:wtilde}
\end{equation} 
In such case the evolution of the energy-momentum tensor can be characterized by $\mathcal P_L/\e$ as a function of $\wtilde$~\cite{Heller:2011ju,Heller:2016rtz}.
Because $\wtilde^{-1}$ is proportional
to the Knudsen number (the natural expansion parameter for deviations from equilibrium) one would expect that for large $\wtilde$ the kinetic theory would agree with the viscous hydrodynamic result
$\mathcal P_L/\e=\frac{1}{3}-\frac{16}{9}\frac{\upeta/s}{\tau T}$. 
Surprisingly, the simplest viscous constitutive relation is  already satisfied for $\wtilde\approx 1$, when viscous correction is comparable to the equilibrium pressure. Such an effective hydrodynamic description of systems  substantially away from equilibrium is now called the hydrodynamic attractor. This notion is in fact much richer and its further aspects are discussed in Sec.~\ref{sec.strongattractors}.

\begin{figure}
\centering
\includegraphics[width=0.8\columnwidth]{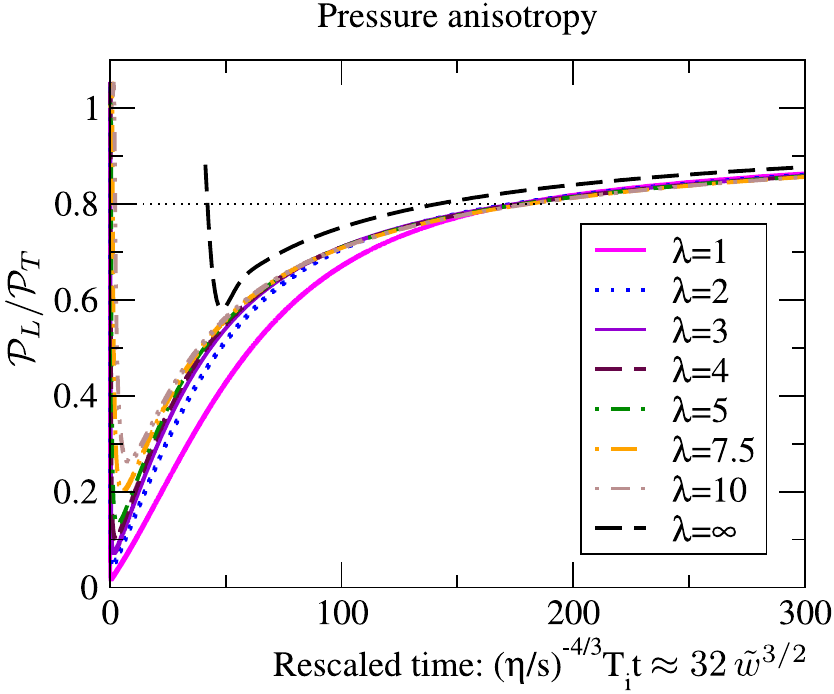}
\caption{Pressure anisotropy evolution in expanding geometry. 
Gluon kinetic theory simulations $\lambda=1,\ldots 10$ are compared 
to a  supersymmetric Yang-Mills holographic model ($\lambda=\infty$ ). Note that here $T_i$ is the initial temperature, so at late times $(\eta/s)^{4/3}T_it\approx 32\,\wtilde^{3/2}$. 
Adapted from Ref.~\cite{Keegan:2015avk}\label{fig:fig5_vanilla}.}
\end{figure}
Figure \ref{fig:fig5_vanilla} shows the pressure anisotropy $\mathcal P_L/\mathcal P_T$ as a function
of rescaled time in an expanding homogeneous system for different values of the coupling constant.
The system is prepared in an equilibrium state at initial time
and then is allowed to undergo a boost invariant expansion which drives the system away
from equilibrium. However, as the expansion slows down it
relaxes back to isotropy, satisfying $\mathcal P_L/\mathcal P_T=1$.

Note that the kinetic simulations for different couplings (which correspond to very different
kinetic relaxation times) collapse onto each other even when the pressure anisotropy $\mathcal P_L/\mathcal P_T$
is significant. Overall, the  kinetic evolution is very close to
that of an infinitely strongly coupled system. Although neither a weakly coupled kinetic theory, nor an infinitely strongly coupled supersymmetric Yang-Mills theory
is an exact description of QCD in heavy-ion collisions,
\Fig{fig:fig5_vanilla} gives some indication that in the rescaled time units $\wtilde$
the final stages of QCD equilibration could follow a very similar hydrodynamic attractor curve.

To map the hydrodynamic attractor evolution in dimensionless time $\wtilde$ to that in physical units, one needs to
fix the interaction strength by setting the shear viscosity over entropy ratio
$\upeta/s$ and the dimensionful temperature scale. Extensive hydrodynamic model comparisons to data constrain the shear viscosity to rather
small values of $4\pi\upeta/s\sim 2$ close to $T_c\approx 155\,\text{MeV}$, although its value at higher temperatures is not well determined~\cite{Bernhard:2019bmu, Devetak:2019lsk}.
The characteristic temperature scale in the hydrodynamic stage
is well constrained by the transverse entropy density  
per rapidity $(s\tau)_{\rm hydro}\sim (T^3\tau)_{\rm hydro}$, which is directly proportional to the produced particle multiplicity,
and hence can be inferred from the experimental measurements~\cite{Hanus:2019fnc}.
Inverting \Eq{eq:wtilde}, we can relate the dimensionless time $\wtilde$ in a longitudinally
expanding conformal plasma to Bjorken time~$\tau$ via 
\begin{equation}
  \label{eq:wt2t}
\tau =\kappa^{1/2}\, \wtilde^{3/2}(4\pi\upeta/s )^{3/2} \left(s\tau \right)_{\rm hydro}^{-1/2}\,.
\end{equation}
  The proportionality coefficient
  $\kappa= (s\tau)_{\rm hydro}/(\tau T^3)$
becomes a numerical constant in thermal equilibrium, where $\kappa = \nu_\text{eff}4\pi^2/90 $.
Because the kinetic simulations converge toward conventional viscous hydrodynamic predictions
for $\wtilde \gtrsim 1$, 
 it was estimated in Ref.~\cite{Kurkela:2018wud, Kurkela:2018vqr}
that the hydrodynamic description becomes applicable for times $\tau\gtrsim 1\,\text{fm}/c$ for $\upeta/s\approx 0.16$ and typical entropy densities found in central Pb-Pb collisions~\cite{Kurkela:2018wud, Kurkela:2018vqr}. This is consistent with the early hydrodynamization picture employed in the modeling of heavy-ion collisions.

\begin{figure}
  \centering
  \includegraphics[width=0.75\linewidth]{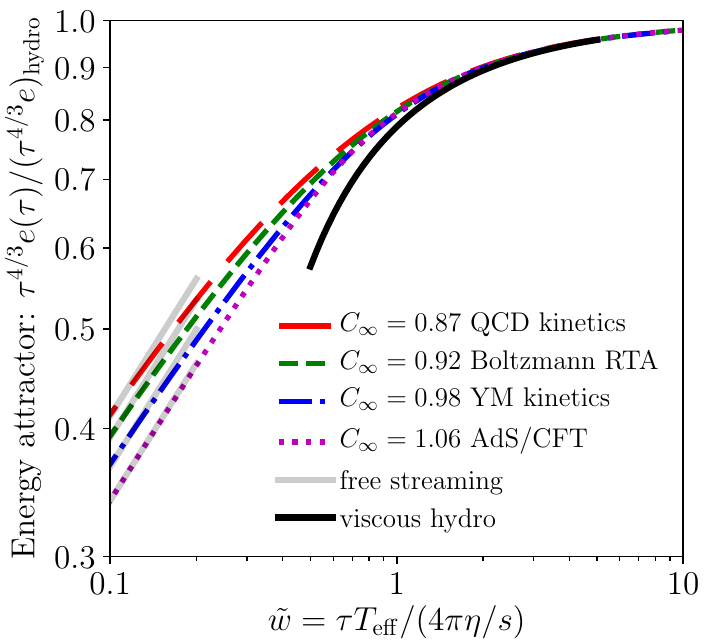}
  \caption{Hydrodynamic attractors for pre-equilibrium evolution of energy
    density for different microscopic theories.
From \cite{Giacalone:2019ldn}.}
  \label{fig:-figures-figure1}
\end{figure}

\subsubsection{Entropy production and initial energy density}

At even earlier times $\wtilde \lesssim 1$, kinetic simulations with very different initial conditions
might  not have collapsed yet onto a single curve~\cite{Kurkela:2015qoa, Almaalol:2020rnu}.
Nevertheless, one may employ the hydrodynamic attractor curve, which is regular for $\wtilde \to  0$, for a macroscopic fluid dynamic description far from equilibrium~\cite{Romatschke:2017vte};
see also \Sec{sec.strongBjorken}. 
In kinetic theory at early times, such an attractor curve has
vanishingly small longitudinal pressure ($P_L\approx 0$) and constant energy density per rapidity  ($\e \tau =\text{const}$). Such
initial conditions are typical for kinetic evolution in the bottom-up
picture discussed in \Sec{subsec:bottomup}. 
Figure \ref{fig:-figures-figure1} shows the
energy density $\e$ normalized by the equilibrium evolution $(\e \tau^{4/3})_\text{hydro}/\tau^{4/3}$ for different hydrodynamic attractors obtained from QCD and YM kinetic theory~\cite{Kurkela:2018vqr,Kurkela:2018wud,Kurkela:2018xxd,Kurkela:2018oqw}, AdS/CFT~\cite{Heller:2011ju,Heller:2015dha,Romatschke:2017vte}, and Boltzmann
RTA~\cite{Heller:2016rtz,Strickland:2017kux,Blaizot:2017ucy,Strickland:2018ayk,Behtash:2019txb}.
All attractors approach the universal viscous hydrodynamic description at late times $\tilde{w}>1$, while at
early times they follow  $\e \sim \tau^{-1}$, corresponding to  ``free-streaming" behavior\footnote{The presence of scattering terms in \Eq{eq:Boltz} is crucial for the early time  anisotropy evolution, but not for the energy density. According to the equations of motion $\partial_\tau(\tau \e) = -P_L$, and $\tau \e \approx \text{const}$ as long as $P_L/\e \ll 1$. }, which can be expressed as
\begin{equation}
  \frac{\e \tau^{4/3}(\wtilde \ll 1)}{(\e \tau^{4/3})_\text{hydro}}  = C_\infty^{-1} \wtilde^{4/9}
.\end{equation}
Here the dimensionless constant $C_\infty$ quantifies the amount of work done.  

A directly observable consequence of the equilibration process is the particle multiplicity, which is a measure of the entropy produced in heavy-ion collisions~\cite{Muller:2011ra}.
For a given hydrodynamic attractor,  the final entropy for boost invariant expansion
is proportional to the initial energy and is given by 
the following simple formula~\cite{Giacalone:2019ldn}
\begin{align}
\label{eq:Entropy}
(s\tau)_{\rm hydro} &= \frac{4}{3} C_\infty^{3 /4} \left(4\pi \frac{\upeta}{s}\right)^{1/3} \kappa^{1/3} \left(\e\tau\right)^{2/3}_{0},
\end{align}
\Ref{Giacalone:2019ldn} showed that combining the
entropy production from hydrodynamic attractors with initial state energy deposition in the CGC framework gives a good description of the centrality dependence of measured particle multiplicities. 
In particular, one can extend the original Bjorken estimate~\cite{Bjorken:1982qr} of the initial energy density in heavy-ion collisions to much earlier times. For central Pb-Pb collisions at $\sqrt{s_\text{NN}}=2.76\,\text{TeV}$ one finds that $\e(\tau_0)=270\,\text{GeV/fm}^3$ at $\tau_0=0.1\,\text{fm}/c$, which is a nearly 1000 times larger energy density than at the QCD crossover temperature.

\subsubsection{Chemical equilibration of QGP\label{sec:ektchemical}}

The early quark production from strong gauge fields was discussed in \Sec{sec:Diracfermions}.
However, once the gluon fields are no longer overoccupied, chemical equilibration has to be described using
QCD effective kinetic theory. A study of  light quark flavor ($up$, $down$ and $strange$) chemical equilibration in isotropic and longitudinally expanding systems were
recently presented in \cite{Kurkela:2018oqw, Kurkela:2018xxd}.
At leading order, there are two fermion production
channels: gluon fusion $gg\to q \bar{q}$ and splitting $g\to q \bar{q}$.
It was found that quark production processes are slower than gluon self-interactions. Therefore the gluon self-similar energy cascade seen in non-expanding isotropic systems
is over well before an appreciable number of fermions is produced. 
Similarly, gluons maintain an approximate kinetic equilibrium among themselves, while
fermions attain a Fermi-Dirac distribution at much later times.

\begin{figure}
\centering
\includegraphics[width=0.8\linewidth]{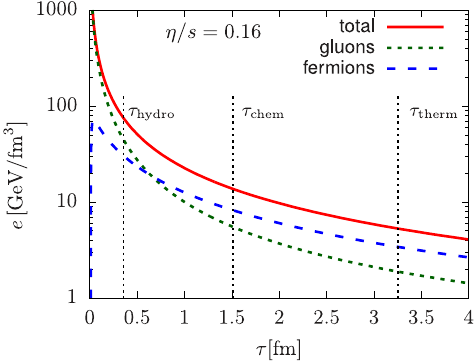}
\caption{Energy density evolution in a chemically equilibrating quark-gluon plasma. The vertical lines indicate the times of approximate hydrodynamic, chemical and thermal equilibriums. From Ref.~\cite{Kurkela:2018xxd}.}
\label{fig:ploteprlv2}
\end{figure}

The longitudinal expansion drives both gluons and fermions from the kinetic equilibrium, ensuring
that equilibrium distributions can  be approached only at late times when the expansion rate slows down. However, the expansion does not seem to affect fermion production; therefore,  chemical equilibrium is achieved \emph{before} thermal equilibrium.
For massless quarks, the quark-gluon plasma satisfies the conformal equation of
state
$\mathcal P= \frac{1}{3} \e$ and the
chemical composition of the plasma has little effect on the total evolution
of the energy-momentum tensor. 
Therefore, hydrodynamization and chemical and
thermal equilibrium are achieved sequentially~\cite{Kurkela:2018oqw, Kurkela:2018xxd}, satisfying 
\begin{equation}
  \tau_\text{hydro}< \tau_\text{chem} < \tau_\text{therm}\,.\end{equation}

Figure \ref{fig:ploteprlv2}
shows the total energy density (red solid line), gluon energy density (green dotted line)
and quark energy density (blue dashed line) as a function of time.
 Gluons, which dominate initially, are quickly overtaken by quarks and 
the approximate chemical equilibrium energy ratios are reached by $\tau = 1.5\, \text{fm}/c$. This supports an assumption of  chemical equilibrium in the lattice equation of state  used in hydrodynamic simulations of the quark-gluon plasma.

Finally, an important piece of evidence for the formation of a chemically equilibrated QGP in heavy-ion collisions is the 
enhanced production of hadrons carrying strange quarks~\cite{Andronic:2017pug}. It is believed that in small collision systems such as proton-proton collisions, strange quarks are not produced thermally in sufficient numbers and therefore  that strange
hadron production is suppressed. Although in the previous kinetic description the three light flavors are all taken to be massless, the chemical equilibration rate can be used to estimate the necessary life time (and system size) for the creation of the chemically equilibrated QGP. The results in Ref.~\cite{Kurkela:2018xxd} showed 
that the plasma may reach chemical equilibrium for particle multiplicities down to $dN_\text{ch}/d\eta \sim 10^2$.
Strange hadron production in such high multiplicity proton-proton collisions will be tested in future runs of the LHC~\cite{Citron:2018lsq}.

\subsubsection{Equilibration of spatially inhomogeneous systems \label{sec.kineqinh}}

Thus far we have discussed the equilibration of longitudinally expanding but otherwise
homogeneous systems. Realistic heavy-ion collisions create initial conditions that are not homogeneous in
the transverse plane. Such geometric deformations are strongly believed to be the source of the 
multi-particle correlations that have been observed experimentally~\cite{Heinz:2013th}.
In the weak coupling picture discussed in \Sec{sec:Glasma}, the spatial fluctuations are the result of the uneven color
charge distributions in the colliding nuclei. On the largest scale ($\sim 10\,\text{fm}$) it is determined by
the overlap of the average nuclear profiles. On nucleon scales $\sim 1\,\text{fm}$  one can resolve event-by-event
fluctuations of individual colliding nucleons. On yet smaller scales $\sim 0.1 \text{fm}$ 
one has stochastic fluctuations of color charges in the  internal structure of a nucleon.

Equilibration in kinetic theory, 
of small transverse perturbations around the homogeneous far-from-equilibrium background, has been investigated in several works \cite{Keegan:2016cpi, Kurkela:2018vqr, Kurkela:2018wud}. Relevant information on the complicated kinetic evolution of the particle distribution $f^s_\p$ can be captured by the linearized energy-momentum tensor response functions $G^{\mu\nu}_{\alpha \beta}$
\begin{align}
	\delta T^{\mu\nu}_\x(\tauhydro,\x) &=\int 
	d^2\xt'~G^{\mu\nu}_{\alpha 
		\beta}\left(\xt,\xt',\tauhydro,\tauekt\right)\nonumber\\
&\times\delta 
	T_\x^{\alpha\beta}(\tauekt,\xt') \frac{\TBg^{\tau \tau}_\x(\tauhydro)}{\TBg^{\tau\tau}_\x(\tauekt)}\label{eq:pert_evol}.
\end{align}
Here the Green's functions $G^{\mu\nu}_{\alpha
\beta}\left(\xt,\xt',\tauekt,\tauhydro\right)$ describe the evolution and
equilibration of energy-momentum tensor perturbations from an early time $\tauekt$ to a later time
$\tauhydro$.

Remarkably, the linearized response functions are to a good approximation  universal functions
of the dimensionless time $\wtilde$, which is similar to the hydrodynamic attractor describing the background equilibration. 
This provides a 
practical tool, the linearized pre-equilibrium propagator \kompost{}, for a  pre-equilibrium kinetic description of heavy-ion collisions based on QCD kinetic theory-\cite{Kurkela:2018vqr, Kurkela:2018wud}.
For the first time, the combination of the initial state IP-Glasma model discussed in \Sec{sec:ipglasma}, 
kinetic equilibration and viscous hydrodynamics evolution make it possible to
describe all the early stages of heavy-ion collisions in a theoretically complete setup. Experimental signatures of such setups are currently being investigated~\cite{Schenke:2020uqq, Gale:2020xlg}.

As with the evolution of the background, the equilibration of linearized perturbations in QCD kinetic theory
shares universal features with other microscopic descriptions~\cite{Broniowski:2008qk,Liu:2015nwa,vanderSchee:2013pia,Romatschke:2015gxa}. Thanks to this universal behavior, 
 ``universal pre-flow"
is guaranteed to grow linearly with time for small gradients  $\nabla \e/\e \ll1$~\cite{Vredevoogd:2008id, Keegan:2016cpi, Kurkela:2018vqr}:
\begin{equation}
  \vec{v}\approx -\frac{1}{2} \frac{\vec \nabla \e}{\e+\mathcal P_T}\tau
,\end{equation}
where for long wavelength perturbations ${\vec \nabla \e}/{(\e+\mathcal P_T)}=\text{const}$ in conformal theories~\cite{Keegan:2016cpi}. These  response functions have been  directly compared in Yang-Mills and RTA kinetic theories~\cite{Kamata:2020mka}.

QCD kinetic theory simulations beyond the linearized regime have not yet been
accomplished, albeit there exist phenomenological studies of  parton transport  simulations based on perturbative QCD matrix elements~\cite{Greif:2017bnr}.
To what extent the macroscopic description in terms of hydrodynamics can be applied to inhomogeneous systems with non-linear transverse expansion is still an open question; see \Sec{sec.strongtransverse} for a discussion of holography. 
However, encouragingly, the results of several  works~\cite{Kurkela:2019set,Kurkela:2019kip,Kurkela:2018ygx, Kurkela:2018qeb} 
have demonstrated that for transversely expanding systems the hydrodynamic attractor remains 
a good description of local equilibration until the evolution time becomes comparable to the transverse system size.

\section{Ab initio holographic description of strong coupling phenomena\label{sec:strongcoupling}}

\subsection{Holography and heavy-ion collisions \label{sec.strongbeginning}}

Sections \ref{sec:hadrons}-\ref{sec:kinetictheory} were concerned with the description of heavy-ion collisions in a weak coupling QCD framework. Here we will present what currently constitutes the only approach capable of describing real time phenomena in genuinely strongly coupled (1+3)-dimensional quantum field theories in a fully \emph{ab initio} manner: holography~\cite{Maldacena:1997re,Gubser:1998bc,Witten:1998qj}. 

The available description in this case does not make visible use of the gauge field degrees of freedom. Instead, it is based on the notion of a correspondence to higher dimensional geometries, which arise as solutions of Einstein's equations with a negative cosmological constant and appropriate matter fields.

The guiding principle for our presentation will be universality. We will be interested in phenomena shared across strongly-coupled quantum field theories and seek in them theoretical lessons and phenomenological implications for thermalization in QCD.

A prime example of such a quantity is the aforementioned $\upeta/s=1/(4 \, \pi)$ in \emph{all} holographic QFTs, as long as they are described by two-derivative gravity theories. One purpose of this review is to examine other kinds of universalities that exist in the genuine non-equilibrium regime. 

\subsection{Controlled strong coupling regime \label{sec.holooverview}}

The best-known holographic gauge theory is the ${\cal N} = 4$ super Yang-Mills theory. At the Lagrangian level, it can be viewed as the gluon sector of $SU(N_{c})$ QCD coupled in a maximally supersymmetric way to four Weyl fermions and six real scalars, both in the adjoint representation of the gauge group~\cite{Ammon:2015wua}. This theory, as opposed to QCD, is conformally invariant; the coupling constant does not run with the energy and becomes an external parameter that defines the theory.

In the planar $N_{c} \rightarrow \infty$ limit for asymptotically large values of the ’t Hooft coupling constant
\begin{equation}
    \lambda \equiv 4 \pi \alpha_{S} N_{c} \rightarrow \infty\,,
\end{equation} 
the degrees of freedom in the ${\cal N} = 4$ super Yang-Mills theory reorganize themselves in such a way that correlation functions of certain operators, including the energy-momentum tensor in an entire class of interesting states, can be computed using a 5-dimensional Einstein gravity action with a negative cosmological constant
\begin{equation}
\label{eq.Sgrav}
S_\text{grav} = \frac{1}{16 \pi G_{N}} \int d^5 x \sqrt{\mathrm{det} g} \left\{ R - 2 \left( - \frac{6}{L^2}\right) \right\}
\end{equation}
and supplemented by matter fields. In \Eq{eq.Sgrav} $R$ is the Ricci scalar and $L$ is the length scale set by the cosmological constant. For the ${\cal N} = 4$ super Yang-Mills theory at $\lambda \rightarrow \infty$ one has
\begin{equation}
\frac{L^3}{G_{N}} = \frac{2 \, N_{c}^2}{\pi}
\end{equation}
and a particular matter sector. They both follow from relevant string theory considerations~\cite{Maldacena:1997re}.

One should view the Einstein gravity description as applicable only when $\lambda \rightarrow \infty$. The QFT coupling constant does not appear in any form in \Eq{eq.Sgrav}, indicating that the coupling constant dependence drops from all the QFT quantities that one can describe in this way for $\lambda \rightarrow \infty$. When the coupling constant is large but not infinite, the relevant description becomes Einstein gravity supplemented by higher-curvature terms like the fourth power of the curvature. The form of these terms follows again from string theory considerations, and in controllable situations they should  necessarily be treated as small corrections. Because equations of motion become generically higher order in derivatives, the uncontrollable extrapolation of the kind that one performs in kinetic theory can be done here in only a very limited number of cases~\cite{Woodard:2015zca}. We will discuss these topics in~Sec.~\ref{sec.weakcouplingholo}. 

The ``vanilla'' setting in holography is five-dimensional gravity with a negative cosmological constant, encapsulated by \Eq{eq.Sgrav}, which provides a consistent dual holographic description of an infinite class of strongly coupled conformal field theories (CFTs) with a  large number of microscopic constituents~\cite{Bhattacharyya:2008mz}. Specifically, it describes a class of states in strongly coupled CFTs in which the only local operator acquiring an expectation value is the energy-momentum tensor~$T^{\mu \nu}$. The most comprehensive holographic results on heavy-ion collisions concern this case.

A generic five-dimensional metric can always be brought to the form
\begin{equation}
\label{eq.bulkmetric}
ds^2 = \frac{L^{2}}{u^2}  \left[- du^2 + g_{\mu \nu}(u,x) \, dx^{\mu}dx^{\nu} \right].
\end{equation}
 Here $u$ is an additional direction emerging on the gravity side interpreted as a scale in a dual QFT. Einstein's equations put conditions on acceptable forms of $g_{\mu \nu}(u,x)$. The most symmetric solution for gravity with a negative cosmological constant has $g_{\mu \nu}(u,x) = \eta_{\mu \nu}$, which is the four-dimensional Minkowski metric. This is the empty AdS$_{5}$ (anti-de Sitter) solution, which represents in gravitational language the time development of the vacuum in holographic CFTs. The surface $u = 0$ acts as a boundary of AdS$_{5}$ and, more generally, $g_{\mu \nu} (u = 0,x)$ has the interpretation of a metric in which the corresponding QFT lives. 
 
 The expectation value of the energy-momentum tensor arises by looking 
 at the subleading behavior of $g_{\mu \nu}(u,x)$ close to the boundary~\cite{Balasubramanian:1999re,deHaro:2000vlm}. This is particularly simple for CFTs living in Minkowski space:
\begin{equation}
\label{eq.NBmetric}
g_{\mu \nu}(u,x) = \eta_{\mu \nu} + \frac{4 \pi \, G_{N}}{L^3} \langle T_{\mu \nu} \rangle (x) \, u^{4} + \ldots
\end{equation}
The ellipsis denotes higher order terms in the small-$u$ expansion that turn out to contain only even powers of $u$ with the coefficients being polynomials in $\langle T^{\mu \nu} \rangle$ and its derivatives. One cannot a priori exclude terms like $\exp{(-1/u)}$ that were considered in  Ref.~\cite{Heller:2013oxa}, but a general understanding of such terms is lacking. In the following, we will refer to the interior of AdS spacetimes as ``bulk physics" and the QFT physics as ``boundary physics."

 We are interested here in discussing time dependent states in Minkowski spacetime that model the dynamics of heavy-ion collisions. Given \Eq{eq.NBmetric}, such states can be probed through their expectation value of the energy-momentum tensor by solving the equations of motion of \Eq{eq.Sgrav} as an initial value problem. This is achieved using numerical relativity techniques~\cite{Heller:2012je,Chesler:2013lia,Liu:2018crr} and requires one to specify initial conditions, and the solutions are subject to boundary conditions at~$u = 0$. 

\begin{figure}
\includegraphics[width=0.7\linewidth]{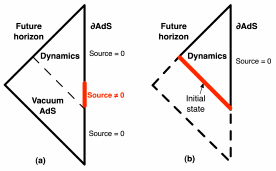}
\caption{Penrose diagrams dual to far-from-equilibrium states in strongly coupled QFTs. 
(a) The system starts in the vacuum with known bulk geometry and is perturbed by a non-trivial source, which appears as an asymptotic boundary condition in gravity. After the source is turned off, the QFT is in a non-equilibrium state modeled by a time dependent geometry. (b) The sources are always off, but one instead specifies non-trivial initial conditions for the bulk metric. Adapted from Ref.~\cite{Heller:2012km}.}
\label{fig:plotinivalueinAdSCFT}
\end{figure}

There are two natural ways (with pros and cons) of studying the  non-equilibrium physics of quantum field theories using holography: see~\Fig{fig:plotinivalueinAdSCFT}. 
The first approach circumvents the problem of finding initial conditions, a key reason for its use in early works on the subject~\cite{Chesler:2008hg,Chesler:2009cy}. Moreover, this approach allows one to compare equilibration across theories by starting with the same kind of initial state (such as the vacuum or a thermal state) and perturbing  it in a defined manner. In particular, it underlies a significant body of research on understanding features of linear response theory in different microscopic models~\cite{Kovtun:2005ev,Romatschke:2015gic,Grozdanov:2016vgg,Kurkela:2017xis}. As an example, Ref.~\cite{Keegan:2015avk}  discussed in \Sec{sec:extrastrong} (see \Fig{fig:fig5_vanilla}) compared the approach to hydrodynamics across models (including holography) using fully non-linear kicks. The drawbacks to perturbing simple states are, first, that the approach to hydrodynamics is so rapid that it is difficult to disentangle exciting the system from its subsequent relaxation and, second, that the class of states that one obtains in this way is rather limited.

The second method, in which one solves gravity equations for different initial conditions, allows one to access a larger range of transient behavior. In particular, since we do not know which initial conditions are closest to the physics realized in experiment, one may want to scan as many of these initial conditions as possible to obtain a comprehensive picture. The downside is that in most cases this way of phrasing the problem is very specific to the geometric language of describing strongly coupled QFTs similarly to the one-particle distribution function being very specific to the weak coupling language.  It does not allow for controllable comparisons with other frameworks akin to Ref.~\cite{Keegan:2015avk}. This can be somewhat ameliorated in holographic collisions in which the initial conditions for gravity originate from superimposing two exact solutions corresponding to individual projectiles approaching each other.

Thermalization at strong coupling is a process in which one starts with an excited geometry in the bulk that after some time becomes locally very close to a black hole geometry. This encapsulates the notion of thermalization of expectation values of local operators. Non-local observables discussed in Sec.~\ref{sec.holononloc} can still show traces of non-equilibrium behavior after local thermalization occurs. This should not come as a surprise since the thermalization of non-local observables is necessarily constrained by causality.

The discussion thus far was quite generic but the explicit formulas were provided for strongly coupled CFTs. While QCD is not a CFT, holography does not pose any conceptual problems in studying strongly coupled gauge theories with a non-trivial RG, provided that the theory remains strongly coupled at all scales. This can be realized by introducing relevant deformations to holographic CFTs, modifying their Lagrangian by $\int d^{4} x \, J \, O(x)$ with the scaling dimension $\Delta < 4$ of $O(x)$. This triggers a non-trivial bulk metric dependence on $u$ providing the gravitational counterpart of a RG flow. 

In holography, the bulk object corresponding to $O$ is a scalar field $\phi$ appearing in the matter sector that supplements the universal sector in \Eq{eq.Sgrav}. This scalar field is non-zero because the $J$ of the QFT translates into its asymptotic boundary conditions;  the latter generates a non-trivial profile for $\phi$ when solving the bulk equations of motion. Of course, the action for the bulk matter fields equips $\phi$ with a potential and the form of the potential determines the physics of the RG flow in the corresponding QFT (including the information about~$\Delta$). We will review representative results in \Sec{sec.strongnonconf}.

To close, holography provides an \emph{ab initio} window to study strongly-coupled QFTs, which include conformal and non-conformal gauge theories. The conceptual problem of fully non-perturbative real time evolution of an entire class of QFTs reduces in this setting to a technical challenge of solving a set of coupled partial differential equations in higher number of dimensions, which is well within reach of the existing numerical relativity methods. 

The holographic approach is very general and can be equally well applied to the problem of time evolution of the nuclear medium in heavy-ion collisions, as well as to problems originating in branches of physics~\cite{Ammon:2015wua,Hartnoll:2016apf}. Finally, we stress again that holography as a tool for QFT comes with its own limitations illustrated by the fact that one needs to work in regimes where the gravity description is classical or semi-classical.

\subsection{Early times in Bjorken flow at strong coupling \label{sec.strongBjorken}}

Bjorken flow~\cite{Bjorken:1982qr} without transverse expansion in a CFT setting is arguably the best studied example of a nonlinear non-equilibrium phenomenon in holography.\footnote{Recently devised hyperbolic quenches~\cite{Mitra:2018xob} adopt an effectively (1+1)-dimensional boost invariant geometry of heavy-ion collisions in the context of condensed matter physics.} Because of the conservation of the energy-momentum tensor, all the non-trivial information about the dynamics can be extracted from $\langle T^{\tau \tau} \rangle \equiv  {\cal E}(\tau)$. This parametrization is useful for describing the early time physics relevant for modeling initial stages of ultra-relativistic heavy-ion collisions. 

\begin{figure}
\includegraphics[width=0.75\linewidth]{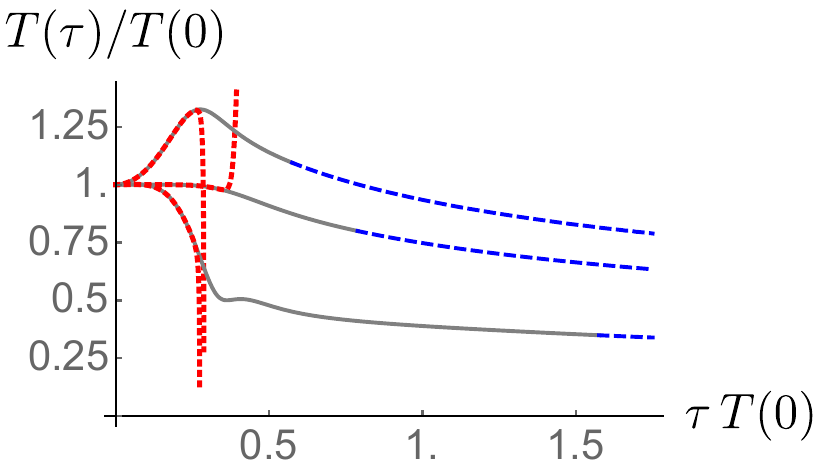}
\caption{Evolution of the effective temperature as a function of time for three different states with nonzero initial energy density. The gray curves denote the far-from-equilibrium regime. The blue dashed area extending indefinitely to the right mark the applicability of viscous hydrodynamic relations truncated at the third order in derivatives~ \eqref{eq.Ahydroholo}. The red dotted curves denote the series in Eq.~\eqref{eq.epsilonETbif} extracted using the method of Ref.~\cite{Beuf:2009cx}. Adapted from Ref.~\cite{Heller:2012je}.}
\label{fig:bifETpossibilities}
\end{figure}

Toward this end, Ref.~\cite{Beuf:2009cx} noticed that combining \Eq{eq.NBmetric} (expanded to sufficiently high order in $u$) with a general Taylor series ansatz for ${\cal E}(\tau)$ around $\tau = 0$ does not lead to singular bulk metric coefficients in the limit $\tau \rightarrow 0$ as long at the early time expansion contains only positive even powers of proper time:
\begin{equation}
\label{eq.epsilonETbif}
{\cal E}(\tau \approx 0) = {\cal E}_{0} + {\cal E}_{2} \tau^2 + {\cal E}_{4} \tau^{4} + \ldots \, .
\end{equation}
The coefficients in \Eq{eq.epsilonETbif} are not entirely arbitrary, but they are related one-to-one to the near-boundary expansion of the bulk metric that satisfies the constraints on the initial time slice, as encapsulated by \Eqs{eq.bulkmetric} and \eq{eq.NBmetric}. The early time series \eq{eq.epsilonETbif} turns out to have a non-zero but finite radius of convergence, which allows one to reliably study the initial dynamics of the system. However, as shown in Ref.~\cite{Beuf:2009cx}, and as later corroborated in Ref.~\cite{Heller:2012je} using the full numerical solution of bulk Einstein's equations, the radius of convergence of \Eq{eq.epsilonETbif} is much too small for us to see the transition to hydrodynamics. This point is illustrated in \Fig{fig:bifETpossibilities} using the effective temperature (see footnote \text{\ref{foot.effT}}). Furthermore, simple analytic continuations of the series~\eqref{eq.epsilonETbif} based on the Pad{\'e} approximants method provide unreliable extrapolations.

One lesson therefore is that the only method for obtaining $\langle T^{\mu \nu} \rangle$ in strongly coupled QFTs beyond the early time limit examples is to use numerical relativity. Before we proceed in that maner, a few more comments related to \Eq{eq.epsilonETbif} are in order. First, the analysis of Ref.~\cite{Beuf:2009cx} uses regularity of the initial metric on a particular constant time slice of the bulk geometry, namely, the one dictated by the coordinates chosen in \Eq{eq.bulkmetric}. It is therefore logically possible\footnote{Ref.~\cite{Jankowski:2014lna} chose initial surfaces in the bulk as in \Fig{fig:plotinivalueinAdSCFT}, with results being consistent with \Eq{eq.epsilonETbif}.} that there are initial metrics defined on other bulk constant time slices that give rise to energies densities of the form other than those dictated by \Eq{eq.epsilonETbif}. Second, note that in \Eq{eq.epsilonETbif} any number of the lowest order terms can vanish and the energy density at early time can behave like ${\cal E}{\big|}_{\tau \approx 0} \sim \tau^2$~\cite{Grumiller:2008va}. 

Another point is that there are various reasons why one may not want to start the evolution at $\tau = 0$. The most obvious one is related to creating either non-equilibrium initial states from the vacuum or thermal states, as discussed in \Fig{fig:plotinivalueinAdSCFT}. In these cases, the sources will need some non-zero time to act~\cite{Chesler:2009cy}. The other reason is more conceptual and is related to the observation that while one should not expect the infinitely strongly coupled approach to be a  phenomenologically viable description at $\tau = 0$, it may become one from some $\tau > 0$ onward. Note that from the gravity point of view, it is not clear that all the initial conditions set in the bulk for $\tau > 0$ are extendable to $\tau = 0$ and, as a result, one can view them as \emph{a priori} containing richer behavior. 

Because of this issue, it is unclear whether all well behaved initial conditions for numerical relativity simulations actually describe genuine states in underlying QFTs. Unlike Refs.~\cite{Heller:2011ju,Heller:2012je,Jankowski:2014lna},  Refs.~\cite{Wu:2011yd,Romatschke:2017vte,Kurkela:2019set} initialized their codes at later times with turned off sources. In particular, Ref.~\cite{Romatschke:2017vte} found initial conditions at some early but non-zero $\tau$ such that ${\cal E} \sim \frac{1}{\tau}$ initially, which is clearly very different from \Eq{eq.epsilonETbif}.

As discussed in Sec.~\ref{sec:extrastrong}, the transition to hydrodynamics can be observed in the cleanest way upon introducing the scale-invariant time variable~$\wtilde$ defined in \Eq{eq:wtilde} and using ${\cal P}_{T}/{\cal P}_{L}$, ${\cal P}_{L} / {\cal E}$ or any reasonable function of this ratio such as 
\begin{equation}
\mathscr{A} = \frac{{\cal P}_{T} - {\cal P}_{L}}{{\cal E}/3}= \frac{3 \frac{\mathcal P_{T}}{\mathcal P_{L}} - 3}{2 \frac{\mathcal P_{T}}{\mathcal P_{L}} + 1},
\end{equation} 
which was introduced in Refs.~\cite{Heller:2011ju,Jankowski:2014lna,Florkowski:2017olj} as a function of $w\equiv\tau T$. Note that in the strongly coupled limit of holography $4\pi \, \upeta/s=1$, and we will simply denote $\wtilde$ as~$w$.

It is well understood by now that at late time $\mathscr{A}(w)$ acquires the form of a trans-series~\cite{Heller:2013fn,Heller:2015dha,Aniceto:2015mto,Aniceto:2018uik} known from the studies of asymptotic expansions in mathematical and quantum physics: see Refs.~\cite{Dorigoni:2014hea,Aniceto:2018bis} for reviews. The hydrodynamic part is a series in inverse powers of $w$ and has a vanishing radius of convergence\footnote{The same applies to Gubser~\cite{Denicol:2018pak} and cosmological~\cite{Buchel:2016cbj} flows but is not the case for Bjorken flow with fine-tuned transport coefficients~\cite{Denicol:2019lio}. Furthermore, Ref.~\cite{linearhydrodiv} used the results of~\cite{Withers:2018srf,Grozdanov:2019kge,Grozdanov:2019uhi} to show that divergence of the hydrodynamic gradient expansion is a generic feature of linear response theory.}. Its first few terms read
\begin{eqnarray}
\label{eq.Ahydroholo}
&&\mathscr{A}(w) = \frac{2}{\pi} w^{-1} + \frac{2 - 2 \log{2}}{3 \pi^2} w^{-2} \nonumber\\ &&+ 
\frac{15-2\pi^2 - 45 \log{2} + 24 \log^2{2}}{54 \pi^3} w^{-3} + \ldots , \quad
\end{eqnarray}
see Refs.~\cite{Nakamura:2006ih,Janik:2006ft,Heller:2007qt,Heller:2008mb,Kinoshita:2008dq,Kinoshita:2009dx,Booth:2009ct,Heller:2011ju,Jankowski:2014lna,Florkowski:2017olj}. Equation \eq{eq.Ahydroholo} should be understood as expressing the energy-momentum tensor in terms of hydrodynamic constitutive relations to the third lowest order. The first term carries information about the first derivative of flow velocity and the shear viscosity, while the second term is a contribution from second derivatives of velocity and associated transport coefficients. The third term is the last one that is known analytically. The current state of the art was set by Ref.~\cite{Casalderrey-Solana:2017zyh} which, improving on the earlier efforts of Ref.~\cite{Heller:2013fn},  numerically computed the lowest 380 terms in the expansion given in \Eq{eq.Ahydroholo}. On top of the power law late time~($w$) expansion come exponentially suppressed terms that represent transient phenomena that are also visible in linear response theory~\cite{Janik:2006gp,Heller:2013fn,Heller:2015dha,Heller:2018qvh}.

\begin{figure}
\includegraphics[width=0.8\linewidth]{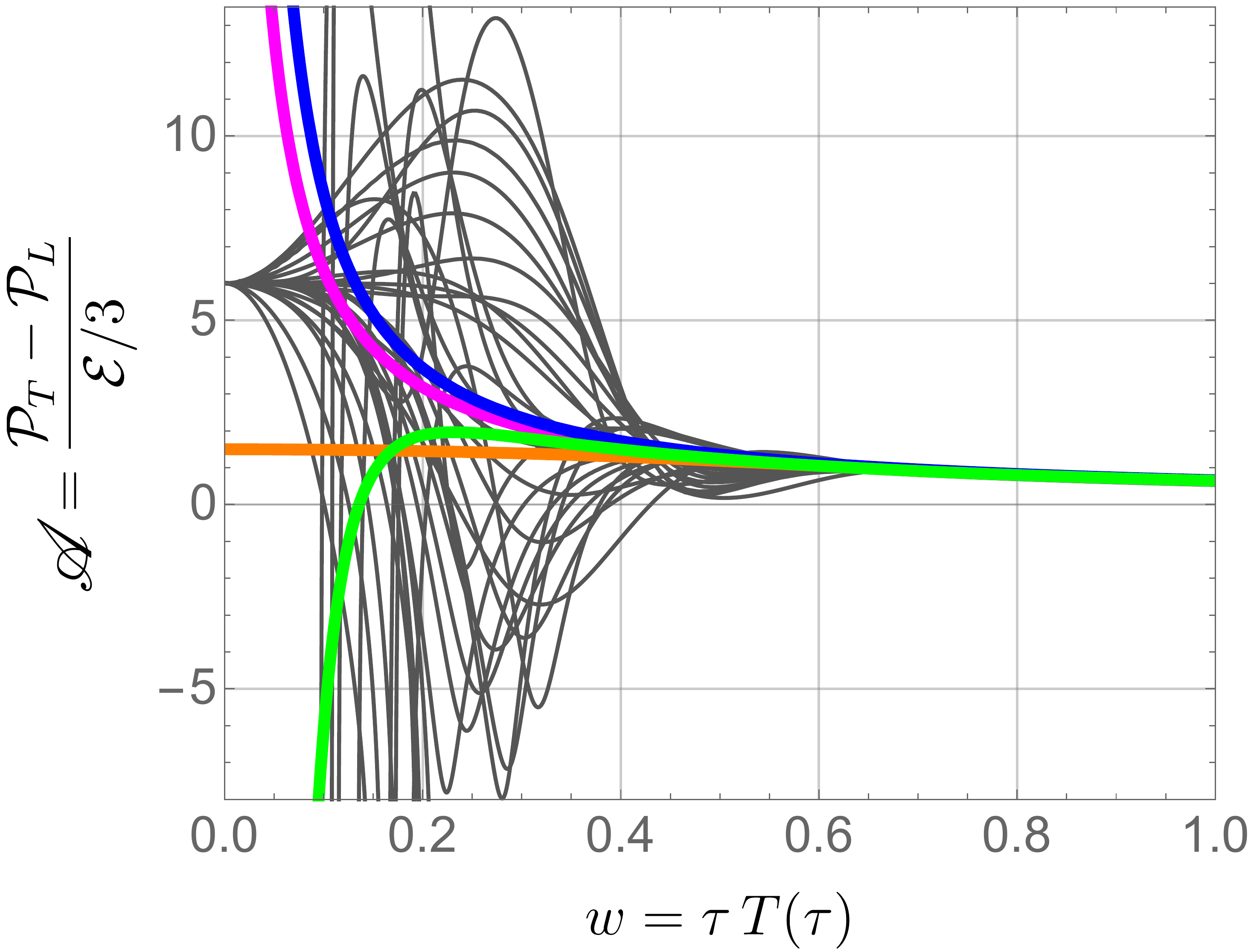}
\caption{$\langle T^{\mu \nu} \rangle$ in a holographic CFT as a function of the dimensionless clock variable $w$ for 29 different initial states (gray curves). Magenta, blue and green curves denote predictions of hydrodynamic constitutive relations truncated, respectively, at first, second, and third order [\Eq{eq.Ahydroholo}]. The orange curve is the hydrodynamic attractor~\cite{Romatschke:2017vte}. Adapted from Refs.~\cite{Heller:2011ju,Heller:2012je,Romatschke:2017vte}.}
\label{fig:bifAholo}
\end{figure}

 \Fig{fig:bifETpossibilities} illustrates time evolution of the effective temperature $T(\tau)$. Hydrodynamics is applicable at a time after which the pressure anisotropy deviates only slightly from \Eq{eq.Ahydroholo}. As discussed in detail in Ref.~\cite{Heller:2012je}, the precise moment of applicability of hydrodynamics depends on the desired accuracy of the match to \Eq{eq.Ahydroholo} and the order of the truncation. Of course, the latter aspect should be understood in the sense of an asymptotic series. 
 
 The main message from the studies in Refs.~\cite{Chesler:2009cy,Heller:2011ju,Wu:2011yd,Heller:2012je,Jankowski:2014lna,Romatschke:2017vte,Kurkela:2019set} and related works is that low order hydrodynamic constitutive relations (see \Eq{eq.Ahydroholo}) become applicable at strong coupling after $\tau = {\cal O}(1/T)$. This is the regime where the pressure anisotropy in the system is sizable, as illustrated in \Fig{fig:bifAholo}. Since the system is  still far away from local thermal equilibrium, the word hydrodynamization was coined in~\cite{CasalderreySolana:2011us} to distinguish the applicability of viscous hydrodynamics constitutive relations from local thermalization. The latter phenomenon occurs at strong coupling for times that can  even be 10 times larger than the  hydrodynamization time.

The modern perspective on hydrodynamics,  viewing in  particular the gradient expansion as a part of a trans-series, was reviewed in detail in Ref.~\cite{Florkowski:2017olj}. In the following, we will discuss an alternative way of thinking about the applicability of hydrodynamics using the concept of hydrodynamic attractors. These objects made their appearance in Sec.~\ref{sec:extrastrong} and bear a structural similarity to the non-thermal attractors (fixed points) discussed in Sec.~\ref{sec.nonthatt}.

\subsection{Hydrodynamic attractors in holography \label{sec.strongattractors}}

Hydrodynamic attractors proposed in Ref.~\cite{Heller:2015dha}, and developed in many works including Refs.~\cite{Basar:2015ava,Aniceto:2015mto,Romatschke:2017vte,Spalinski:2017mel,Strickland:2017kux,Romatschke:2017acs,Denicol:2017lxn,Florkowski:2017jnz,Behtash:2017wqg,Casalderrey-Solana:2017zyh,Blaizot:2017ucy,Almaalol:2018ynx,Denicol:2018pak,Behtash:2018moe,Spalinski:2018mqg,Strickland:2018ayk,Strickland:2019hff,Blaizot:2019scw,Jaiswal:2019cju,Kurkela:2019set,Denicol:2019lio,Brewer:2019oha,Behtash:2019qtk,Chattopadhyay:2019jqj,Dash:2020zqx,Shokri:2020cxa,Heller:2020anv,Almaalol:2020rnu} 
can be viewed as a way of approaching the problem of information loss about the underlying state from the point of view of observations restricted to the energy-momentum tensor~$\langle T^{\mu \nu} \rangle$. 

Reexamining \Fig{fig:bifAholo} through these lenses, we see that a set of different states considered there follows to a good approximation a single profile of $\mathscr{A}(w)$ from a certain value of $w$ onward. This is the notion of attraction between different initial conditions as seen by an \emph{effective} phase space covered by $\mathscr{A}$ at a fixed value of $w$. While this observation does not call for invoking a truncated gradient expansion, the emerging universality seen in \Fig{fig:bifAholo} agrees very well with a hydrodynamic gradient expansion truncated at low order. These observations lie behind the name hydrodynamic attractor and parallel the discussion in Sec.~\ref{sec.hydroattractorkin}.

We now step back and review this phenomenon from a broader perspective advocated recently in Ref.~\cite{Heller:2020anv}. To proceed, we will utilize the aforementioned notion of phase space introduced in this context in Ref.~\cite{Behtash:2017wqg}. Specifically, one should think of $\mathscr{A}$ as a particularly clean scale invariant way of representing information about $\langle T^{\mu \nu} \rangle$ and $w$ as a useful way of parametrizing time evolution, adjusted to the fact that transient phenomena in conformal theories occur over time scales set by the energy density.

\begin{figure}
\includegraphics[width=0.75\linewidth]{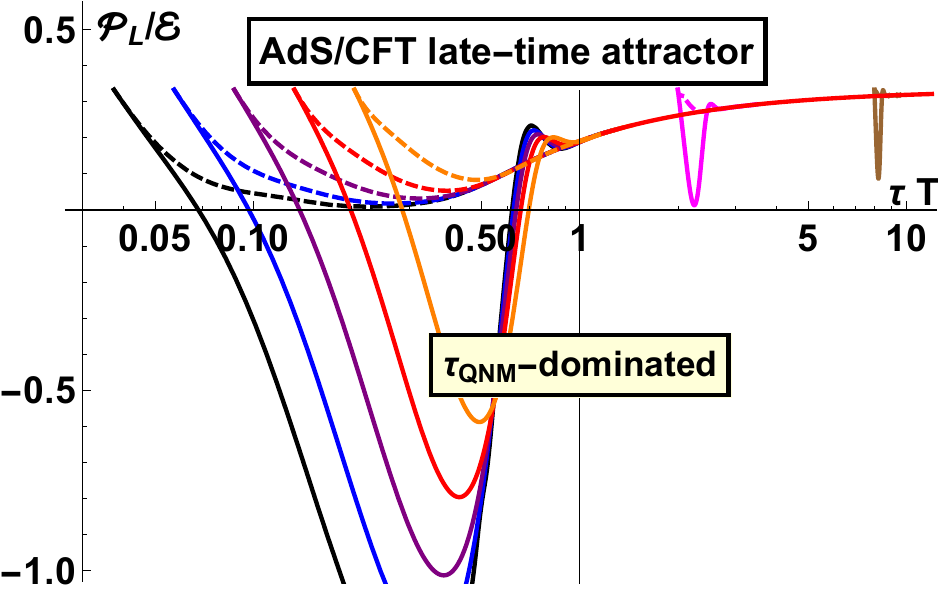}
\caption{Hydrodynamization of states whose gravity dual initially has support close to the boundary (dashed curves) or deep in the bulk (solid curves) initialized at different times (different colors); see the text for details. Adapted  from Ref.~\cite{Kurkela:2019set}.}
\label{fig.approachtoattractor}
\end{figure}

Of course, knowing $\mathscr{A}$ at a given value of $w$ does not allow one to predict its value later, since the true microscopic variable is the bulk metric. A larger chunk of information is provided by considering $\mathscr{A}$ and some of its derivatives with respect to $w$ (or $\cal E$ and its derivatives with respect to $\tau$). Such sets of variables form the notion of an effective phase space. In fact, there is a limit to how large such phase space needs to be: the numerical solutions of Einstein's equations displayed in \Fig{fig:bifAholo} typically require one to specify a few functions on several dozen grid points.

One can then assign a metric to an effective phase space, i.e., the  distance between points representing  classes of solutions here, and track how such a distance changes as time evolves. The loss of information is expected to make a set of solutions reduce its volume in the effective phase space. For example, in \Fig{fig:bifAholo} one introduces the notion of proximity between two solutions $|\mathscr{A}_{1}(w) - \mathscr{A}_{2}(w)|$. With respect to this notion, various solutions from the chosen set eventually  collapse to approximately a point in $\mathscr{A}$ at a fixed value of $w$. It should be clear that the hydrodynamic attractor at a given value of $w$ is not a notion relevant to all states. It needs to be regarded as a statement about properties of some class of states initialized prior to that.

Furthermore, assigning a distance measure to phase space allows one to define the notion of slow evolution. This topic was introduced in Ref.~\cite{Heller:2015dha} under the name slow roll approximation, which originates from the field of inflationary cosmology~\cite{Liddle:1994dx}. The previously discussed distance notion leads to the magnitude of velocity of a given state being $| \mathscr{A}'(w)|$, and slowly evolving solutions (note that Ref.~\cite{Heller:2020anv} instead defined regions of slow evolution) are those that lead to the flattest form of $\mathscr{A}(w)$. In \Fig{fig:bifAholo}, such a solution given in Ref.~\cite{Romatschke:2017vte} using fine tuning initial conditions is denoted by an orange curve. Note that this solution at early times has $\mathscr{A}$ very close to $\frac{3}{2}$. This corresponds to free streaming ${\cal P}_{L} = 0$, which evades the study of initial conditions behind \Eq{eq.epsilonETbif} reported in Ref.~\cite{Beuf:2009cx}.

We stress that the notion of slowly evolving solutions is a priori independent from the notion of convergence (attraction). However, in full phase space, or at least a representative projection of it, one can make a thermodynamic-like argument, as in Ref.~\cite{Heller:2020anv}, in favor of typical states residing in the slow roll region. One can think of slow evolution as a generalization of the notion of the gradient expansion that does not involve an expansion with individual terms badly behaving at very early times, namely, as inverse powers of $w$ in \Eq{eq.Ahydroholo}.

Finally, the approach to the hydrodynamic attractor at strong coupling and mechanisms that govern it were examined in Ref.~\cite{Kurkela:2019set} by looking at results of simulations with different initialization times. This is depicted in \Fig{fig.approachtoattractor}. The idea behind it, building on earlier results in Refs.~\cite{Blaizot:2017ucy,Blaizot:2019scw}, is that information loss can be driven by at least two distinct mechanisms. The first one involves exponentially suppressed corrections to \Eq{eq.Ahydroholo}, which stem from linear response theory physics. The characteristic feature of them is that their decay rates do not depend on $w$. The second mechanisms driving the information loss is expansion, which for the comoving velocity $u^{\mu} \partial_{\mu} \equiv \partial_{\tau}$ gives $\nabla_{\mu} u^{\mu} = \frac{1}{\tau}$. What one therefore expects is that information loss predominantly driven by the expansion is going to be faster at earlier times (smaller $w$) and slower at later times. Indeed, such a feature was seen in Ref.~\cite{Kurkela:2019set} for hydrodynamic models and for the kinetic theory for early initialization times. However, in holography this does not seem to be the case and the approach to the hydrodynamic attractor takes roughly a fixed amount of time regardless of the chosen initialization time (see \Fig{fig.approachtoattractor}), which is consistent with it being governed by transients.

\subsection{Holographic collisions \label{sec.holographiccollisions}}

In CFTs, Bjorken flow in the absence of transverse expansion has a high degree of symmetry that  allows for comprehensive studies of hydrodynamization and associated phenomena. In particular, the numerical approach pursued in Refs.~\cite{Chesler:2009cy,Chesler:2013lia,Jankowski:2014lna} fully determines the evolution of $\langle T^{\mu \nu} \rangle$ as a function of proper time~$\tau$ upon specifying one positive number (initial energy density $\cal E$) and a single function of the AdS direction~$u$, see \Eq{eq.NBmetric}. As a result, it was possible to comprehensively scan over initial states in search of universal behavior. 

If one relaxes these symmetry assumptions and allows for dynamics in the  transverse plane~\cite{vanderSchee:2012qj,vanderSchee:2013pia}, the space of initial conditions becomes too large to allow for a comprehensive analysis. Therefore, one wants to have another guiding principle to arrive at interesting configurations for modeling non-equilibrium evolution of $\langle T^{\mu \nu} \rangle$ in holographic heavy-ion collisions. The key idea is to study holographic collisions of localized lumps of matter~\cite{Chesler:2010bi,Grumiller:2008va,Casalderrey-Solana:2013aba,Casalderrey-Solana:2013sxa,Chesler:2015fpa,Chesler:2015wra,Chesler:2015bba,Chesler:2016ceu}.

The localized objects (shockwaves) in question move at the speed of light and are characterized by the following non-zero components of~$\langle T^{\mu \nu} \rangle$,
\begin{equation}
\label{eq.holographicshock}
\langle T^{00} \rangle = \langle T^{33} \rangle = \pm \langle T^{03} \rangle =  \mu_{\pm} (\mathbf{x}_{\perp}) h( x^{0} \mp x^3)\,,
\end{equation}
where $x^{0}$ is the lab-frame time, $x^3$ is the direction along which the object is moving (specified by $\mp$ in the argument of $h$), $\mu_{\pm} (\mathbf{x}_{\perp}) \geq 0$ is an arbitrary function specifying the transverse profile and $h( x^{0} \mp x^3) \geq 0$ is another arbitrary function specifying the longitudinal profile~\cite{Chesler:2015bba}. While a single projectile defined by \Eq{eq.holographicshock} is exact, the superposition of two projectiles approaching each other and overlapping in the transverse plane leads to a non-trivial collisional process. 

 Such collisions should not be regarded as literal models of the early stages of heavy-ion collisions, since the projectiles do not originate from QCD. 
(See, however, \cite{Gubser:2008pc,Lin:2009pn,vanderSchee:2015rta}.) Instead, one should treat holographic shockwave collisions as illustrating  possible far-from-equilibrium phenomena accessible in a fully \emph{ab initio} way at strong coupling that goes well beyond the previously discussed Bjorken flow geometry. 

\subsubsection{Planar shocks \label{sec.planarshocks}}

\begin{figure}
\includegraphics[width=0.9\linewidth]{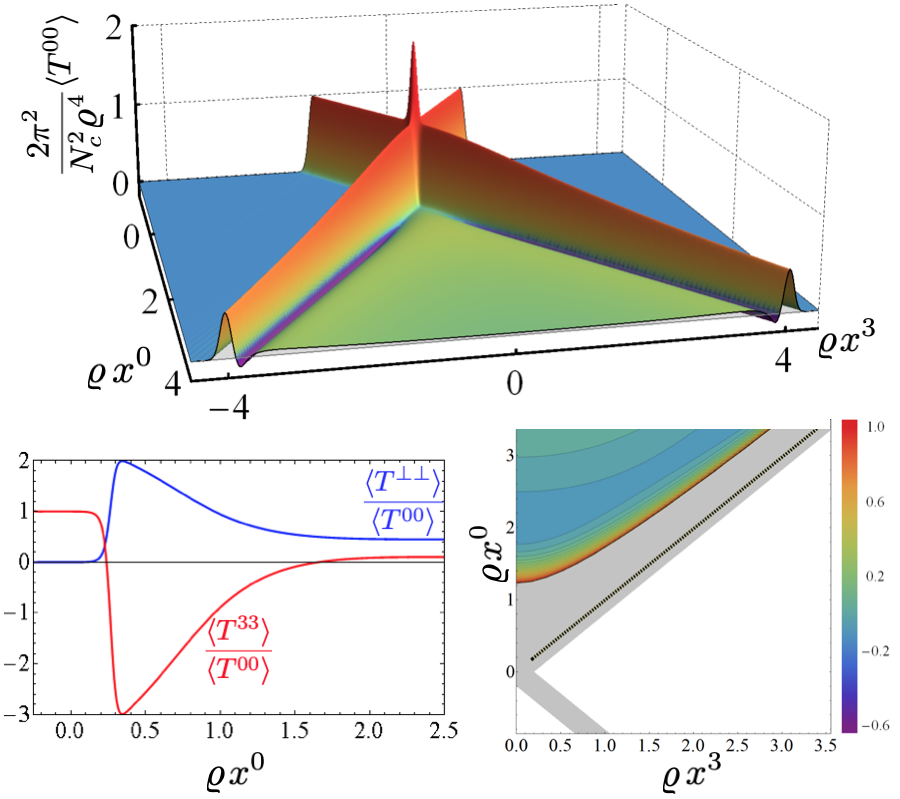}
\caption{$\langle T^{\mu \nu} \rangle$ resulting from a collision of thin planar shocks with $\varrho \, d = 0.08$~\eqref{eq.gaussianshocks}. Top panel: lab-frame energy density as a function of time $x^0$ and longitudinal position $x^3$. Between the remnants and the central rapidity region, there are small regions of negative energy density. Bottom-left panel: at mid-rapidity, the transverse and longitudinal pressure after the collision are consistent with~$\langle T^{0 0} \rangle \sim \tau^2$ in \Eq{eq.epsilonETbif}. Bottom-right panel: the color encoding denotes deviations from 
constitutive relations and points to the applicability of hydrodynamics. 
The post-collision~$\langle T^{\mu \nu} \rangle$ does not have a rest frame in the gray region~\cite{Arnold:2014jva}. Adapted from Ref.~\cite{Casalderrey-Solana:2013aba}.}
\label{fig.thinshocks}
\end{figure}

The simplest settings to consider are collisions of planar shockwaves: objects defined by \Eq{eq.holographicshock} with $\mu_{\pm}$ constant. Following Ref.~\cite{Casalderrey-Solana:2013aba}, one can consider a Gaussian longitudinal profile for $h$ of the form,
\begin{equation}
\label{eq.gaussianshocks}
h (x^{0} \mp x^3) = \frac{N_{c}^2}{2 \pi^2} \varrho^4 e^{-\frac{(x^0 \mp x^3)^2}{2 d^2}}\,,
\end{equation}
and recognize that, in heavy-ion collisions, the dimensionless product of the amplitude $\varrho$ (not to be confused with the previously discussed charge density) and the width~$d$ decreases as $\gamma^{-1/2}$ as the total center-of-mass energy of the collision ($\sqrt{s} = 2 \gamma M_{\mathrm{ion}}$) increases. 

Within this analogy, high energy collisions correspond to collisions of very thin shockwaves\footnote{The problem of colliding planar projectiles in \Eq{eq.holographicshock} with $h(x^0 \mp x^3) \sim \delta(x^0 \mp x^3)$ was originally posed in \cite{Janik:2005zt} and addressed in an early time expansion akin to \Eq{eq.epsilonETbif} in \cite{Grumiller:2008va}.}.  The collisions of projectiles defined by \Eq{eq.holographicshock} do not lead to longitudinal boost invariance since the initial state of the two projectiles is not boost invariant even when they are infinitely thin. 
The extent to which this is the case was explored in~\cite{Casalderrey-Solana:2013aba} and, quite remarkably, the results fit well~\cite{Gubser:2014qua} with complex deformations of the purely boost invariant flow introduced in~\cite{Gubser:2012gy}.

As it turns out, the features of the collision change as a function of~$\gamma$. First, the collision of ``low-$\gamma$'' thick shockwaves proceed such that the two blobs of matter first merge and their subsequent evolution is approximated well by viscous hydrodynamics. This is referred to \cite{Casalderrey-Solana:2013aba} as to the Landau scenario~\cite{Landau:1953gs,Belenkij:1956cd}. As  seen in \Fig{fig.thinshocks}, the ``high-$\gamma$'' regime of thin shocks leads to a rich set of transient physics before hydrodynamics becomes applicable. Another important phenomenon, discussed in \cite{Casalderrey-Solana:2013sxa,Waeber:2019nqd,Muller:2020ziz}, is the notion of longitudinal coherence. This notion applies to the  ``centre-of-mass'' frame of high energy collisions 
and states that the longitudinal structure of projectiles does not leave an imprint on the transient form of the energy-momentum tensor in the post-collision region provided that it is sufficiently localized. Finally, despite the differences between thin and thick shocks' collisions at transient times after the remnants dissolve, which take a much longer time than shown in \Fig{fig.thinshocks}, the structure of the late time hydrodynamic flow is very similar in the two cases~\cite{Chesler:2015fpa}.

\subsubsection{Transverse dynamics in holography \label{sec.strongtransverse}}

\begin{figure}
\includegraphics[width=0.8\linewidth]{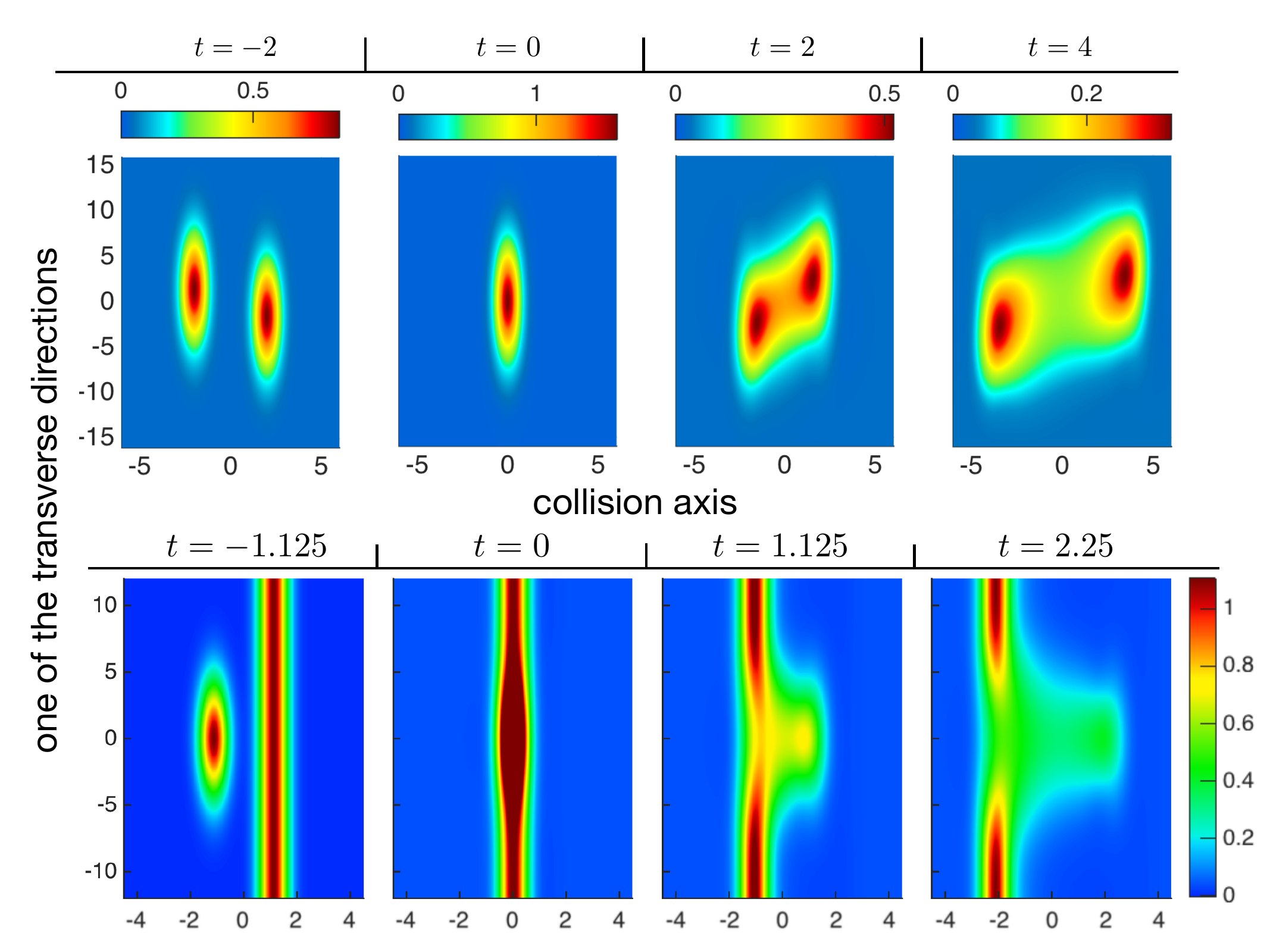}
\caption{Energy density in holographic heavy-ion collisions with transverse dynamics. Top panels: off-central collision with modest elliptic flow. Bottom panels: proton-nucleus collision as modeled with a shockwave with a small Gaussian extent in the transverse plane (left projectile) and a planar shock (right projectile). The smaller projectile punches out a hole in the larger projectile and excites matter at mid-rapidity, leading to substantial radial flow. Adapted from~\cite{Chesler:2015wra} and~\cite{Chesler:2015bba}.}
\label{fig.shockstransv}
\end{figure}

Studies of hydrodynamization in the presence of transverse expansion in~\cite{Chesler:2015wra,Chesler:2015bba,Chesler:2016ceu} still define the state-of-the-art in numerical applied holography.  Figure~\ref{fig.shockstransv} illustrates the profile of the energy density in such collisions. The main lesson from these works is the early applicability of viscous hydrodynamics not only for very large longitudinal gradients of the energy-momentum tensor (as for   Bjorken flow and planar shocks) but also in the presence of large transverse gradients generating transverse expansion. 

From the perspective of these strong coupling results, the applicability of hydrodynamics in pA and even pp collisions~\cite{Chesler:2016ceu} is as natural as the applicability of hydrodynamics in  Bjorken flow and can be explained in terms of fast decaying contributions to the trans-series for $\langle T^{\mu \nu} \rangle$. Further, these works corroborate studies in \cite{vanderSchee:2012qj} by providing successful tests of the early time radial expansion model proposed in \cite{Vredevoogd:2008id}. Toward this end, Ref.~\cite{Chesler:2015wra} found very small elliptic flow despite off-central collision and confirmed that near mid-rapidity the energy flux grew linearly with proper time, as predicted in Ref.~\cite{Vredevoogd:2008id}.

As discussed in \Sec{sec.kineqinh}, such ``universal flow" at small wavenumber is also reproduced by weak coupling kinetic theory. It would be interesting to see whether the full transverse response functions of the energy-momentum tensor in strong coupling agrees with those discussed in \Sec{sec.kineqinh} in the context of kinetic theory.

\subsection{Other aspects of thermalization at strong coupling \label{sec.strongother}}

\subsubsection{Non-conformal strongly-coupled QFTs \label{sec.strongnonconf}}

All the strong coupling results reviewed thus far have concerned well defined QFTs without a scale. As reviewed in \Sec{sec.holooverview}, in holography there are no conceptual obstacles to breaking  conformal symmetry. However, considering QFTs with non-trivial renormalization group flows does make gravitational calculations more involved due to the presence of  field(s) in addition to gravity that one needs to solve for and due to the more involved near-boundary analysis that generalizes \Eq{eq.NBmetric}. All in all, the number of results on this front relevant for thermalization in QCD is significantly lower than in the conformal case but still allows one to draw conclusions. 

Broadly speaking, there are two approaches to this problem. The first is top-down and studies renormalization group flows originating from turning on a relevant deformation in a known holographic CFT. The prime example is the so-called $\mathcal{N} = 2^{*}$ gauge theory arising as a deformation of $\mathcal{N} = 4$ super Yang-Mills theory by adding masses to half of its fields~\cite{Buchel:2007vy}. The advantage of this approach is that one makes sure that one is studying well defined features of a  strongly coupled QFT. The drawback is that such well understood examples are scarce and might have rigid features that do not exist in QCD. 

The other class are so-called bottom-up models that couple AdS gravity to a bulk scalar field or fields whose Lagrangian is chosen by insisting that it  reproduce some desired feature of QCD. One such approach was introduced in \cite{Gursoy:2007cb,Gursoy:2007er} using the QCD $\beta$-function as a guideline; another model \cite{Gubser:2008yx} uses as a benchmark reproducing the QCD equation of state at vanishing baryon density. 

Furthermore, one can also introduce confinement by making the geometry end smoothly in the bulk~\cite{Witten:1998zw}. One can think of it as the manifestation of a mass gap, with no excitations below the lowest bound state energy.

The breaking of conformal symmetry introduces an additional scale in the problem of thermalization and changes hydrodynamization times, although in none of the setups explored to date by an order of magnitude or more with respect to the strong coupling CFT prediction of~$\sim 1/T$~\cite{Buchel:2015saa,Janik:2015waa}. This also indicates that the $\tilde{w}$ defined in \Eq{eq:wtilde} plays a less prominent role in non-conformal QFTs than it does in strongly coupled CFTs. 

Furthermore, the hydrodynamic gradient expansion acquires new transport terms, most notably the bulk viscosity $\zeta$. Hydrodynamization and on a much later time scale isotropization still do occur, but there are now two more emergent time scales related to i) the applicability of the equation of state and ii) the expectation value of the operator breaking  conformal symmetry reaching its thermal value. The relation between these scales depends on the details of the model~\cite{Attems:2016tby,Attems:2017zam,Attems:2018gou}.

Finally, confinement represented holographically as the appearance of an infrared wall leads to the new physical effect in which excitations of the bulk geometry and matter fields bounce back and forth as in a cavity~\cite{Craps:2015upq,Bantilan:2020pay}. Such an effect was not present in the studies reviewed earlier and has not yet been explored in the context of expanding plasmas. 

\subsubsection{Away from the strong coupling regime \label{sec.weakcouplingholo}}

Another important direction studied in the context of thermalization  in strongly coupled gauge theories concerns corrections  from finite values of the coupling constant. In the context of the ${\cal N} = 4$ super Yang-Mills theory, the leading correction in the inverse power of the ’t~Hooft coupling constant behaves as $\lambda^{-3/2}$; on the gravity side, it arises at least in part due to a particular expression quartic in the  curvature~\cite{Gubser:1998nz}. Such a higher curvature gravity action when  treated exactly is ill-behaved due to the Ostrogradsky instability~\cite{Woodard:2015zca}. It is, however, not meant to be considered as such, since it is just an effective field theory truncated at a fixed order in the derivative expansion. 

Treating these higher curvature terms as small contributions to the Einstein's equations with negative cosmological constant allows one  to derive the leading order corrections to various holographic predictions at $\lambda \rightarrow \infty$. For example, they increase the shear viscosity of the ${\cal N} = 4$ super Yang-Mills theory from $\upeta/s = 1/(4 \pi)$ at $\lambda \rightarrow \infty$~\cite{Policastro:2001yc} to $\upeta/s = 1/(4\pi) \times \left( 1+ 15\, \zeta(3)\, \lambda^{-3/2} \right)$ for large but finite $\lambda$~\cite{Buchel:2008ac,Buchel:2008sh}. 

The previously discussed quartic term is the first higher order term appearing for the ${\cal N} = 4$ super Yang-Mills theory, but  remember that the Einstein-Hilbert action with negative cosmological constant describes infinitely many strongly coupled CFTs. For some of these~\cite{Buchel:2008vz}, the leading correction to Eq.~\eqref{eq.Sgrav} is quadratic in curvature and can be written as the so-called Gauss-Bonnet term
\begin{equation}
\label{eq.SGB}
\delta S_\text{grav}^{GB} = \frac{\lambda_{GB}}{2} L^2 \left( R^{2} - 4 R_{a b} R^{a b} + R_{a b c d} R^{a b c d} \right)\,.
\end{equation}
This contribution has $|\lambda_{GB}| \ll 1$ in top-down settings and the sign of $\lambda_{GB}$ can be either positive or negative. 

As a result, there are bona fide holographic CFTs for which the ratio of shear viscosity to entropy density is slightly lower than $1/(4 \pi)$~\cite{Kats:2007mq,Buchel:2008vz}. This important result showed that the celebrated value of $1/(4 \pi)$ \emph{is not} the lower bound in nature as originally conjectured in Ref.~\cite{Kovtun:2004de}, although the existence of another lower bound cannot be excluded. 

Furthermore, the combined gravity action of \Eqs{eq.Sgrav} and~\eq{eq.SGB} leads to, at least superficially, second order equations of motion. While it is known that microscopically this does not correspond to a well behaved QFT outside the regime $|\lambda_{GB}| \ll 1$~\cite{Camanho:2014apa}, in the spirit of bottom-up models discussed in \Sec{sec.strongnonconf} one can treat it, at least in some cases, as a model of QFT at a finite value of the ``coupling constant.'' 

In the context of the planar shockwave collisions discussed in Sec.~\ref{sec.holographiccollisions}, perturbative calculations in $\lambda_{GB}$ predict less stopping and more energy deposited close to the lightcone~\cite{Grozdanov:2016zjj,Folkestad:2019lam}. There also appears to be a correlation between the shear viscosity and hydrodynamization times, as encapsulated by \Eq{eq:wtilde}.

Furthermore, linear response calculations performed exactly in $\lambda_{GB}$ reveal that the singularity structure of real time correlators in equilibrium can change drastically as the coupling is varied~\cite{Grozdanov:2016vgg}. In particular, the results seem to mimic features expected from a kinetic theory, such as the appearance of branch cuts~\cite{Romatschke:2015gic,Kurkela:2017xis}, rather than single pole singularities known in strongly coupled QFTs~\cite{Kovtun:2005ev}.

The situation at a nonlinear level is more complicated. While the equations of motion are second order, the coefficients in front of the highest derivative terms are complicated and can vanish in regions of spacetime. This signals a breakdown of the initial value problem. Overcoming this obstacle is currently an active topic of research in the relativity community~\cite{Cayuso:2017iqc,Ripley:2019irj,Ripley:2019aqj,Ripley:2020vpk}.

Finally, we mention a more phenomenological set of hybrid approaches~\cite{Iancu:2014ava,Mukhopadhyay:2015smb,Kurkela:2018dku,Ecker:2018ucc} in which gravity is used to model the IR of a QFT and a weak coupling framework is put to work to represent the UV. Both frameworks are coupled to each other and predictions rely on a subtle interplay between the two combined models. Such a setting bears structural similarity to \cite{Gursoy:2007cb,Gursoy:2007er}, as discussed in Sec.~\ref{sec.strongtransverse}. However it uses the gravitational description only where it can be trusted, which is the regime where the coupling constant is large.

\subsubsection{Non-local correlators
\label{sec.holononloc}}

All the quantities we have discussed at strong coupling  concerned one-point functions of gauge invariant operators. Because of the underlying large-$N_{c}$ hierarchy, the problem of finding connected two- and higher-point functions correlation functions decouples from the problem of finding the one-point functions discussed thus far. Such correlation functions can be thought of as correlation functions of the bulk free (for two-point functions) or weakly interacting (for higher-point functions) quantum fields\footnote{They should not be confused with the underlying strongly coupled QFT for which both the classical bulk background and free bulk quantum fields are  effective descriptions.} living on top of gravitational backgrounds when the insertion points of the bulk correlators are taken to the boundary~\cite{Banks:1998dd}. In the following, we will focus on two-point functions.

Since we are talking about time dependent setups and, hence, Lorentzian correlators, the distinction between Wightman, retarded, or other correlators is appropriate~\cite{Son:2002sd,Herzog:2002pc,Skenderis:2008dh,Skenderis:2008dg}. Toward this end, the retarded correlator depends only on the gravitational background and captures the response of the strongly coupled QFT to sources. However, the Wightman correlator depends on both the constructed gravitational background and the state of the bulk quantum field. Therefore, its calculation is challenging in time dependent processes and, unless one creates a non-equilibrium state using sources exciting the vacuum or a thermal state~\cite{Chesler:2011ds,Chesler:2012zk,Keranen:2014lna,Keranen:2015mqc}, one has to deal with an additional freedom of initial conditions to scan.

It should perhaps not come as a surprise that to date there have been no studies of such correlators in an expanding plasma. Noteworthy works in this area were Refs.~\cite{Chesler:2011ds,Chesler:2012zk,Keranen:2014lna,Keranen:2015mqc}, which studied equilibration of scalar operator two-point functions under a spatially uniform quench. 

Many researchers use a proxy for correlators being a bulk geodesic spanned between the insertion points appropriate for operators of large scaling dimension in the Euclidean signature. However, in Lorentzian signature this is an uncontrollable approximation~\cite{Louko:2000tp,Headrick:2014cta,Keranen:2014lna}. On the other hand, the comparison between Wightman functions calculated according to the correct microscopic prescription and the geodesic proxy led to qualitatively similar results~\cite{Keranen:2014lna,Keranen:2015mqc}. 

If one takes this as an indication of the geodesic proxy as capturing the relevant physics, then one lesson following from such studies is that the symmetrized correlator with small spacelike separation between its insertion points thermalizes sooner than the one with larger separation~\cite{Balasubramanian:2010ce,Balasubramanian:2011ur}. This is also natural from the point of view of causality.

Furthermore, Ref.~\cite{Keranen:2015mqc} observed a relation between the equilibration time scale of the spatially Fourier transformed Wightman function and the equilibration time scale of~$1/T$ governing hydrodynamization at strong coupling and discussed in Sec.~\ref{sec.strongBjorken}. This study was done for a scalar operator, which does not exhibit a hydrodynamic tail. 

It is natural to conjecture that the energy-momentum tensor or a $U(1)$ current Wightman function would take longer to equilibrate due to the presence of hydrodynamic modes, but such studies have not been yet performed. Finally, as noted in Ref.~\cite{Keranen:2015mqc}, we should stress that the aforementioned momentum space features of equilibration do not  easily  translate to the real-space properties. This is so because sharp features in the correlator do not necessarily reside at small distances.

\section{Signatures of non-equilibrium QCD\label{sec:signatures}}

The experimental heavy-ion collision programs at BNL and CERN, combined with advances in theory
and empirically motivated models have, over the last couple of decades, greatly advanced our understanding of 
deconfined QCD matter. Successful multi-observable data-to-model comparisons have provided ample evidence that a new phase of matter is created with the thermodynamic
properties predicted by lattice QCD~\cite{Ding:2016qdj,Andronic:2017pug,Gardim:2019xjs,Bellwied:2019pxh,Bazavov:2019lgz,Bernhard:2019bmu,Pang:2016vdc}.
While thermodynamic features of QCD can also possibly be extracted from neutron star physics, with a spectacular recent example being the gravitational radiation pattern of neutron star mergers~\cite{Weih:2019xvw}, heavy-ion collisions are likely the only place in the Universe where the \emph{non-equilibrium} many-body properties of QCD can be explored. 

We will not discuss here signatures of high parton density matter in the hadron wavefunctions that were discussed elsewhere~\cite{Blaizot:2016qgz}. Uncovering definitive evidence for 
and systematic study of gluon saturation is a major goal of the Electron-Ion Collider (EIC)~\cite{Accardi:2012qut,Aschenauer:2017jsk}. We note that diffractive and exclusive signatures of gluon saturation at the EIC are especially promising~\cite{Mantysaari:2019hkq,Mantysaari:2017slo}.

Our focus here will be on quark-gluon matter formed after the collision. In the high parton density framework of the CGC EFT, the Glasma matter at the earliest times is most sensitive 
to the physics of gluon saturation. Indeed, if the contributions of the initial state can be isolated from that of the final state, heavy-ion collisions could present definitive evidence for gluon saturation. 

However, as we later discuss, a clean separation of initial and final state effects in the complex spacetime evolution of the heavy-ion collision is challenging~\cite{Adolfsson:2020dhm}. Nevertheless, data from both light and heavy-ion collisions at RHIC and the LHC can help constrain key features of gluon saturation, with an example being the energy and nuclear dependence of the saturation scale $Q_S$. 

\subsection{Electromagnetic and hard probes}
Since the Glasma matter is likely to be far off-equilibrium at the earliest instants of the heavy-ion collision, its features can be extracted most directly in probes that are the least sensitive to the later stages of the collision. The primary candidates here are electromagnetic probes of the medium such as photons and dileptons that, once emitted, do not interact with the medium. 

The problem here is that photons and di-leptons are produced continuously throughout the spacetime evolution of the quark-gluon matter and from the subsequent hadronic phase as well. 
Current models of heavy-ion collisions, which include photon yields from the pre-hydro kinetic theory phase tend to under predict the produced photon yields~\cite{Gale:2020xlg,Churchill:2020yny}; for an alternative mechanism, see \cite{Oliva:2017pri}.  

Photons emitted from highly occupied Glasma have been suggested as an additional source of radiation~\cite{Berges:2017eom}. While phenomenological model comparisons show a significant Glasma contribution~\cite{Garcia-Montero:2019vju}, the theoretical modeling of photon rates currently carries sizable uncertainty. 

Besides photons and di-leptons, inclusive yields of high momentum strongly interacting final states are also sensitive to gluon saturation and to early time dynamics in the heavy-ion collision. These include hadrons at high transverse momenta, jets, and heavy quarkonia. Gluon saturation influences the production rates for these processes and rescattering in the Glasma influences their 
dynamics. These effects are most pronounced for $p_\perp \sim Q_S$. We discussed heavy quark pair production in the Glasma in Section~\ref{sec:classicalstatistical}. The diffusion coefficient of 
these heavy quarks was computed recently in this framework and scales as $Q_S^3$~\cite{Boguslavski:2020tqz}. Heavy quark diffusion in Glasma-like environments and their subsequent evolution were also explored recently in several works~\cite{Mrowczynski:2017kso,Carrington:2020sww,Liu:2019lac}. A non-trivial problem is distinguishing this early-time evolution of heavy quarks from their late time evolution~\cite{Rapp:2009my,Brambilla:2019tpt,Akamatsu:2018xim}. Similar considerations also hold for the propagation of jets\footnote{The final stage of ``bottom up" thermalization corresponds 
to the ``jet quenching" of partons of momentum $\sim Q_S$ that are quenched to the thermal medium; this framework also explains key features of the quenching of very high momentum jets in the QGP~\cite{Blaizot:2016bsg,Blaizot:2013hx}.} in the Glasma~\cite{Dumitru:2007rp,Asakawa:2010xf,Carrington:2016mhd,Ipp:2020mjc}.

Higher point correlations of hard probes, add significant sensitivity to the dynamics of quark-gluon matter off-equilibrium. An example is the potential of  two-particle Hanbury-Brown--Twiss (HBT) photon interferometry to study early time dynamics~\cite{Garcia-Montero:2019kjk}. Such measurements are sensitive to the large longitudinal-transverse anisotropies that are not reflected in photon yields. However experimental measurements of soft photon correlations are very challenging experimentally and high statistics would be needed to disentangle the signal.

\subsection{Long-range rapidity correlations}
Long-range rapidity correlations are an important tool in disentangling initial and final state effects in hadron-nucleus and nucleus-nucleus collisions. This is because causality 
dictates that the latest time that a correlation can be induced between two particles $A$ and $B$ that freeze-out is given by 
\begin{equation}
\tau=\tau_{\rm freeze-out} \exp\left(-\frac{|y_A - y_B |}{2}\right)\,.
\end{equation}
Thus two particles that are long-range in rapidity $|y_A - y_B|\gg 1$ would be correlated at very early times in the collision~\cite{Dumitru:2008wn}. A particular example is the so-called ``ridge" effect, reviewed in~\cite{Dusling:2015gta}, which correlates two particles not only in rapidity but also in relative azimuthal angle~\cite{Dumitru:2010iy}. A recent summary of the physics of 
initial state correlations was given in Ref.~\cite{Altinoluk:2020wpf}. 

However, if hydrodynamic flow also sets in early, this ridge could be a final state effect~\cite{Shuryak:2013ke} due to the underlying boost-invariance of the hydrodynamic fluid. A way forward to disentangling 
initial state physics of CGCs and the Glasma  at early times from late time dynamics is to look at the evolution of two-particle correlations with their rapidity separation~\cite{Bzdak:2016aii} Another approach is 
to study the long range correlations of particles with large transverse momenta that do not follow hydrodynamically~\cite{Dusling:2013oia,Martinez:2018tuf}.

\subsection{Bulk observables}
We  previously discussed limiting fragmentation of hadron distributions and its potential to distinguish initial and final state effects in hadron-hadron collisions~\cite{Goncalves:2019uod}. 
We will now discuss other bulk observables in high energy nucleus-nucleus, hadron-nucleus, and hadron-hadron collisions that can help constrain the properties of 
saturated gluons and their early-time evolution. In the smaller systems, even if the system hydrodynamizes quickly, the large shape fluctuations of partons will provide 
insight into multi-parton correlations in the initial state~\cite{Mantysaari:2020axf}; understanding these from first principles is a challenging problem~\cite{Dumitru:2020fdh} that may also require the EIC to resolve. 

A number of works have explored applications of holographic ideas to the study of bulk observables in heavy-ion collisions.
A universal prediction of holography is that of hydrodynamization being distinct from local thermalization. A specific phenomenological investigation implementing this idea  used holographic boost invariant dynamics with transverse expansion as a successful model of preflow~\cite{vanderSchee:2013pia}. Another development was discussed in Ref.~\cite{vanderSchee:2015rta}, which treated the planar shockwave collisions discussed in Sec.~\ref{sec.planarshocks} as an explicit model of initial state physics. While this study recovered qualitative features of soft particle spectra, the rapidity distribution of produced particles is too narrow relative to the experimental data. It would be very interesting to explore more complicated holographic models of heavy-ion collisions and constrain them with experimental data.

In a thermalizing system, the loss of information of the initial conditions
manifests itself as the production of entropy. Therefore, if the system locally thermalizes and its flow is nearly isentropic, the measured number of particles probes the entropy produced during the non-equilibrium 
evolution of quark-gluon matter. The CGC framework accounts for the increase of particle
multiplicity with increasing collision energy with the growth of the saturation scale $Q_S$~\cite{Albacete:2014fwa}.
Recent calculations of entropy production in the equilibration processes using
 hydrodynamic attractors provide a quantitative relation between
 the energy deposition in the CGC picture and the final particle numbers~\cite{Giacalone:2019ldn}.

On the other hand, the energy of the observed particles depends on the work done
during the entire expansion and therefore has different dependencies on the dynamics of the pre-equilibrium stage.
Comparing these two robust experimental measurements (energies and multiplicities) 
already casts doubts on the complete equilibration of QGP in peripheral nucleus-nucleus collisions~\cite{Kurkela:2019kip, Giacalone:2019ldn}.

Many of the experimental signatures of QGP (strangeness enhancement, jet suppression, flow harmonics, etc.) show a smooth dependence on system size from central to peripheral nucleus-nucleus, proton-nucleus, and proton-proton collisions. As the system size shrinks, so also does its lifetime, corresponding to an increase in the relative importance of non-equilibrium QCD process increases.

Equilibration studies in large systems already put a lower bound below which the system will not reach hydrodynamization or chemical equilibrium~\cite{Kurkela:2018xxd,Kurkela:2018wud}.
Therefore, explaining observed signals of collectivity (or the absence thereof) in small
collisions systems requires a proper treatment of non-equilibrium QCD dynamics. Some recent examples of work in this direction include studies of flow harmonics~\cite{Schenke:2019pmk,Kurkela:2018qeb}, parton energy loss~\cite{Andres:2019eus} and 
heavy-quark evolution~\cite{Mrowczynski:2017kso}. Furthermore, as discussed in Sec.~\ref{sec.strongtransverse}, hydrodynamization without equilibration of small systems is very natural in holography. 

Also noteworthy is recent phenomenological work~\cite{Huang:2017tsq} quantifying the role of non-equilibrium dynamics in the Chiral Magnetic Effect, which we discussed in Section~\ref{sec:classicalstatistical}. A topic that demands further investigation is the origin of the very large vorticities measured in off-central heavy-ion collisions, as extracted from measurements of the polarization of $\Lambda$-baryons~\cite{Becattini:2020ngo}. The vorticities are introduced on macroscopic scales of the order of the system size; how these propagate efficiently down to the microscopic scales of $\Lambda$ is not yet understood.

\subsection{Future prospects}
A recent recommendation from the  European Strategy for Particle Physics report emphasized that
the main physics goal of future experiments with heavy-ion and proton beams at the LHC will be 
\emph{a detailed, experimentally tested dynamical
understanding of how out-of-equilibrium evolution occurs and equilibrium properties arise in a
non-Abelian quantum field theory}~\cite{Strategy:2019vxc,Citron:2018lsq}. The scheduled runs 3 and 4 of the LHC will mark a decade of high-statistics data across system sizes at the highest achievable collision energies. 

In the United States, continued operation of RHIC will provide further insight into several of the signatures that we have discussed. In particular, with the anticipated commissioning of the sPHENIX detector~\cite{Roland:2019cwl}, hard probes of QCD off-equilibrium will be studied in a dynamical range that is complementary to that of the LHC.

Looking further into the future, the EIC project has received Critical Decision Zero (CD0) approval from the U.S. Department of Energy. The EIC will explore with high 
precision the landscape of hadron structure at high energies~\cite{Accardi:2012qut,Aschenauer:2017jsk}.

One may therefore anticipate that this decade and the next will bring many opportunities to exploit the signatures that we have articulated here, and likely several novel ones, of the properties of QCD off-equilibrium.

\section{Interdisciplinary connections\label{sec:interdisciplinary}}

Understanding the thermalization process in QCD associated with heavy-ion collisions addresses some of the most fundamental questions in quantum dynamics, with exciting interdisciplinary connections made to very different many-body systems. The transient ``fireball" expanding in vacuum explores far-from-equilibrium conditions at early times, followed by a series of characteristic stages that are finally expected to lead to
a fluid-like behavior governing the approach to local thermal equilibrium. Very similar questions of equilibration and the emergence of collective behavior from the underlying unitary quantum dynamics are relevant for diverse applications ranging from high-energy and condensed matter physics to practical quantum technology. For reviews in the context of condensed matter physics, see~\cite{Gogolin_2016,D_Alessio_2016,Borgonovi_2016}.

Several non-equilibrium phenomena were first proposed in the context of QCD matter in extreme conditions and then explored and experimentally probed in alternative quantum many-body systems. For instance, the phenomenon of prethermalization~\cite{Berges:2004ce} with the rapid establishment of an effective equation of state during the early stages of heavy-ion collisions~\cite{Arnold:2004ti,Dusling:2012ig} has been explored for early Universe inflaton dynamics~\cite{Podolsky:2005bw} and condensed matter systems~\cite{Moeckel_2008,Langen_2016,Mori_2018}, and experimentally discovered in ultracold quantum gases on an atom chip~\cite{Smith:2012jea}. 

In turn, aspects of entanglement represent one of the major overarching schemes in contemporary physics of quantum-many body systems, and gravity in and out of equilibrium, while investigations about its relevance to the thermalization process in QCD are relatively recent. There are many excellent topical reviews on entanglement and we refer the reader to  Refs.~\cite{Eisert:2008ur,Calabrese:2009qy,Casini:2009sr,Rangamani:2016dms}. We discuss some aspects of entanglement in our context in more detail later.

To capture the thermalization dynamics in QCD related to heavy-ion collisions, detailed comparisons take into account the fact that the coupling of non-Abelian gauge theories is not a constant but changes with characteristic energy or momentum scale in a particular way. While strong at low scales, the coupling becomes weak at sufficiently high energies because of the phenomenon of asymptotic freedom~\cite{Gross:1973ju,Politzer:1973fx}. Even in the high-energy limit, where the gauge coupling is weak, one faces a strongly correlated system because a plasma of gluons with high occupancy [$f(Q_S) \sim 1/\alpha_S(Q_S)$] is expected to form; see Sec.~\ref{sec:Glasma}. Such a transient over-occupation leading to strong correlations even for weakly coupled systems can be found in a variety of physical applications that are far from equilibrium. Examples include the pre-heating scenario for the very early stages of our Universe after a period of strongly accelerated expansion called inflation~\cite{Kofman:2008zz} and the relaxation dynamics in table-top setups with ultracold quantum gases following a sudden change in external control parameters such as magnetic fields~\cite{Prufer:2018hto}.

The very high level of control in experiments with synthetic quantum systems, such as ultracold quantum gases, enables dedicated quantum simulations. These systems provide very flexible testbeds, which can realize a wide range of Hamiltonians with variable interactions and degrees of freedom based on atomic, molecular and optical physics engineering~\cite{Bloch_2008}. Since these setups can be well isolated from the environment, they offer the possibility of studying fundamental aspects such as the thermalization process from the underlying unitary quantum evolution. 

While digital quantum simulations based on a Trotterized time evolution on a universal quantum computer are challenging to scale up, present large scale analog quantum simulators using ultracold quantum gases already explore the many-body limit described by quantum field theory~\cite{Bloch_2008, haller2010, Gring12, Hung2013, Langen15, Navon2015, Navon2016, parsons2016site, schweigler2017experimental, bernien2017probing, Prufer:2018hto, Erne:2018gmz, Eckel2018, Hu2019, Feng2019, Murthy2019, Keesling2019, prufer2019experimental, Zache:2019xkx}. In principle, with quantum simulators  non-universal aspects of the dynamics of gauge theories can be studied. This was first achieved for Abelian gauge theory with digital quantum simulations, such as those using trapped ions~\cite{Martinez_2016} or superconducting qubits~\cite{Klco_2018}. 

An interesting possibility to consider is the application of a hybrid quantum-classical framework to real time problems. This has been discussed in a ``single particle" digital strategy for scattering problems whereby higher loop quantum contributions can be simulated digitally and the background gauge field treated in principle on a quantum simulator~\cite{Mueller:2019qqj,Mueller:2020vha}. 
It is also important to note that scalable analog systems for the quantum simulations of gauge theories using ultracold atoms have been reported~\cite{Kokail_2019,Mil_2020}.  We anticipate significant progress in all of these approaches to quantum computation of real time problems in the decade ahead. 

\subsection{Strong interactions: Unitary Fermi gas}\label{sec:fermi}

A paradigmatic example of the interdisciplinary cross-fertilization among the different physical applications is the work on collective motion of a unitary Fermi gas. Near unitarity, the $s$-wave scattering length, which characterizes the two-body interaction strength, becomes very large and the effective scale invariance of the interaction at unitarity can lead to universal behavior~\cite{Chin_2010}, which can also be accessed out of equilibrium~\cite{Eigen_2018}. Many similarities for dynamical properties, such as a low ratio of shear viscosity to entropy density, have been discussed in this context in comparison to QCD.  See the discussion in Sec.~\ref{sec:kinetictheory}. 

We noted that heavy-ion experiments indicate that the hot quark-gluon plasma may be described as the most perfect fluid realized in nature
~\cite{Adams:2005dq,Adcox:2004mh,Aamodt:2010pa,ATLAS:2012at,Chatrchyan:2011sx}. The only serious experimental competitors are  ultracold quantum gases at temperatures that differ by 20 orders of magnitude! Strong interactions also play a central role in holographic approaches, a concept that is addressed in Sec.~\ref{sec:strongcoupling}, and there are concrete proposals on how to realize holographically systems resembling unitary Fermi gases starting with Refs.~\cite{Son:2008ye,Balasubramanian:2008dm}. A comprehensive review of common aspects of QCD, unitary Fermi gases and holography was provided in Ref.~\cite{Adams:2012th}. 

\subsection{Highly occupied systems I:  Preheating in the early Universe}\label{sec:earlyuniverse}

The dilution of matter and radiation during the inflationary period of the early Universe leads to an extreme condition that may be well characterized by a pure state with vacuum-like energy density carried by a time dependent coherent inflaton field with large amplitude~\cite{Kofman:2008zz}. A wide class of post-inflationary models with weak couplings exhibit the subsequent decay of the inflaton field amplitude via non-equilibrium instabilities~\cite{Traschen:1990sw,Kofman:1994rk}. Detailed mechanisms for the origin of an instability and the scattering processes are different than in QCD with strong color fields. 

However, the rapid growth of fluctuations from the inflaton decay leads to a non-linear time evolution that follows along lines similar to those outlined in Sec.~\ref{sec:classicalstatistical} for QCD. For instance, for scalar fields with weak quartic interaction $\lambda \ll 1$, a corresponding overoccupation $\sim 1/\lambda$ up to a characteristic momentum scale is achieved after the instability. Likewise, at this stage the prethermalization~\cite{Berges:2004ce,Arnold:2004ti} of characteristic properties, including an effective equation of state, is observed in these scalar models~\cite{Podolsky:2005bw}.

Moreover, a self-similar attractor solution is approached subsequently, as discussed in Sec.~\ref{sec:scalar-NTA}. Compared to the longitudinally expanding QCD plasma, a major difference stems from the isotropic expansion of the Universe. Some aspects of isotropic expansion can be lifted for the inflaton field dynamics by introducing suitably rescaled conformal time and field amplitudes, such that the dynamics is essentially that of Minkowski spacetime without expansion~\cite{Micha:2002ey}. In fact if compared to QCD dynamics without expansion, then characteristic dynamical properties such as the values of scaling exponents in the attractor regime agree with what is found for self-interacting scalar field dynamics with quartic interactions in the absence of spontaneous symmetry breaking~\cite{Berges:2016nru}. 

This concerns the gauge theory's direct energy cascade toward the perturbative high-momentum regime~\cite{Kurkela:2011ti,Schlichting:2012es,Kurkela:2012hp}, as well as the inverse particle cascade toward low momenta in the non-perturbative regime associated with non-thermal fixed points~\cite{Berges:2019oun}. In turn, scalar fields with longitudinal expansion seem to exhibit several universal features shared with QCD dynamics in the transient scaling regime~\cite{Berges:2015ixa}. In particular, the inverse cascade essentially follows the behavior of the corresponding non-expanding system because of the strong Bose enhancement of rates at low momenta~\cite{Berges:2015ixa}; see also Sec.~\ref{sec:ECD}.  

\subsection{Highly occupied systems II: Bose gases far from equilibrium}\label{sec:bosegas}

Although the inflaton dynamics is described by a relativistic field theory, the self-similar scaling behavior at sufficiently low momenta below the screening mass scale is predicted to exhibit universal properties of a non-relativistic system~\cite{Orioli:2015dxa}. The non-equilibrium infrared dynamics for scalars starting from overoccupation
has been theoretically studied in great detail~\cite{Berges:2008wm,Scheppach:2009wu,Berges:2010ez, Nowak:2010tm,Nowak:2011sk,Berges:2012us,Berges:2014bba,Moore:2015adu,Walz:2017ffj,Chantesana:2018qsb,Deng:2018xsk,PineiroOrioli:2018hst,Shen:2019jhl,Boguslavski:2019ecc}. However important aspects of this far-from-equilibrium dynamics can be probed experimentally using Bose gases in an optical trap. For the example of an interacting, non-relativistic Bose gas of density $n$ in three spatial dimensions, this concerns the dilute regime ($\sqrt{n a^3} \ll 1$), with a characteristic inverse coherence length given by the momentum scale $Q = \sqrt{16 \pi a n}$. Here $Q$ plays a similar role as the saturation scale for gluons in the gauge theory case, and the diluteness $\sqrt{n a^3}$ provides the dimensionless coupling parameter. An overoccupied Bose gas then features large occupancies $\sim 1/\sqrt{n a^3}$ for modes with momenta of the order of $Q$~\cite{Orioli:2015dxa}. 
 
\begin{figure}[t]
\includegraphics[width=\columnwidth]{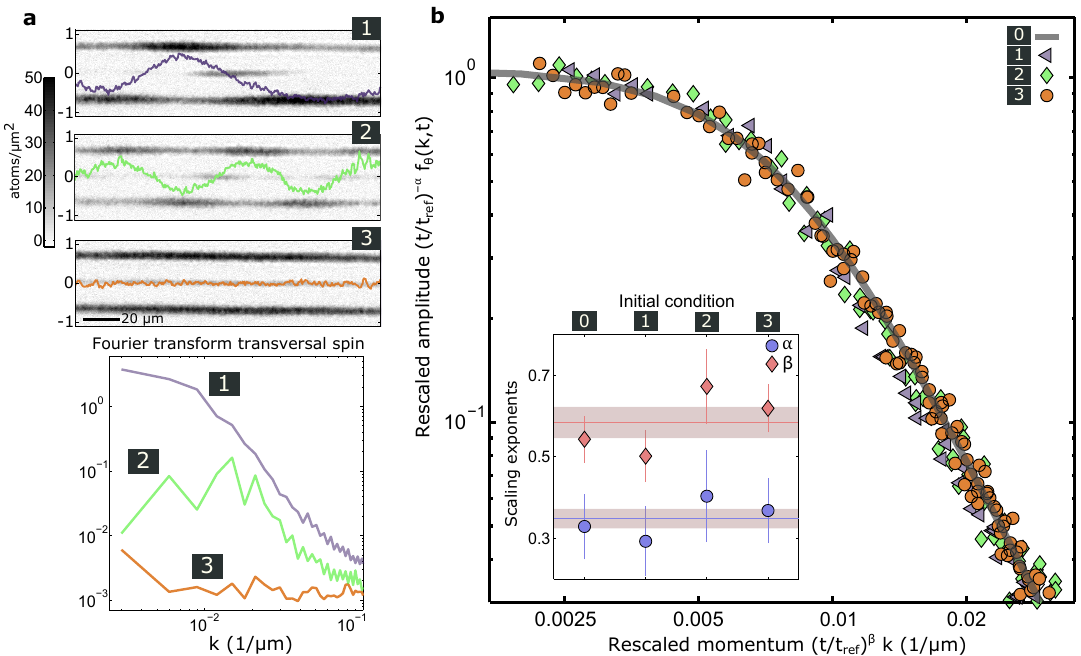}
\caption{\label{fig:BoseGasScaling} (a) Absorption images of different magnetic hyperfine states of a spin-1
Bose gas with the extracted transversal spin (solid lines) for three different far-from-equilibrium initial conditions. (b) All initial conditions lead to the same universal scaling behavior, such that all data points collapse onto a single curve after rescaling with time using the universal exponents $\alpha$ and $\beta$. From Ref.~\cite{Prufer:2018hto}.}
\end{figure}
 
Universal scaling far from equilibrium associated with non-thermal fixed points has been experimentally discovered using different cold atom systems~\cite{Prufer:2018hto, Erne:2018gmz}. For instance, in Refs.~\cite{Prufer:2018hto, prufer2019experimental} the non-equilibrium dynamics of magnetic hyperfine excitations of a spin-1  Bose gas was studied in an elongated trap, following a sudden change in the applied magnetic field as an external control parameter. Figure~\ref{fig:BoseGasScaling} exemplifies the scaling dynamics of the measured transversal spin for three different initial conditions. After an initial non-equilibrium instability regime, all data in the self-similar scaling regime are seen to collapse to a single curve after rescaling with time using universal scaling exponents. While this example concerns infrared scaling, bi-directional scaling including a self-similar evolution toward higher momenta with subsequent thermalization was experimentally analyzed in Ref.~\cite{glidden2020bidirectional}.  

\subsection{Highly occupied systems III: Classicalization and unitarization of gravitational amplitudes}\label{sec:gravity-CGC}
An intriguing idea is that of black holes as long lived states of highly occupied gravitons ($f\gg 1$) that satisfy the condition $\alpha_{\rm grav} f = 1$ \cite{Dvali:2011aa}. Here $\alpha_{\rm grav}=L_P^2/R_S^2$, where $L_P$ is the Planck length and $R_S$ denotes the Schwarzchild radius. A dynamical picture of the formation of such a black hole state is in $2\rightarrow N$ scattering of gravitons at trans-Planckian energies. In the Regge limit, as first discussed in Ref.~\cite{Lipatov:1991nf} and subsequently in Ref.~\cite{Amati:1987wq}, the scattering is dominated by the formation of $N-2$ soft quanta. The argument of Dvali and collaborators was that the copious production of soft gravitons leads to perturbative unitarization of the scattering cross-section precisely when $\alpha_{\rm grav} f = 1$. 

This ``classicalization of amplitudes" was shown explicitly~\cite{Dvali:2014ila} using the tree level Kawai-Lewellen-Tye (KLT) relations~\cite{Kawai:1985xq} that express $N$-point tree level gravity amplitudes in terms of sums of products of Yang-Mills $N$-point tree amplitudes. These results are in remarkable agreement with computations in Lipatov's EFT approach~\cite{Addazi:2016ksu}.

The ideas of the classicalization and unitarization of $2\rightarrow N$ gravitational amplitudes are remarkably similar to the discussion of the CGC EFT in Secs.~{sec:hadrons} and \ref{sec:Glasma}. The BFKL results on $2\rightarrow N$ gluon scattering are likewise reproduced in the semi-classical CGC EFT. 
A path forward is to employ so-called ``double copy" methods that exploit a color-kinematics duality between gravity and QCD amplitudes~\cite{Bern:2019prr}. Such a correspondence was prefigured in the high energy limit in Ref.~\cite{Lipatov:1991nf} and  further discussed more recently~\cite{SabioVera:2011wy,Liu:2018lam}. 

Of particular interest in our context is the ``classical double copy" between classical Yang-Mills equations and classical gravity~\cite{Monteiro:2014cda,Goldberger:2016iau}. This points to a concrete correspondence between collisions of the classical gluon shock waves producing the Glasma and that of gravitational shock waves that produce black holes~\cite{Dvali:2021ooc}. It would also be interesting to understand whether this correspondence shares universal features at the unitarity limit with that of the holographic gravitational shock waves discussed in \Sec{sec:strongcoupling}.

\subsection{Anomalous currents in non-equilibrium QED: Condensed matter systems and strong laser fields}\label{sec:QED}

Strong color fields as well as strong electromagnetic fields are an essential
ingredient for the understanding of the early stages of the plasma's space-time evolution in off-central heavy-ion collisions. Strong gauge fields lead to a wealth of intriguing phenomena related to quantum anomalies, such as the chiral magnetic effect~\cite{Kharzeev:2015znc,Skokov:2016yrj} described in Sec.~\ref{sec:classicalstatistical}. As discussed, there are strong connections between the transport properties of anomalous currents in hot QCD and in strongly correlated condensed matter systems, in particular  Dirac and Weyl semimetals with applied fields~\cite{Li:2016vlc}.

Here we note that the similar questions could also be addressed in future strong laser field experiments that will be able to explore QED dynamics in extreme conditions~\cite{DiPiazza:2011tq}. For instance, for QED field strengths exceeding the Schwinger limit for pair production, a highly absorptive medium with quantum anomaly-induced dynamical refractive properties related to the chiral magnetic effect was predicted~\cite{Mueller:2016aao}. 

\subsection{Thermalization and entanglement}\label{sec:entanglement}

While the time evolution of isolated quantum systems is unitary, relevant observables in non-equilibrium quantum field theory can approach thermal equilibrium values at sufficiently late times, without the need for any coarse-graining or reference to a reduced density operator. Thermalization in quantum field theory has been demonstrated for scalar quantum field theories in various spatial dimensions~\cite{Berges:2000ur,Berges:2001fi,Juchem:2004cs,Arrizabalaga:2005tf} and with fermions~\cite{Berges:2002wr,Shen:2020jya}; see Ref.~\cite{Berges:2004yj} for an introductory review\footnote{For thermalization studies in classical-statistical field theories for given regularization, see Ref.~\cite{Aarts:2000wi}.}. In gauge theories at strong coupling, thermalization from unitary dynamics was observed using holographic approaches, as discussed in~Sec.~\ref{sec:strongcoupling}.

It has been analyzed in detail how, in particular, locally defined quantities of isolated quantum many-body systems can exhibit thermal features~\cite{Deutsch1991,Srednicki1994,Rigol2008}. 
In such time-dependent processes, entanglement entropy of spatial subregions (the von Neumann entropy of spatially reduced density matrices) was seen to reach the value predicted by thermal states after exhibiting a period of growth; see, e.g., Refs.~\cite{Calabrese:2005in,AbajoArrastia:2010yt,Liu:2013qca,Cotler:2016acd,Kaufman2016,Alba2017}. Understanding why and how this happens has been an active sub-field of research in lattice systems, quantum field theory and holography.

Ref.~\cite{Berges:2017zws,Berges:2017hne} applied similar considerations to a model of $e^{+}e^{-}$ collisions and pursued the idea of viewing entanglement as a source of an apparent thermal behavior seen in multiparticle production in such events, as discussed in Refs.~\cite{Becattini:1995if,Andronic:2008ev}. Recently an entanglement entropy measure devised for proton-proton collisions at the LHC was argued to be consistent with the data; the latter is at variance with expectations from Monte-Carlo simulations~\cite{Tu:2019ouv}. In the same vein, Ref.~\cite{Ecker:2016thn} explored the behavior of the entanglement entropy in a holographic model of heavy-ion collisions discussed in Sec.~\ref{sec.holographiccollisions} and found it can serve as an order parameter distinguishing between the Landau (full stopping) and Bjorken (transparency) scenarios. 

The notion of entanglement plays a key role in tensor network methods that represent quantum-many body wave functions and density matrices of physical interest yet with low enough entanglement to allow for their efficient manipulation on classical computers; see Ref.~\cite{Orus:2013kga} for a review. Such methods are robust in describing ground states and low-lying excited states in 1+1 dimensions~\cite{Hastings2006,Vidal:2008zz}, and considerable progress has been  made in the past few years with using them for condensed-matter physics applications in 2+1 dimensions~\cite{Corboz2016a,Corboz2016,Vanderstraeten2016a,Corboz2018prx,Rader2018prx}. 

In the context of this review, we highlight a number of recent developments in applying tensor networks to QCD and heavy-ion collision motivated problems in (1+1)-dimensional settings ranging from the applications to gauge theories reviewed in Ref.~\cite{Banuls:2019rao} to non-equilibrium processes in interacting QFTs on a lattice~\cite{Pichler:2015yqa,Buyens:2015tea,Buyens:2016hhu,Banuls:2019qrq}. In the last cases, the aforementioned growth of entanglement with time is a bottleneck preventing simulations from reaching late times. 

Finally, entanglement entropy in holography arises as a Bekenstein-Hawking entropy of a special class of surfaces~\cite{Ryu:2006bv,Hubeny:2007xt,Lewkowycz:2013nqa,Dong:2016hjy}. This discovery has led to new insight into quantum gravity by bringing quantum information tools to the mix. An impressive result in this direction is the quantitative understanding of the time evolution of the entropy of Hawking radiation from an evaporating black hole~\cite{Penington:2019npb,Almheiri:2019psf,Almheiri:2019hni,Penington:2019kki,Almheiri:2019qdq}. The cited works point to a new mechanism toward resolving Hawking's information paradox~\cite{Hawking:1976ra,Page:1993wv}. From the point of view of this review, they can be thought of as including finite-$N_{c}$ effects in holographic studies of a class of thermalization processes at very late times.

\section{Summary and Outlook\label{sec:conclusions}}

In 1974, T.D.\ Lee suggested that {\it it would be interesting to explore new phenomena by distributing a high amount of energy or high nuclear density over relatively large volume}~\cite{osti_4061527}.  We are beginning to come to grips with the richness of many-body QCD dynamics 46 years later owing to experimental programs in nucleus-nucleus collisions in the decades since, 
culminating in the discovery of the quark-gluon plasma at RHIC and the LHC. As demonstrated at these colliders, the non-Abelian QGP is a nearly perfect fluid showing little resistance to pressure gradients. 

This conclusion is a consequence of the remarkable and apparently unreasonable success of relativistic viscous hydrodynamics in the description of the heavy-ion data from RHIC and the LHC. However the quantitative phenomenological success of hydrodynamical models also owes a great deal to our improved understanding of the initial conditions for hydrodynamic evolution, in 
particular, in the modeling of event-by-event fluctuations in the nuclear geometry, as well as a deepening understanding of how the quark-gluon matter is released in the heavy-ion collisions and thermalizes to form the QGP. 

With regard to the latter, comparisons of the hydrodynamical models to data require thermalization to occur very rapidly on time scales on the order of 3 yoctoseconds: approximately a tenth of the lifetime of the nuclear collision. These very short lifetimes and the nearly perfect fluidity of the subsequent flow of the QGP suggest that the non-equilibrium matter formed is very strongly correlated. The quest to understand {\it ab initio} the structure of strongly correlated QCD matter in nuclear wavefunctions at high energies, and how this matter is released, decoheres, and thermalizes, has motivated a large body of work over the last couple of decades, from the inception of the RHIC program to the present. 

Strongly correlated QCD matter can arise either in weak coupling when the occupancies of the constituents are very large or in strong coupling. Further, since the coupling runs towards strong coupling as the system evolves, both weak and strong couplings may be realized in the fluid. In this review, we summarized the theoretical ideas and techniques in both strong and weak coupling frameworks that address the thermalization process in heavy-ion collisions. 

We emphasized the emergence of attractors in both the weak coupling EFT and  holographic approaches that may be universal across a wide range of energy scales. We also noted concomitantly the very concrete interdisciplinary connections of strongly correlated QCD (and QCD-like) matter off-equilibrium to dynamical features of phenomena ranging from 
pre-heating in inflationary cosmology to pair-production in laser induced strong QED fields to to non-equilibrium dynamics in ultracold atomic gases. 

In particular, we discussed an intriguing universality in the non-thermal attractor discovered in simulations of overoccupied expanding Glasma to that discovered in identically prepared simulations of the self-interacting scalar fields that model the ultracold systems. Remarkably, cold atom experiments have discovered such a non-thermal attractor, albeit with a different geometry than that of a heavy-ion collision. 
This opens up the exciting prospect of extending the program underway of  the ``tabletop engineering" of ultracold atom systems as analog quantum simulators of the ground state properties of gauge theories to uncover far-from-equilibrium properties of non-Abelian gauge theories. 

We also discussed the signatures for QCD matter off-equilibrium and the challenges of disentangling these from contributions at later stages of the heavy-ion collision. Ongoing and near-term experiments at both RHIC and the LHC will greatly enhance these prospects  through both novel measurements and larger datasets than are currently available. The EIC will provide information complementary to those of the heavy-ion experiments to further tease out and make more precise our understanding of the initial state.
Further progress will also depend on theoretical developments in the weak and strong coupling frameworks and the convergence between the two when extrapolated to the realistic couplings of the heavy-ion experiments. 

Computations of the properties of saturated gluons in the CGC EFT are now at next-to-leading-order and next-to-leading log accuracy for a few processes. We expect this trend to continue,  which will allow for very precise extractions of the saturation scale in DIS and proton-nucleus collisions. A more conceptual challenging problem is to understand the large fluctuations in the large~$x$ initial conditions that may generate very anisotropic shape distributions of small~$x$ partons. As we noted briefly, such studies may benefit from the universality between the non-linear equations that describe high energy QCD evolution and those that describe reaction diffusion processes in statistical mechanics. 

In the description of the Glasma, a straightforward but technically challenging problem is to extend several of the computations in fixed box geometries to the more realistic longitudinally expanding case. 
A more difficult challenge is to implement fully quantum contributions beyond the classical-statistical approximation. While there is considerable insight gained from ongoing studies of scalar field theories in this regard, further progress will require additional conceptual breakthroughs. A noteworthy feature of the overoccupied Glasma is the emergence of infrared structures that may have non-trivial topological features~\cite{Spitz:2020wej}. This may be universal to other many-body systems, leading to novel potential synergies in addition to those discussed in this review.

Recent numerical simulations using QCD effective kinetic theory have painted a detailed
picture of the different equilibration stages 
in longitudinally expanding, albeit homogeneous, QCD matter. 
However the kinetic description of inhomogeneous systems with rapid radial expansion needs further development. This is especially important for studies of 
collisions of light nuclei or in proton-nucleus collisions, where tantalizing signals of collective behavior have been seen. It will be interesting 
within this framework to understand whether a unified many-body description emerges that smoothly interpolates from a few parton scatterings in the smallest collision systems
to the emergent fluid-like behavior in the largest systems. 

On the more formal side, computations of various transport properties
of the QGP beyond leading order have higher order corrections that are large for all but extremely small values of the coupling constant. Finite temperature 
resummation techniques may help improve the convergence of the perturbative expansion. A potential path forward is to combine a non-perturbative description of the infrared sector with 
kinetic theory in the UV.

A key part of our review was devoted to developments in holographic approaches to off-equilibrium dynamics in QCD like theories. 
An important discovery is that the hydrodynamic gradient expansion is an asymptotic series, which allows one to view the applicability of hydrodynamics through the emergent universal behavior of a hydrodynamic attractor. 

An open problem is the existence of hydrodynamic attractors for flows with transverse expansion and/or broken conformal symmetry. 
It would be very interesting to make a clear-cut statement as to what extent these phenomena appear in a tracktable manner outside idealizations of the geometry of ultrarelativistic heavy-ion collisions or highly-symmetric cosmologies. 
Another important future direction is to address collisions in holographic models that incorporate confinement following recent promising work in this direction. 
Not least, it would be  interesting to reconsider expanding plasma setups and, more broadly, thermalization at strong coupling in the context of the Gauss-Bonnet gravity discussed in Section~\ref{sec.weakcouplingholo}. First steps in this direction relied on treating the Gauss-Bonnet term as a small correction. 
Going beyond this regime, which is challenging from many perspectives, can reveal genuinely new effects in holographic setups like non-thermal fixed points discussed in Section~\ref{sec.nonthatt}.
Finally, an important open question in holography is to understand whether long-range ``ridge-like" correlations can naturally arise at strong coupling and whether they can survive until late time.

\begin{acknowledgments}
We have all benefited greatly from the combined wisdom on this topic of our collaborators and colleagues over the years. We thank, in particular, Gert Aarts, Peter Arnold, Rudolf Baier, Adam Ball, Guillaume Beuf, Jean-Paul Blaizot, Kirill Boguslavski, Szabolcs Borsányi, Alex Buchel, Jorge Casalderrey-Solana, Paul Chesler, Jian Deng, Adrian Dumitru, Kevin Dusling, Gia Dvali, Thomas Epelbaum, Sebastian Erne, Stefan Floerchinger, Wojciech Florkowski, Charles Gale, Oscar Garcia-Montero, Thomas Gasenzer, Daniil Gelfand, François Gelis, Jacopo Ghighlieri, Giuliano Giacalone, Philipp Hauke, Florian Hebenstreit, Edmond Iancu, Jamal Jalilian-Marian, Romuald Janik, Fred Jendrzejewski, Sangyong Jeon, Valentin Kasper, Dmitri Kharzeev, Alex Krasnitz, Aleksi Kurkela, Tuomas Lappi, Nicole Löhrer, Mark Mace, David Mateos, Aleksandr Mikheev, Guy Moore, Alfred Mueller, Niklas Mueller, Swagato Mukherjee, Larry McLerran, Rob Myers, Yasushi Nara, Markus Oberthaler, Robert Ott, Jean-François Paquet, Monica Pate, Asier Piñeiro Orioli, Jan M.\ Pawlowski, Robi Peschanski, Rob Pisarski, Maximilian Prüfer, Ana Raclariu, Klaus Reygers, Paul Romatschke, Alexander Rothkopf, Kaushik Roy, Björn Schenke, Sören Schlichting, Jörg Schmiedmayer, Julien Serreau, Dénes Sexty, Linda Shen, Vladimir Skokov, Michal Spalinski, Daniel Spitz, Andy Strominger, Viktor Svensson, Naoto Tanji, Derek Teaney, Robin Törnkvist, Prithwish Tribedy, Wilke van der Schee, Benjamin Wallisch, Qun Wang, Christof Wetterich, Paul Wiesemeyer, Przemek Witaszczyk, Larry Yaffe, Yi Yin and Torsten Zache.

We would like to thank Paul Romatschke for correspondence about previous work~\cite{Romatschke:2017vte} and for providing us with the numerical data needed for \Fig{fig:bifAholo}, 
Bj\"{o}rn Schenke for the plot in Fig.~\ref{fig:JIMWLK-soln}, which was adapted from Ref.~\cite{Dumitru:2011vk} and Wilke van der Schee for the plot in Fig.~\ref{fig.approachtoattractor}, which was adapted from Ref.~\cite{Kurkela:2019set}.

We also thank Kirill Boguslavski, Jacopo Ghighlieri, Michal Spalinski, Bj\"{o}rn Schenke, Viktor Svensson, and Wilke van der Schee for reading the manuscript and offering their suggestions.

The work of JB and AM  is part of and supported by the Deutsche Forschungsgemeinschaft [(DFG) German Research Foundation] Collaborative Research Center ``SFB 1225 (ISOQUANT)." JB's work is also partly supported by the DFG under Germany's Excellence Strategy EXC 2181/1 - 390900948 (the Heidelberg STRUCTURES Excellence Cluster), by DFG Grant No. BE 2795/4-1, and by the Bundesministerium für Bildung und Forschung [ (BMBF) German Federal Ministry of Education and Research] Grant No. 05P18VHFCA. 

M.P.H and the Gravity, Quantum Fields and Information group at AEI are generously supported by the Alexander von Humboldt Foundation and the Federal Ministry for Education and Research through the Sofja Kovalevskaja Award.

RV is supported by the U.S. Department of Energy, Office of Science, Office of Nuclear Physics, under Contract No. DE- SC0012704, and within the framework of the Beam Energy Scan Theory (BEST) DOE Topical Collaboration. He  also  acknowledges the Humboldt Foundation for its generous support through a Humboldt Prize,  ITP Heidelberg for their kind hospitality, and the DFG Collaborative Research Centre ``SFB 1225 (ISOQUANT)" for supporting his research collaboration with Heidelberg University. 

\end{acknowledgments}

\bibliography{master.bib,new_master.bib}

\end{document}